\documentclass[epj]{svjour}
\usepackage{graphicx}
\usepackage{amsmath}
\usepackage{amsfonts}
\usepackage{amssymb}
\usepackage{bm}

\newcommand{\gea}{\raisebox{-.3ex}{\small $ \
\stackrel{\textstyle >}{\sim} $ }}

\newcommand{\beq}{\begin{equation}}
\newcommand{\eeq}{\end{equation}}
\newcommand{\beqa}{\begin{eqnarray}}
\newcommand{\eeqa}{\end{eqnarray}}
\newcommand{\eq}[1]{eq.~(\ref{#1})}
\newcommand{\be}{\begin{eqnarray}}
\newcommand{\ee}{\end{eqnarray}}

\begin{document}
\title{Can chiral EFT give us satisfaction?}
\author{R. Machleidt\thanks{\email{machleid@uidaho.edu}} \and 
F. Sammarruca\thanks{\email{fsammarr@uidaho.edu}}
}                     
%
%
\institute{Department of Physics, University of Idaho, Moscow, Idaho 83844, USA}
\date{Received: date / Revised version: date}
%
\abstract{
We compare nuclear forces derived from chiral effective field theory (EFT)
with those obtained from traditional (phenomenological and meson) models.
By means of a careful analysis of paralleles and differences, we show that
chiral EFT is superior to all earlier approaches 
in terms of both formal aspects and successful applications in
{\it ab initio} calculations. 
However, in spite of the considerable progress made possible by chiral EFT,
complete satisfaction cannot be claimed until outstanding 
problems---the renormalization issue being the most 
important one---are finally settled.
%
} 
\maketitle
%
%
\section{Introduction}
\label{sec_intro}

As the Editors of this Topical Issue point out in the Preface, the nuclear theory of the past was nothing but an omnium-gatherum of models. This is very unsatisfactory in view 
of the traditional goal of theoretical physics, namely, to develop theories that are reductionist, 
unifying, and fundamental. However, the gap between the jumble of nuclear models and 
the holy grail of theory is so wide that there is no hope to overcome it any time soon. This is where 
the notions of emergence and effectiveness (effective theories) enter the picture. They provide a
compromise as well as a more realistic aim. Beyond that, it may even be true that a field as
 complex as nuclear physics may, by its intrinsic nature, never be amenable to the ideals of the 
 epistomological purist. Thus, effective theories may represent the highest level of understanding that 
 we may ever be able to achieve for nuclear physics phenomena. In this spirit, the past quarter century
 has seen progress in nuclear theory in terms of the development of effective theories.
 The chaos of the models of the past has been redesigned and absorbed into the organized structures of effective (field) theories.
 Ideas and mechanisms already contained in some of those models are put on more
 fundamental grounds and arranged within the proper order that effective field theories (EFTs) typically provide. 
 Abandoning pure phenomenology and reordering valid ideas within a systematic scheme
 are the novel and progressive steps.
 
Nuclear theory has essentially two ingredients: nuclear forces and many-body methods/models.
This contribution will be about the nuclear force part of the story. We will explain how,
within an EFT, the plurality of past nuclear force models is replaced 
by a systematic scheme reflecting the essential phenomenology that the models tried to catch.
EFT discards unacceptable phenomenology and retains and reformulates the remainder in the framework of proper order---this order being characterized by symmetries and some form of systematic expansion
 based upon an appropriate scale.

The bottom line question will be: Is the EFT approach to nuclear forces more satisfying from the
theoretical point of view than the previous multitude of models? We will address this question at the end of this contribution.

To facilitate ease of understanding, we subdivide the flow of information into the following three historical eras:
\begin{itemize}
\item
Era I $(1935-1960)$:
``Fundamental'' theories for nuclear forces,
\item
Era II $(1960 - 2000)$:
Diverse nuclear force models,
\item
Era III ($1990 - {\rm today})$:
Chiral EFT of nuclear forces.
\end{itemize}
Era III overlaps with Era II, because the dawn of EFT occured during the dusk of intense model construction.

The irony in the history of the theory of nuclear forces is that, originally (during Era I),
the goal was the traditional one, namely, to pursue a unifying and fundamental (field) theory.
Meson theory offered to be the best candidate, but
could ultimately not satisfy the basic requirements for a valid field theory. That should have 
suggested, early on (already around 1960), to switch to the concept of EFT. However, this framework did not get established until the 1980's, although ideas that in modern language could be called EFT concepts were already advanced in the 1960's~\cite{Wei67,Wei68,Wei79,Wei97,Wei09}. The ultimate reason for the failure 
of meson theory is, of course, that the fundamental theory of strong interactions (QCD) involves quarks and gluons rather than nucleons and mesons. 
 
This paper is organized such that
further sections follow the above stated historical phases and
end with conclusions in sect.~\ref{sec_concl}.

\section{Era I (1935 -- 1960): ``Fundamental'' theories for nuclear forces}
\label{sec_erai}

  In 1935, the Japanese physicist Hideki Yukawa~\cite{Yuk35} suggested that nucleons would exchange particles between each other and this mechanism would create the nuclear force. Yukawa constructed his theory in analogy to the theory of the electromagnetic interaction where the exchange of a (massless) photon is the cause of the force. However, in the case of the nuclear force, Yukawa assumed that the ``force makers'' (which were eventually called ``mesons'') carry a mass equal to a fraction of the nucleon mass. This would limit the effect of the force to a finite range.
Similar to other theories that were floating around in the 1930's (like the Fermi-field theory~\cite{Fer34}), Yukawa's meson theory was originally meant to represent a unified field theory for all interactions in the
atomic nucleus (weak and strong, but not electromagnetic). 
But after about 1940, it was generally agreed that strong and and weak nuclear forces should be treated separately. 

Yukawa's proposal did not receive much attention
until the discovery of the muon in cosmic ray~\cite{NA37} in 1937 after which the interest in meson theory escalated.
In his first 1935 paper, Yukawa had envisioned a scalar field theory, but when the spin of the deuteron ruled that out, he contemplated vector fields~\cite{YS37}. Kemmer considered the whole variety of non-derivative couplings for spin-0 and spin-1 fields (scalar, pseudoscalar, vector, axial-vector, and tensor)~\cite{Kem38}. By the early 1940's, the pseudoscalar theory was gaining popularity, since it appeared
more suitable for the deuteron (quadrupole moment).
In 1947, a strongly interacting meson was found in cosmic ray~\cite{LOP47} and, in 1948, in the laboratory~\cite{GL48}: the isovector pseudoscalar pion with mass around 138 MeV. It appeared that, finally, the right quantum of strong interactions had been found.

Originally, the meson theory of nuclear forces was perceived as a fundamental relativistic quantum field theory (QFT), similar to quantum electrodynamics (QED), the exemplary QFT that was so successful.
In this spirit, much effort was devoted to pion field theories in the early 
1950's~\cite{TMO52,BW53,Mar52,SBH55,BH55,Mor63}. Ultimately, all of these meson QFTs failed.
 In retrospect, they would have been replaced anyhow, because mesons and nucleons are not elementary particles and QCD is the correct QFT of strong interactions. However, the meson field concept failed long before QCD was proposed since, even 
 when considering mesons as elementary, 
the theory was beset with problems that could not be resolved. 
Assuming the renormalizable pseudoscalar ($\gamma_5$) coupling between pions and nucleons, 
large virtual pair terms emerged from the theory, but were not confirmed experimentally in
pion-nucleon ($\pi N$) or nucleon-nucleon ($NN$) scattering. Using the pseudo-vector or derivative coupling ($\gamma_5 \gamma^\mu \partial_\mu$), these pair terms were suppressed, but this type of coupling was not renormalizable~\cite{SBH55}.
Moreover, the large coupling constant ($g^2_\pi/4\pi \approx 14$) made perturbation theory unsuitable. Last not least, the pion-exchange potential contained unmanageable singularities at short distances.

The above problems led the theorists of the time to abandon quantum field theories for the strong interaction. Instead, $S$-matrix and dispersion theories became
popular, since thay do not start from a Lagrangian.

Ideas which, in today's terminology, would be characterized as EFT inspired,
emerged as early as 1967. Weinberg showed that the results of current algebra
could be reproduced by starting from a  suitable ``phenomenological'' Langrangian 
and evaluating Feynman diagrams at tree level~\cite{Wei67,Wei68}.
At first, this was not taken very seriously as a dynamic theory, because the derivative coupling
contained in that Lagrangian was not renormalizable, such that it did not seem possible to go beyond tree level.
Only about a decade later~\cite{Wei79}, it was realized that, if the Langrangian includes all terms consistent with the assumed symmetries, there will always be a counter 
term to renormalize the result at the given order~\cite{Wei79}. In this sense, ``Non-renormalizable
theories, ..., are just as renormalizable as renormalizable theories.''~\cite{Wei09}

After Weinberg's 1979 paper~\cite{Wei79}, the EFT approach to $\pi\pi$ and $\pi N$
scattering was picked up by Gasser, Leutwyler, and others~\cite{GL84,GSS88}.
Finally, around 1990 and, again, upon the initiative by Weinberg, also the $NN$ problem was
considered in terms of an EFT~\cite{Wei90,Wei91,Wei92,ORK94,ORK96}.

Thus, for the above rather intriguing reasons, we are faced with a large gap in the history of truly fundamental approaches to nuclear forces that lasted from about 1960 to 1990.
Instead, during this time, a plurality of models were proposed.

\section{Era II (1960 -- 2000): Diverse nuclear force models}
\label{sec_eraii}

To bring some order into the large variety of nuclear force models developed during this period, it is convenient to distinguish between pure phenomenology and meson models. However, it should be noted that, because the importance of the one-pion exchange (1PE) was recognized as early as 1956~\cite{Sup56}, 
all nuclear force models include 1PE  to describe the long-range part of the nuclear force
since around 1960.
Thus, the difference between the various models has essentially to do with how they describe the intermediate and short range parts of the force.

\subsection{Purely phenomenological $NN$ potentials}
\label{sec_phen}

The first semi-quantitative phenomenological $NN$ potential is the one by Gammel and Thaler~\cite{GT57}, which was followed by the more quantitative Hamada-Johnston~\cite{HJ62} and Reid~\cite{Rei68} potentials. To describe the short-range, the former two potentials apply a repulsive (infinite) hard core, while the Reid potential comes in hard- and soft-core versions. The Hamada-Johnston potential was used frequently in the 1960's, and the Reid soft-core potential became the most popular potential
of the 1970's. The construction of purely phenomenological $NN$ potentials continued through the 1980's and 90's with 
the Urbana  $v_{14}$ (UV14)~\cite{LP81},
the Argonne  $v_{14}$ (AV14)~\cite{WSA84},
and the Argonne $v_{18}$ (AV18)~\cite{WSS95}
potentials.

\subsection{Meson models for the $NN$ interaction}
\label{sec_mes}

Around 1960, rich phenomenlogical knowledge about the $NN$ interaction had accumulated
due to systematic measurements of $NN$ observables~\cite{Cha57} and  advances in phase shift analysis~\cite{SYM57}.  Clear evidence for 
a repulsive core and a strong spin-orbit force emerged~\cite{GT57}. This lead Sakurai~\cite{Sak60} and Breit~\cite{Bre60} to postulate the existence of a neutral vector meson ($\omega$ meson), which would create both these features.
 Furthermore, Nambu~\cite{Nam57} and Frazer and Fulco~\cite{FF59,FF60} showed that
the $\omega$ meson and  a 2$\pi$ $P$-wave resonance ($\rho$ meson)
would explain the electromagnetic structure of the nucleons.
Soon after these predictions, heavier (non-strange) mesons were found in experiment, notably the vector (spin-1) mesons
$\rho(770)$ and $\omega(782)$~\cite{Mag61,Erw61}.
It then became fashionable to add these newly discovered mesons to 
the meson theory of the $NN$ interaction.
However, to avoid the problems with multi-meson exchanges and higher order corrections encountered during Era~I, 
the models only allowed for single exchanges of
heavy mesons (i.~e., lowest order). In addition, one would multiply the meson-nucleon vertices with form factors (``cutoffs'') to remove the singularities at short distances.
Clearly, these were models motivated by the meson-exchange idea---not QFT.
The simpler versions of this kind became known as the one-boson-exchange (OBE) models, which were started in the early 1960's~\cite{BS64,BS67,BS69}\footnote{The research devoted to the $NN$ interaction during the 1960's
has been thoroughly reviewed by Moravcsik~\cite{Mor72}.} 
and turned out to be very successful in describing the phenomenology of the $NN$ interaction.
Their popularity extended all the way into the 1990's~\cite{Sto94,MSS96,Mac01}.

\begin{figure}[t]\centering
\scalebox{0.32}{\includegraphics{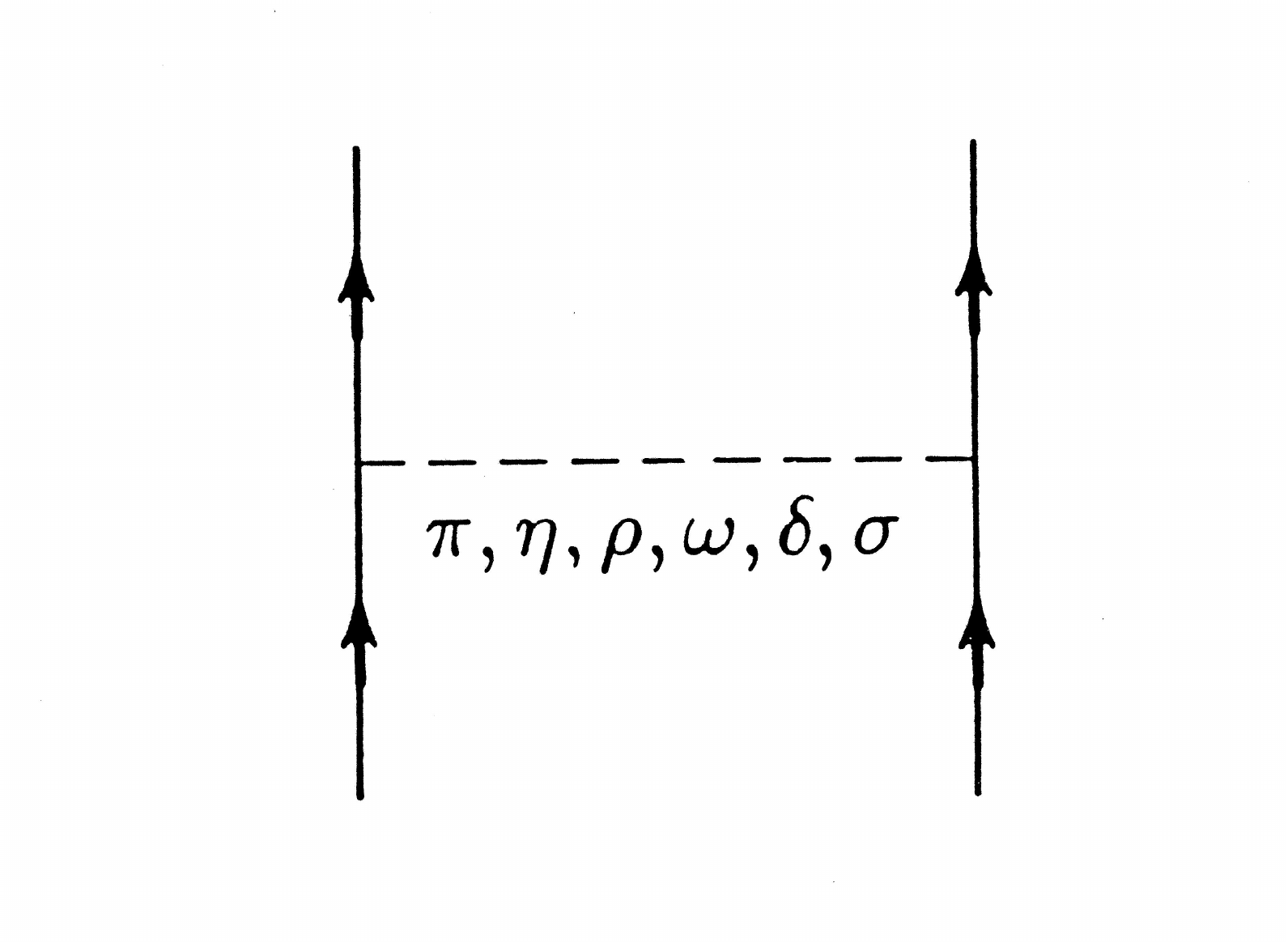}}
\vspace*{-0.5cm}
\caption{The one-boson-exchange model for the $NN$ interaction.
Solid lines denote nucleons and the dashed line represents mesons.
The $\delta$ signifies the $a_0$ meson.}
\label{fig_obep}
\end{figure}

\subsubsection{The One-Boson-Exchange Model}
\label{sec_OBE}

A typical one-boson-exchange model includes
about half a dozen of bosons with masses up to about 1 GeV, fig.~\ref{fig_obep}.
Not all mesons are equally important. The 
leading actors are the following four particles: 
\begin{itemize}
\item The pseudoscalar pion with a mass of about 138 MeV and isospin $I=1$ (isovector).
It is the lightest meson and provides the long-range part of the potential 
and most of the tensor force.
\item The isovector $\rho$ meson, a 2$\pi$ $P$-wave resonance of about 770 MeV.
Its major effect is to cut down the pion
tensor force at short range.
\item The isoscalar $\omega$ meson, a 3$\pi$ resonance of 782 MeV and spin 1.
It creates a strong repulsive central 
force of short range (`repulsive core') and the nuclear spin-orbit force.
\item The scalar-isoscalar $f_0(500)$ or $\sigma$ boson with a mass around 500 MeV. It
 provides the crucial intermediate 
range attraction necessary for nuclear binding.
Its interpretation as a particle is controversial~\cite{PDG}.
It may also be viewed  
as a simulation of effects due to correlated $S$-wave 2$\pi$-exchange.
\end{itemize}
Obviously,
just these four
mesons can produce the major properties of 
the nuclear force.\footnote{The
interested reader can find a pedagogical
introduction into the OBE model in
sections~3 and 4 of ref.~\cite{Mac89}.}
 
 Classic examples for OBE potentials (OBEPs) are the 
Bryan-Scott potentials started in the early 1960's~\cite{BS64,BS67,BS69}, but soon many other researchers got involved~\cite{Sup67,RMP67,NRS78}. 
Since it is suggestive to think of a potential as a function of $r$
(where $r$ denotes the distance between the centers of the two interacting
nucleons), the OBEPs of the 1960's where represented as 
local $r$-space potentials. Some groups continued to hold on to this tradition and, thus, 
the construction of improved $r$-space OBEPs continued well into the 1990's~\cite{Sto94}.
 
An important advance during the 1970's  was the development of the
{\it relativistic OBEP}~\cite{GTG71,Sch72,FT75,FT80,Erk74,HM75,HM76,BG79,GVH92}. In this model, the full, relativistic
Feynman amplitudes for the various one-boson-exchanges
are used to define the potential. These nonlocal expressions do not
pose any numerical problems  when used in momentum space and allow for a more quantitave descripton of $NN$ scattering.
The high-precision CD-Bonn~\cite{Mac01}
 as well as the Stadler-Gross~\cite{GS08} potentials are of this nature.

\begin{figure}[t]\centering
\scalebox{0.35}{\includegraphics{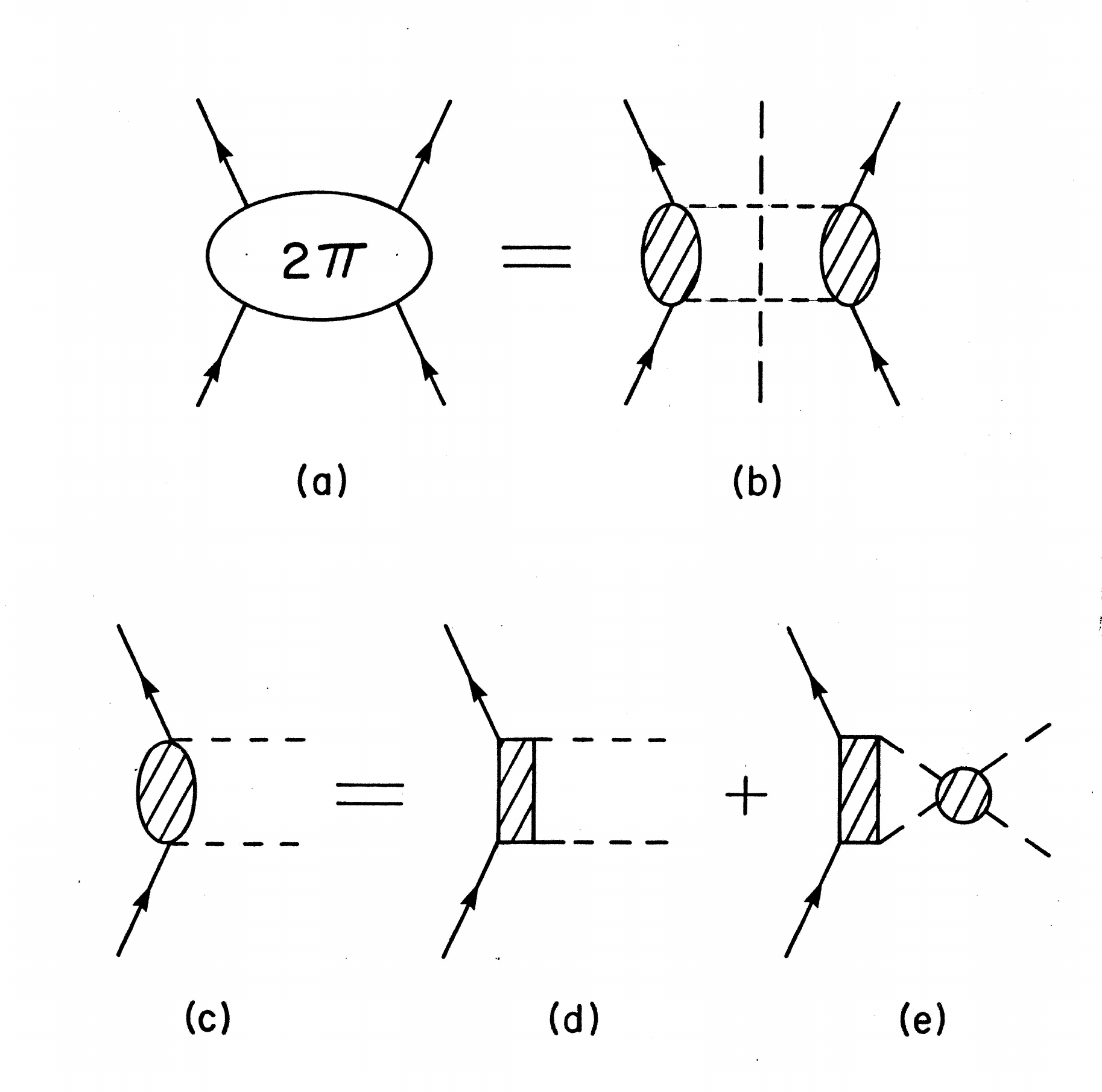}}
\vspace*{-0.2cm}
\caption{The $2\pi$-exchange contribution to the $NN$ interaction as viewed
by dispersion theory. Solid lines represent nucleons and dashed lines pions.
Further explanations are given in the text.}
\label{fig_disp}
\end{figure}

\subsubsection{Beyond the OBE approximation}
\label{sec_2pi}

Historically, one must understand that,
after the failure of the pion theories in the 1950's,
the OBE model was considered a great
success in the 1960's~\cite{Sup67,RMP67}.

On the other hand, one has to concede that
the OBE model is a great simplification of the complicated scenario of a full meson theory for the $NN$ interaction. Therefore, in spite of the
quantitative success of the OBEPs,
there should be concerns about the approximations involved in the model.
Major critical points include:
\begin{itemize}
\item
The scalar isoscalar $\sigma$ `meson' of about 500 MeV,
\item
neglecting all non-iterative diagrams,
\item
the role of meson-nucleon resonances.
\end{itemize}

Two pions, when `in the air', can interact strongly. When in a relative $P$-wave $(L=1)$, they form a proper resonance,
the $\rho$ meson. They can also interact in a relative $S$-wave $(L=0)$, which gives rise to the $\sigma$ boson.
Whether the $\sigma$ is a proper resonance is controversial, even though the Particle Data Group
lists an $f_0(500)$ or $\sigma(500)$ meson, but with a width of 400-700 MeV~\cite{PDG}. 
It is for sure that two pions have correlations,
and if one doesn't believe in the $\sigma$ as a two pion resonance, then one has to take these correlations into account.
There are essentially two approaches that have been used to calculate these 
two-pion exchange (2PE) contributions
to the $NN$ interaction (which generates the intermediate range attraction): dispersion theory and field theory.

\begin{figure}[t]\centering
\scalebox{0.42}{\includegraphics{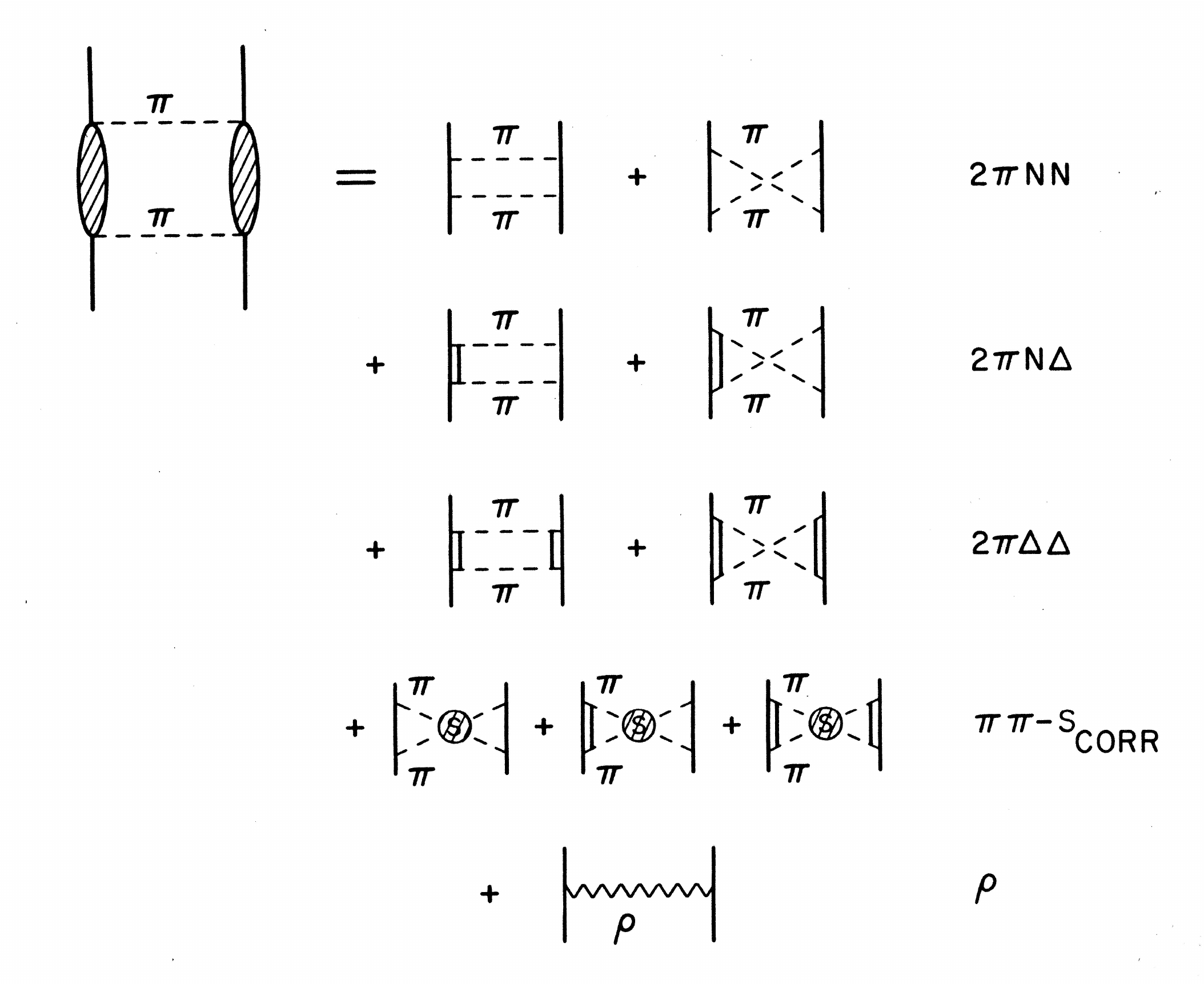}}
\vspace*{-0.2cm}
\caption{Field-theoretic model for the $2\pi$-exchange.
Notation as in fig.~\ref{fig_disp}.
Double lines represent isobars.
The hatched circles are $\pi\pi$ correlations.
Further explanations are given in the text.}
\label{fig_2pi}
\end{figure}

In the 1960's, dispersion theory was developed out of frustration with the failure
of a QFT for strong interactions in the 1950's~\cite{Mor63}.
In the dispersion-theoretic approach, the $NN$ amplitude is connected to the (empirical) $\pi N$ amplitude
by causality (analyticity), unitarity, and crossing symmetry.
Schematically this is shown in 
fig.~\ref{fig_disp}.
The full diagram (a)
is analysed in terms of two `halves' (b). The hatched ovals
stand for all possible  processes
which a pion and a nucleon
can undergo. This is made more explicit in (d) and (e).
The hatched boxes represent 
baryon intermediate states including the nucleon.
(Note that there are also
crossed pion exchanges which are not shown.)
The shaded circle
stands for $\pi \pi$ re-scattering.
Quantitatively, these processes are taken into account
by using  empirical
information from $\pi N$ and $\pi \pi$ scattering
(e.~g., phase shifts) which 
represents the input for such a calculation.
Dispersion relations then provide the on-shell $NN$ amplitude, which
--- with some kind of plausible prescription --- is represented as
a potential.
The Stony Brook~\cite{CDR72,JRV75,BJ76} and Paris~\cite{CV63,Cot73,Vin73} 
groups pursued this approach.
They could show that the intermediate-range part of the nuclear force
is, indeed, decribed fairly well by the $2 \pi$-exchange 
as obtained from dispersion
integrals.
To arrive at a complete potential, the $2\pi$-exchange contribution is
complemented by one-pion
and $\omega$ exchange. Besides this, the Paris potential~\cite{Lac80,Vin79}
 contains a 
phenomenological short-range part for $r< 1.5$ fm to improve the fit to the $NN$ data.
The Paris $NN$ potential was very popular during the 1980's.

\begin{figure}[t]\centering
\vspace*{0.45cm}
\scalebox{0.25}{\includegraphics{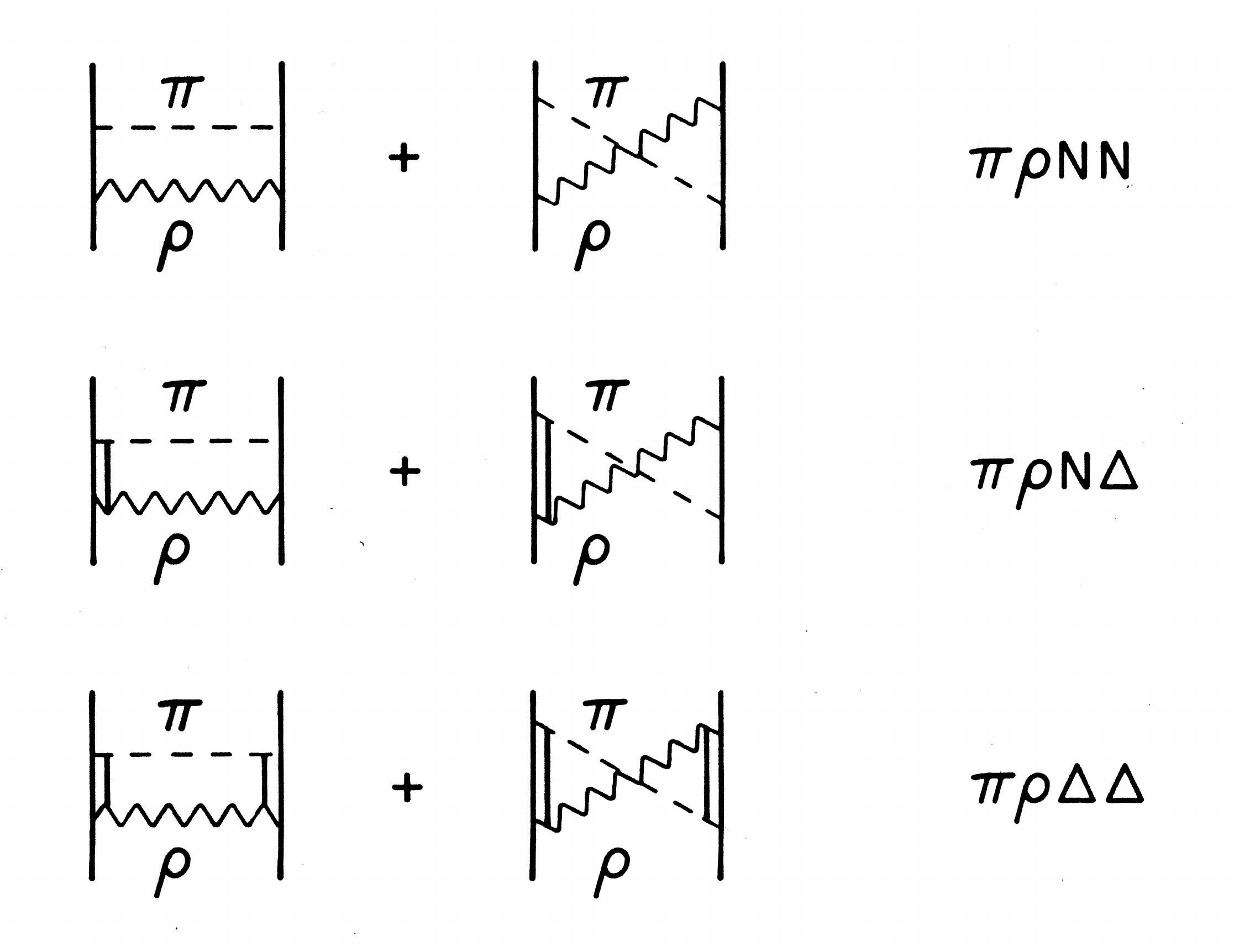}}
\vspace*{0.2cm}
\caption{$\pi\rho$ contributions to the $NN$ interaction.}
\label{fig_pirho}
\end{figure}

A first field-theoretic attempt towards the $2\pi$-exchange was undertaken by Lomon and Partovi~\cite{PL70}.
Later, the more elaborated model shown in fig.~\ref{fig_2pi} was developed by the Bonn group~\cite{MHE87}.
The model includes contributions from isobars as well as
from  $\pi \pi$ correlations.
This  can be understood
in analogy to the dispersion relations picture.
In general, only the lowest $\pi N$
resonance, the
so-called $\Delta$ isobar (spin 3/2, isospin 3/2, mass 1232 MeV),
is taken into account. The contributions from
other resonances have proven to be small for the low-energy $NN$
processes under consideration.
A field-theoretic model treats the
$\Delta$ isobar  as an elementary (Rarita-Schwinger) particle.
The six upper diagrams of fig.~\ref{fig_2pi}
represent uncorrelated $2\pi$
exchange. The crossed (non-iterative)
two-particle exchanges
(second diagram in each row) are important.
They guarantee the proper (very weak) isospin dependence
due to characteristic cancelations in the isospin dependent
parts of box and crossed box diagrams.
Moreover, their contribution is about as large as the one from
the corresponding box diagrams (iterative
diagrams); therefore, they are not negligible.
In addition to the processes discussed, also the correlated $2\pi$
exchange has to be included (lower two rows of 
fig.~\ref{fig_2pi}).
Quantitatively,
these contributions are about as sizable as those from the uncorrelated
processes.
Graphs with virtual pairs are left out, because
the pseudovector (gradient) coupling is used for the pion, in which case pair terms are small.

Besides the contributions from two pions, there are also contributions
from the combination of other mesons. The combination of $\pi$ and $\rho$
is particularly significant, fig.~\ref{fig_pirho}.
This contribution is repulsive and important
to suppress the 2$\pi$ exchange contribution at high momenta (or small distances), which is too strong by itself.

The Bonn Full Model~\cite{MHE87},
includes all the diagrams displayed in
figs.~\ref{fig_2pi} and \ref{fig_pirho}
plus single $\pi$ and $\omega$ exchanges.

Having highly sophisticated models at hand, like the Paris and the Bonn potentials, allows to check
the approximations made in the simple OBE model. As it turns out, the complicated 2$\pi$ exchange
contributions to the $NN$ interaction tamed by the $\pi\rho$ diagrams can well be simulated by the single scalar isoscalar boson, 
the $\sigma$, with a mass around 550 MeV. In retrospect, this fact provides justification for the simple OBE model.

The most important result of Era II is that
meson exchange is an excellent phenomenology for describing the $NN$ interaction.
It allows for the construction of very quantitative models. Therefore, 
the high-precision $NN$ potentials constructed in the mid-1990's are all
based upon meson phenomenology~\cite{Sto94,MSS96,Mac01}.
However, with the rise of QCD to
the ranks of the authoritative theory of strong interactions, 
it became more and more clear that
meson-exchange is to be seen as just a model.

\subsection{Models for nuclear many-body interactions}
\label{sec_mes3nf}

\begin{figure}[t]\centering
\scalebox{0.44}{\includegraphics{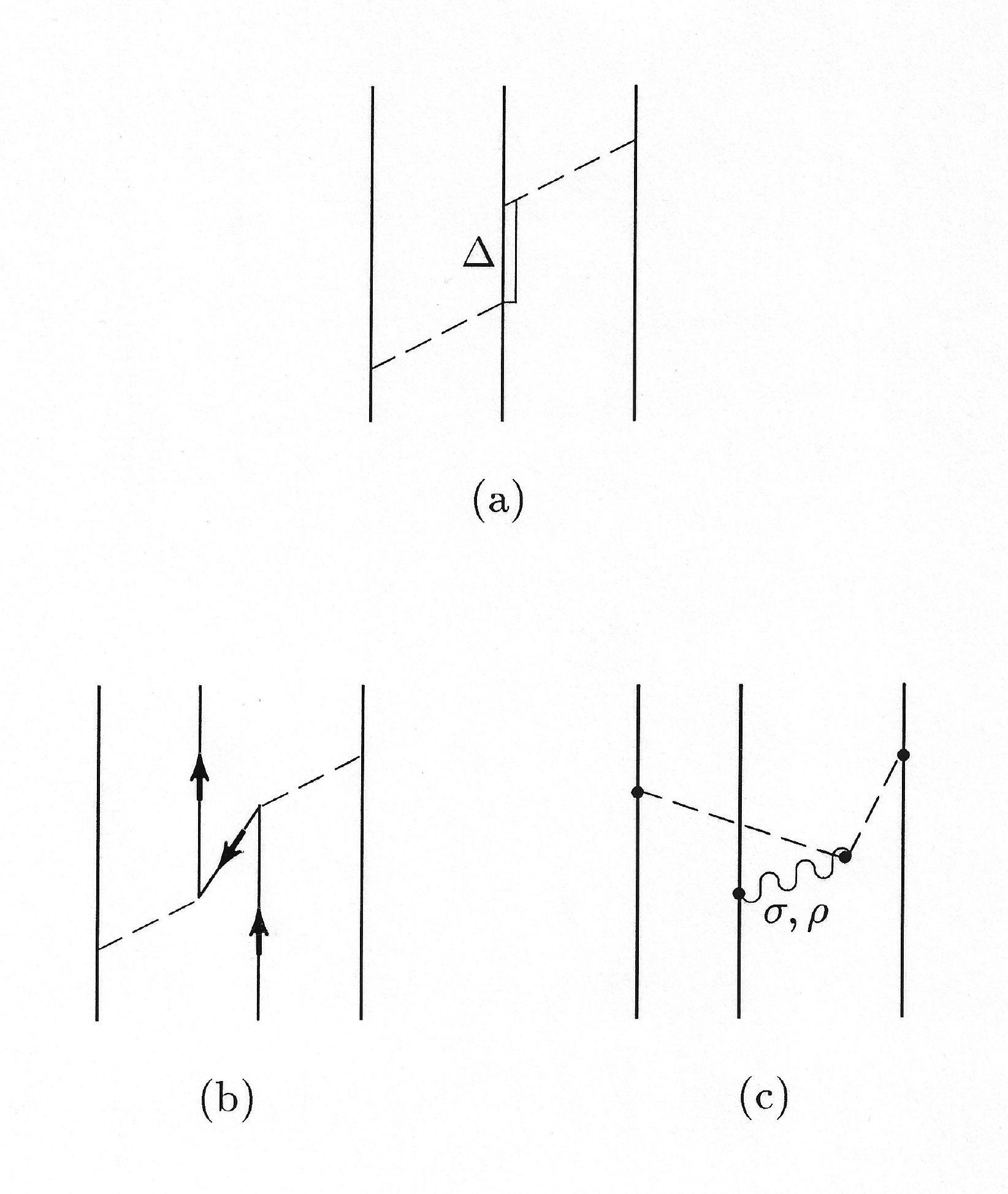}}
\vspace*{-0.5cm}
\caption{Diverse three-nucleon force diagrams. Solid lines (nucleons)
are upward directed unless otherwise noted. Dashed lines represent various mesons, as appropriate.}
\label{fig_3nf1}
\end{figure}

\begin{figure}[t]\centering
\scalebox{0.55}{\includegraphics{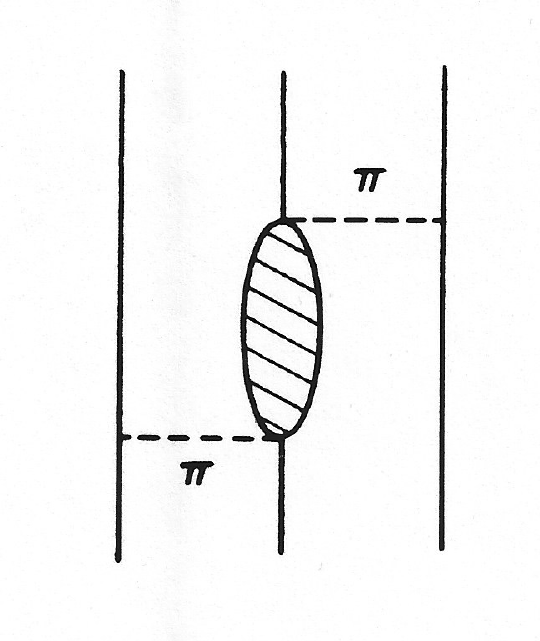}}
\caption{$2\pi$-exchange three-nucleon force of the Tucson-Melbourne type~\cite{CSB75,Coo79,CG81}.
The shaded oval represents the (off-shell) $\pi N$ amplitude with the forward propagating Born term subtracted.}
\label{fig_3nf2}
\end{figure}

Originally, it was hoped that the structure of finite nuclei
could be understood in terms of just the two-nucleon force 
(2NF)~\cite{Neg70}---if one would only find the ``right'' 2NF.
However, in the course of the 1970's, when more reliable microscopic calculations became available, growing evidence accumulated that showed that it was impossible
to saturate nuclear matter at the right energy and density when applying only 2NFs~\cite{Coe70,CDG72,Day83,Mac89}. Another problem was the triton binding energy, which was considerably underpredicted
with the 2NFs available at the time~\cite{BKT74,BSM77}.
These failures were interpreted as an indication for the need of nuclear many-body forces.
 
Strictly speaking, many-nucleon forces are an artefact of theory.
They are created by freezing out non-nucleonic degrees of freedom contained in the
full-fledged problem.
Examples are given in fig.~\ref{fig_3nf1}. In part (a) of the figure, the frozen
degree of freedom is a nucleon resonance (here: the $\Delta(1232)$ isobar with spin and isospin $\frac32$), in (b) an antinucleon, and in (c) meson resonances (here: the $\sigma$ and $\rho$ bosons). 

The oldest examples of three-body forces are those derived by
Primakoff and Holstein~\cite{PH39} in 1939. They arise from particle-antiparticle pairs, fig.~\ref{fig_3nf1}(b), which, in a nonrelativistic model, are represented by three-body potential terms. 
While these forces turned out to be negligible for atomic systems,
they were found to be sizable for nuclear systems~\cite{PH39}.
This fact is demonstrated in
the so-called Dirac-Brueckner approach to nuclear 
matter~\cite{BM84,BM90,HM87,Bro87,AT92,AS03,MSM17}, where
diagram \ref{fig_3nf1}(b) plays a crucial role.
Diagram (a) of fig.~\ref{fig_3nf1} was first considered by
Fujita and Miyazawa (FM)~\cite{FM57} in 1957, and
 both diagrams (a) and (b) were taken into account
 by the Catania group~\cite{Li08}.
 Diagram (c), with all mesons involved of the scalar type,
was evaluated by Barshay and Brown~\cite{BB75}
and found to be fairly large.

 For a reliable approach to three-nucleon forces (3NFs), two aspects need to be considered:
First, as far as possible, one should take into account all processes that may create a 3NF and, second, 
the strength of the contributions should be consistent with their size in other hadronic reactions.
In this context, it was noticed early on that the internal parts of  
fig.~\ref{fig_3nf1}(a)-(c)
are major contributions to pion-nucleon scattering~\cite{BGG68,Bro72}.
This suggests the idea to use the empirical $\pi N$  amplitude as starting point, where
 however the pions are on their mass shell.
On the other hand, within a 3NF diagram, the pions are virtual and space-like and, therefore,
the amplitude must be extrapolated
off mass-shell. In the work of refs.~\cite{CSB75,Coo79,CG81}, this
is accomplished by applying constraints based upon current algebra and partial conservation of axial current (PCAC). This approach has become known as the Tucson-Melbourne (TM) 3NF, shown
symbolically in fig.~\ref{fig_3nf2}. The shaded area in that figure contains everything that contributes to $\pi N$ scattering---except a positive-energy single-nucleon intermediate state to avoid double counting,
since the latter is automatically generated by the iteration of the 1PE two-nucleon force.

It is instructive to note that, for the 2PE 3NF, the same dichotomy exists as for the 2PE
contribution to the 2NF (cf.\ sect.~\ref{sec_2pi}).
The TM 3NF is based upon dispersion theory, while
the alternative, a field-theoretic approach based upon Langrangians, was pursued
by Robilotta and coworker~\cite{CDR83}, known as the Brazilian 3NF.

\begin{figure}[t]\centering
\scalebox{0.4}{\includegraphics{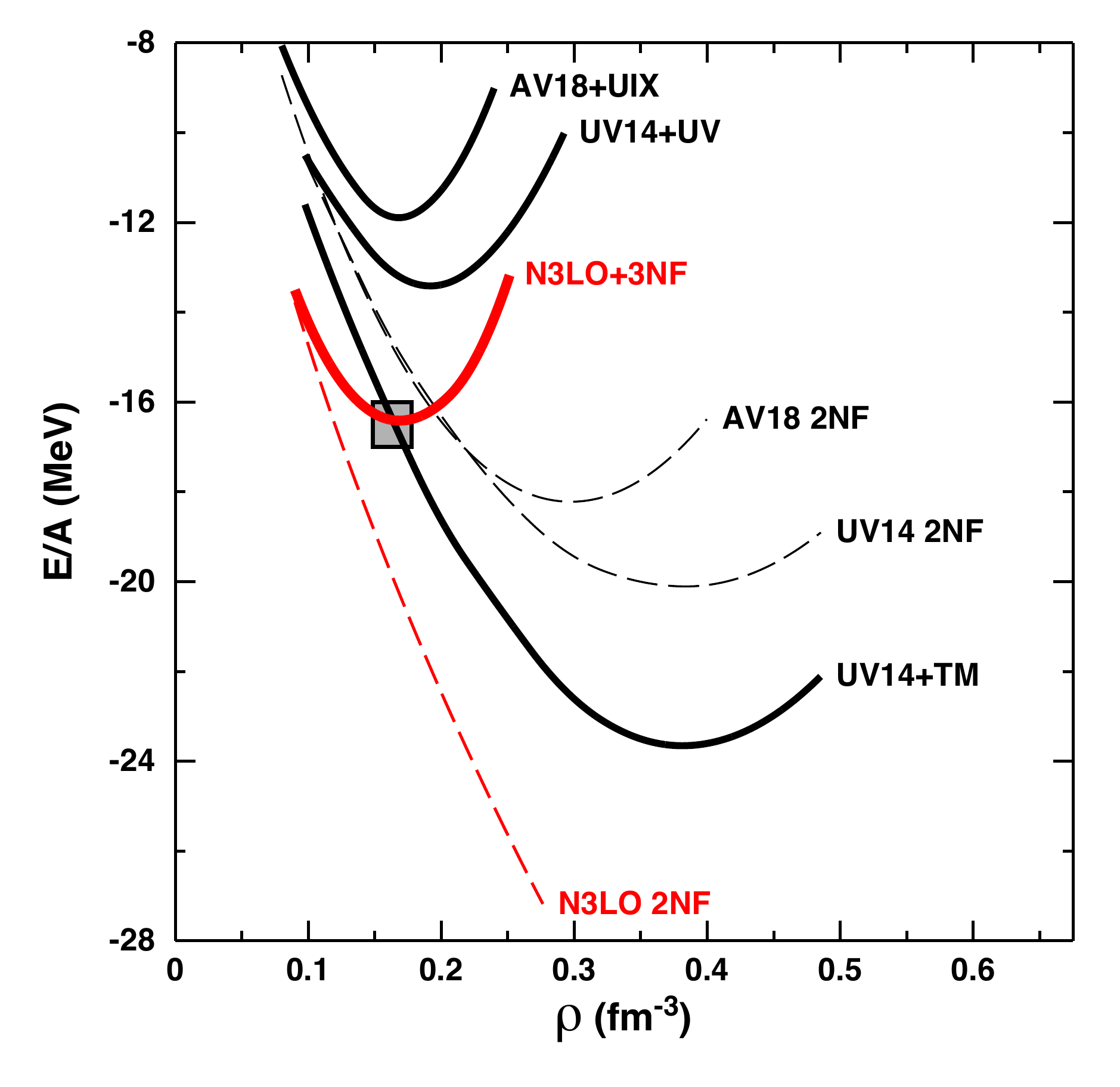}}
\caption{Energy per nucleon, $E/A$, as a function of density, $\rho$, of symmetric nuclear matter. Black dashed curves represent predictions applying purely phenomenologial 2NFs with UV14 referring to the Urbana $v_{14}$ 2NF~\cite{LP81} and AV18 signifying the
Argonne $v_{18}$ 2NF~\cite{WSS95}. Solid black curves include phenomenological 3NFs with TM
 denoting the Tucson-Melbourne~\cite{Coo79}, UV the Urbana V~\cite{CPW83},
 and UIX the Urbana IX~\cite{Pud95} 3NF models.
 Red curves depict results from chiral EFT based forces~\cite{EMN17,Sam18} 
 (sect.~\ref{sec_eraiii}). 
 The shaded box marks the area in which empirical nuclear matter saturation is presumed to occur.}
\label{fig_snm}
\end{figure}

\begin{figure}[t]\centering
\scalebox{1.1}{\includegraphics{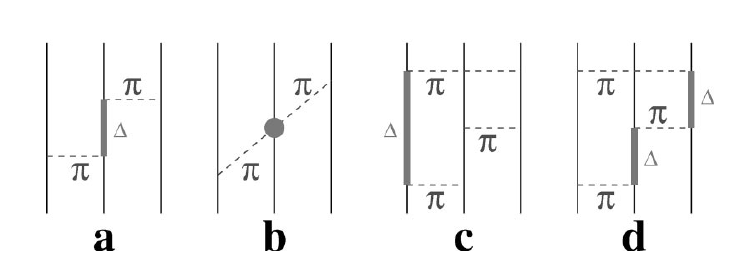}}
\caption{The three-body force Feynman diagrams that define the Illinois 3NF~\cite{Pie01}, with (a) the Fujita-Miyazawa, (b) the two-pion S-wave, and (c) and (d) three-pion exchange ring diagrams with one $\Delta$ isobar in intermediate states.
(Figure reproduced from ref.~\cite{Pie01} with permission.)}
\label{fig_3nf3}
\end{figure}

\begin{figure}[t]\centering
\scalebox{0.55}{\includegraphics{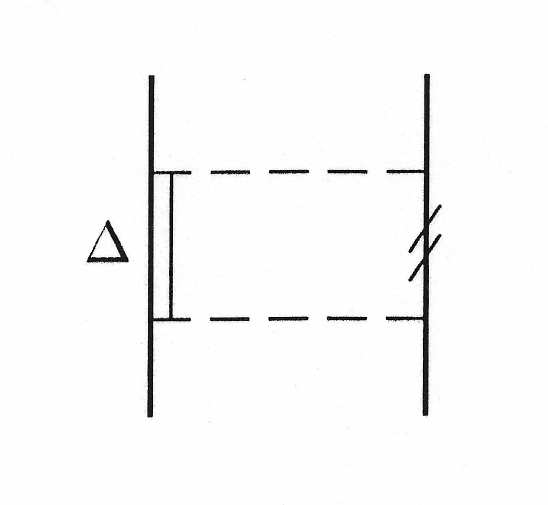}}
\caption{Two-meson exchange 2NF diagram involving one $\Delta$ isobar. The
double slash on the intermediate nucleon line indicates the change of the
propagator in the nuclear medium.}
\label{fig_3nf4}
\end{figure}

Moreover, in analogy to the $\pi\rho$ contributions to the 2NF (fig.~\ref{fig_pirho}), 
later work on the 3NF also included $\rho$-exchange in the diagram of fig.~\ref{fig_3nf2}~\cite{RI84,ECM85,CP93}.

The attraction provided by the FM or TM 2PE 3NFs has proven to be useful in explaining
the binding energies of light nuclei (particularly, $^3$H and $^4$He) which are, in general, 
underbound when only 2NFs are applied.
However, this added attraction leads to overbinding and too high a saturation density
in nuclear matter (cf.\ fig.~\ref{fig_snm}, curve labeled UV14+TM). Therefore, some groups added
a repulsive short-range 3NF which ameliorates the problem,
but does not solve it~\cite{Day83,CPW83} (fig.~\ref{fig_snm}, curve UV14+UV).
In the work by the Urbana group, 
many versions of such 3NF were developed with Urbana IX (UIX) 
being the most popular one---applied in
 light nuclei, nuclear matter (fig.~\ref{fig_snm}, curve AV18+UIX), and neutron matter~\cite{Pud95,Pud97,APR98}.
In later work~\cite{Pie01}, the Urbana group extended their model for the 3NF
by including the $2\pi$-exchange $S$-wave contribution (which according to
ref.~\cite{Fri88} can be sizable) plus three-pion exchange ring diagrams
with one $\Delta$ excitation (fig.~\ref{fig_3nf3}). The peculiar spin and isospin dependencies of $\Delta$-ring diagrams were found to be helpful in the explanation of spectra of light nuclei. This has become known as the Illinois 3NFs~\cite{Pie01}, which so far have
evolved up to Illinois-7 (IL7)~\cite{Pie08}.

 The 3NFs of the Urbana type, adjusted to the ground
state and the spectra of light nuclei, do not saturate nuclear matter
properly~\cite{CPW83,APR98}
(fig.~\ref{fig_snm})\footnote{In fact, the only miroscopic approach developed during Era II that was
able to explain nuclear matter saturation was the relativistic Dirac-Brueckner-Hartree-Fock (DBHF)
method~\cite{BM84,BM90,HM87,Bro87,AT92,AS03,MSM17}. However, the DBHF approach
is unfit to account for the properties of finite nuclei across the nuclear chart.
Alternatively, relativistic~\cite{SW84,Rei89} and nonrelativistic~\cite{Bro98}
mean-field models were constructed to improve the description of nuclear matter and finite nuclei.
However, these models are not {\it ab initio} and, therefore, not a topic of this article.}
 and severely underbind intermediate-mass nuclei~\cite{Lon17}.
The AV18 2NF plus IL7 3NF yield a pathological equation of state
of pure neutron matter~\cite{Mar13a}.
In addition, the so-called $A_y$ puzzle of nucleon-deuteron scattering~\cite{EMW02}
is not resolved by any of the phenomenological 3NFs~\cite{Kie10}.

Last but not least, we also mention that models do exist where the degree of freedom
responsible for the generation of 3NF-like contributions is not frozen out. 
In the work of Amorim and Tjon~\cite{AT92}, the antinucleon degree of freedom is 
treated consistently in a nuclear matter calculation.
Triton calculations in which the nucleon and the $\Delta$ isobar are regarded on an equal footing
have been performed by the Hannover 
group~\cite{HSS83,HSY83,SHS83,DMS03} and 
Picklesimer and coworkers~\cite{PRB92,PRB95}.
In such coupled systems, diagrams of the type displayed in fig.~\ref{fig_3nf3}a, c, and d
(and many more which also include two and three $\Delta$'s)
are generated automatically.
A consistent treatment of the $\Delta$ isobar also affects the two-nucleon force when inserted into the nuclear medium, as indicated in fig.~\ref{fig_3nf4}. This effect is 
repulsive~\cite{HM77,MH80}
and essentially cancels the attraction produced by the
$\Delta$-induced 3NF contributions,
leaving basically no net effect~\cite{Mac89,PRB92,PRB95}.

The bottom line of Era II is that while excellent results were obtained for phenomenological
and meson-theoretic  2NFs,
the 3NFs of the Era did rather poorely.

\section{Era III (1990 -- today): Chiral EFT of nuclear forces}
\label{sec_eraiii}

Quantum chromodynamics (QCD) is the theory of strong interactions.
It deals with quarks, gluons and their interactions and is
part of the Standard Model of Particle Physics.
QCD is a non-Abelian gauge field theory
with color $SU(3)$ as the underlying gauge group.
The non-Abelian nature of the theory has dramatic
consequences. While 
the interaction between colored objects is weak 
at short distances or high momentum transfer
(``asymptotic freedom''),
it is strong at long distances ($\gea 1$ fm) or low energies,
leading to the confinement of quarks into colorless
objects, the hadrons. Consequently, QCD allows for a 
perturbative analysis at high energies, whereas it is
highly non-perturbative in the low-energy regime,
making analytic solutions difficult.
Nuclear physics resides at low energies and
deals with nucleons and mesons, rather than quarks and gluons.
This scenario calls for an EFT, for which
the following steps need to be taken:\footnote{For a more detailed introduction into (chiral) EFT,
see e.g.\ refs.~\cite{ME11,EHM09,HKK19,Dri19}.}
\begin{enumerate}
\item
Identify the low- and high-energy scales.
 \item
 Identify the degrees of freedom active at the low-energy scale.      
\item
Recognize the relevant symmetries and their breakings.                         
\item
Build the most general Lagrangian consistent with those
(broken) symmetries.
\item
Organize an expansion in terms of low over high: Power counting.
\item
Guided by this expansion, evaluate Feynman diagrams for the problem under consideration
to the desired accuracy.
\end{enumerate}

Concerning scales, the large difference between the masses of
the pions and the masses of the vector mesons, like $\rho(770)$ and $\omega(782)$, provides a clue.
From that observation, one is prompted to take the pion mass as the identifier of the soft scale, 
$Q \sim m_\pi$,
while the rho-meson mass sets the hard scale, $\Lambda_\chi \sim m_\rho\sim 1$ GeV, often referred to 
as the chiral-symmetry breaking scale.
The expansion will then be in terms of                                            
$Q/\Lambda_\chi$.

\subsection{Effective Lagrangians}
\label{sec_lagr}

An important approximate symmetry of low-energy QCD is chiral symmetry, because the up and down quarks 
are almost massless. However, this symmetry is spontaneously broken.
The degrees of freedom relevant to nuclear physics are 
pions (the Goldstone bosons of the spontanously broken symmetry) and nucleons.
Since the interactions of Goldstone bosons must
vanish at zero momentum transfer and in the chiral
limit ($m_\pi \rightarrow 0$), the low-energy expansion
of the effective Lagrangian is arranged in powers of derivatives
and pion masses.
This effective Lagrangian is subdivided into the following pieces,
\begin{equation}
{\cal L}_{\rm eff}
=
{\cal L}_{\pi\pi} 
+
{\cal L}_{\pi N} 
+
{\cal L}_{NN} 
 + \, \ldots \,,
\end{equation}
where ${\cal L}_{\pi\pi}$
deals with the dynamics among pions, 
${\cal L}_{\pi N}$ 
describes the interaction
between pions and a nucleon,
and ${\cal L}_{NN}$  contains two-nucleon contact interactions
which consist of four nucleon-fields (four nucleon legs) and no
meson fields.
The ellipsis stands for terms that involve two nucleons plus
pions and three or more
nucleons with or without pions, relevant for nuclear
many-body forces.
The individual Lagrangians are organized in terms of increasing orders:
\begin{eqnarray}
{\cal L}_{\pi\pi} 
 & = &
{\cal L}_{\pi\pi}^{(2)} 
+
{\cal L}_{\pi\pi}^{(4)}
 + \ldots \,, \\
 \label{eq_LpiN}
{\cal L}_{\pi N} 
 & = &
{\cal L}_{\pi N}^{(1)} 
+
{\cal L}_{\pi N}^{(2)} 
+
{\cal L}_{\pi N}^{(3)} 
+
{\cal L}_{\pi N}^{(4)} 
+ \ldots , \\
\label{eq_LNN}
{\cal L}_{NN} &  = &
{\cal L}^{(0)}_{NN} +
{\cal L}^{(2)}_{NN} +
{\cal L}^{(4)}_{NN} + 
\ldots \,,
\end{eqnarray}
where the superscript refers to the number of derivatives or 
pion mass insertions (chiral dimension)
and the ellipses stand for terms of higher dimensions.
In the few-nucleon sector,
it is customary to use the heavy-baryon formulation of the Lagrangians, the
explicit expressions of which can be found in refs.~\cite{ME11,KGE12}.

\subsection{Power counting}
\label{sec_power}
             
An infinite number of Feynman diagrams can be evaluated from 
the effective Langrangians and so one needs to be able to organize these
diagrams in order of their importance. 
Chiral perturbation theory (ChPT) provides such organizational scheme. 

In ChPT, 
graphs are analyzed
in terms of powers of small external momenta over the large scale:
$(Q/\Lambda_\chi)^\nu$,
where $Q$ is generic for a momentum (nucleon three-momentum or
pion four-momentum) or the pion mass, and $\Lambda_\chi \sim 1$~GeV
is the chiral symmetry breaking scale (hadronic scale, hard scale).
Determining the power $\nu$ 
has become known as power counting.

For the moment, we will consider only so-called irreducible
graphs.
By definition, an irreducible graph is a diagram that
cannot be separated into two
by cutting only nucleon lines.
Following the Feynman rules of covariant perturbation theory,
a nucleon propagator carries the dimension $Q^{-1}$,
a pion propagator $Q^{-2}$,
each derivative in any interaction is $Q$,
and each four-momentum integration $Q^4$.
This is known as naive dimensional analysis or Weinberg counting~\cite{Wei91}.
Applying some topological identities, one obtains
for the power of an irreducible diagram
involving $A$ nucleons~\cite{Wei91,ME11}
\begin{equation} \nu = -2 +2A - 2C + 2L 
+ \sum_i \Delta_i \, ,
\label{eq_nu} 
\end{equation}
with
\begin{equation}
\Delta_i  \equiv   d_i + \frac{n_i}{2} - 2  \, .
\label{eq_Deltai}
\end{equation}
In the two equations above: for each
vertex $i$, $C$ represents the number of individually connected parts of the diagram while
$L$ is the number of loops;                  
$d_i$ indicates how many derivatives or pion masses are present 
and $n_i$ the number of nucleon fields.                  
The summation extends over all vertices present in that particular diagram.
Notice also that chiral symmetry implies $\Delta_i \geq 0$. 
Interactions among pions have at least two derivatives
($d_i\geq 2, n_i=0$), while 
interactions between pions and a nucleon have one or more 
derivatives  
($d_i\geq 1, n_i=2$). Finally, pure contact interactions
among nucleons ($n_i=4$)
have $d_i\geq0$.
In this way, a low-momentum expansion based on chiral symmetry 
can be constructed.                   

Naturally,                                            
the powers must be bounded from below for the expansion
to converge. This is in fact the case, 
with $\nu \geq 0$. 

Furthermore, the power formula 
eq.~(\ref{eq_nu}) 
allows to predict
the leading orders of connected multi-nucleon forces.
Consider a $m$-nucleon irreducibly connected diagram
($m$-nucleon force) in an $A$-nucleon system ($m\leq A$).
The number of separately connected pieces is
$C=A-m+1$. Inserting this into
eq.~(\ref{eq_nu}) together with $L=0$ and 
$\sum_i \Delta_i=0$ yields
$\nu=2m-4$. Thus, two-nucleon forces ($m=2$) appear
at $\nu=0$, three-nucleon forces ($m=3$) at
$\nu=2$ (but they happen to cancel at that order),
and four-nucleon forces at $\nu=4$ (they don't cancel).
More about this in sect.~\ref{sec_manyNF}.

For later purposes, we note that, for an irreducible 
$NN$ diagram ($A=2$, $C=1$), the
power formula collapses to the very simple expression
\begin{equation}
\nu =  2L + \sum_i \Delta_i \,.
\label{eq_nunn}
\end{equation}

To summarize, at each order                            
$\nu$ we only have a well defined number of diagrams, 
which renders the theory feasible from a practical standpoint.
The magnitude of what has been left out at order $\nu$ can be estimated (in a 
crude way) from 
$(Q/\Lambda_\chi)^{\nu+1}$. The ability to calculate observables (in 
principle) to any degree of accuracy gives the theory 
its predictive power.

\subsection{The long- and intermediate-range $NN$ potential}
\label{sec_long}

The long- and intermediate-range parts of the $NN$ potential are built up from pion exchanges,
which are ruled by chiral symmetry.
The various pion-exchange contributions may be analyzed
according to the number of pions being exchanged between the two
nucleons:
\begin{equation}
V_\pi = V_{1\pi} + V_{2\pi} + V_{3\pi} +  V_{4\pi} + \ldots \,,
\end{equation}
where the meaning of the subscripts is obvious
and the ellipsis represents $5\pi$ and higher pion exchanges. For each of the above terms, 
we have a low-momentum expansion:
\begin{eqnarray}
V_{1\pi} & = & V_{1\pi}^{(0)} + V_{1\pi}^{(2)} 
+ V_{1\pi}^{(3)} + V_{1\pi}^{(4)} + V_{1\pi}^{(5)} + V_{1\pi}^{(6)} + \ldots 
\label{eq_1pe_orders}
\\
V_{2\pi} & = & V_{2\pi}^{(2)} + V_{2\pi}^{(3)} + V_{2\pi}^{(4)} + V_{2\pi}^{(5)} + V_{2\pi}^{(6)} 
+ \ldots \\
V_{3\pi} & = & V_{3\pi}^{(4)} + V_{3\pi}^{(5)} + V_{3\pi}^{(6)} + \ldots \\
V_{4\pi} & = &  V_{4\pi}^{(6)} + \ldots \,,
\end{eqnarray}
where the superscript denotes the order $\nu$ of the expansion.

At leading order (LO, $\nu=0$), only one-pion exchange (1PE) contributes
which is given by
\begin{equation}
V_{1\pi}^{\rm(0)} ({\vec p}', \vec p) = - 
\frac{g_A^2}{4f_\pi^2}
\: 
\bm{\tau}_1 \cdot \bm{\tau}_2 
\:
\frac{
\vec \sigma_1 \cdot \vec q \,\, \vec \sigma_2 \cdot \vec q}
{q^2 + m_\pi^2} 
\,,
\label{eq_1PEci}
\end{equation}
where ${\vec p}'$ and $\vec p$ denote the final and initial nucleon momenta in the 
center-of-mass system, 
respectively. Moreover, $\vec q = {\vec p}' - \vec p$ is the momentum transfer, 
 and $\vec \sigma_{1,2}$ and $\bm{\tau}_{1,2}$ are the spin 
and isospin operators of nucleon 1 and 2, respectively. The parameters
$g_A$, $f_\pi$, and $m_\pi$ denote the axial-vector coupling constant,
pion-decay constant, and the pion mass, respectively.
Commonly used values are 
$g_A=1.29$,
$f_\pi=92.4$ MeV, and
the average pion mass 
$m_\pi= 138$ MeV.
Higher order corrections to the 1PE  are taken care of by  mass
and coupling constant renormalizations. Note also that, on 
shell, there are no relativistic corrections. Thus, the 1PE,
\eq{eq_1PEci}, applies through all orders
or, in other words, 
$V_{1\pi}  =  V_{1\pi}^{(0)}$.

It is customary to take the charge-dependence of the 1PE due to pion-mass splitting into account,
because it is considerable. Thus, for proton-proton ($pp$) and neutron-neutron ($nn$) scattering, one actually uses
\begin{equation}
V_{1\pi}^{(pp)} ({\vec p}', \vec p) =
V_{1\pi}^{(nn)} ({\vec p}', \vec p) 
= V_{1\pi} (m_{\pi^0})
\,,
\label{eq_1pepp}
\end{equation}
and for neutron-proton ($np$) scattering,
one applies
\begin{equation}
V_{1\pi}^{(np)} ({\vec p}', \vec p) 
= -V_{1\pi} (m_{\pi^0}) + (-1)^{I+1}\, 2\, V_{1\pi} (m_{\pi^\pm})
\,,
\label{eq_1penp}
\end{equation}
where $I=0,1$ denotes the total isospin of the two-nucleon system and
\begin{equation}
V_{1\pi} (m_\pi) \equiv - \,
\frac{g_A^2}{4f_\pi^2} \,
\frac{
\vec \sigma_1 \cdot \vec q \,\, \vec \sigma_2 \cdot \vec q}
{q^2 + m_\pi^2} 
\,,
\end{equation}
with $m_{\pi^0}$ and $m_{\pi^\pm}$ signifying the masses of the neutral and charged pions, respectively.

Two-pion exchange
starts at next-to-leading order (NLO, $\nu=2$), because it involves at least one loop
[cf.\ eq.~(\ref{eq_nunn})], and continues through all higher orders. Each additional pion-exhange requires at least one more loop. Thus,
three-pion exchange (3PE) starts at 
next-to-next-to-next-to-leading order
(N$^3$LO, $\nu=4$) and
four-pion exchange (4PE) at 
N$^5$LO ($\nu=6$).
With every order, the number of diagrams increases dramatically and so do
the mathematical formulas representing them.
A complete collection of all diagrams and formulas concerning the
2PE and 3PE contributions through all orders from 
NLO to N$^5$LO can be found in refs.~\cite{Ent15a,Ent15b}.

\begin{table}
\caption{The $\pi N$ LECs as determined in
the Roy-Steiner-equation analysis of $\pi N$ scattering conducted in  ref.~\cite{Hof15,Hof16}.
The $c_i$, $\bar{d}_i$, 
and $\bar{e}_i$ are the LECs of the second, third, and fourth order $\pi N$ Lagrangian~\cite{KGE12} and are 
 in units of GeV$^{-1}$, GeV$^{-2}$, and GeV$^{-3}$, respectively.
The uncertainties in the last digits are given in parentheses after the values.}
\label{tab_lecs}
\smallskip
\begin{tabular}{crrr}
\hline 
\hline 
\noalign{\smallskip}
              & NNLO & N$^3$LO & N$^4$LO \\
\hline
\noalign{\smallskip}
$c_1$ & --0.74(2) & --1.07(2) & --1.10(3) \\
$c_2$ & ---  & 3.20(3) & 3.57(4) \\
$c_3$ & --3.61(5) & --5.32(5) & --5.54(6) \\
$c_4$ & 2.44(3) & 3.56(3) & 4.17(4) \\
$\bar{d}_1 + \bar{d}_2$ & --- & 1.04(6) & 6.18(8) \\
$\bar{d}_3$ & --- & --0.48(2) & --8.91(9) \\
$\bar{d}_5$ & --- & 0.14(5) & 0.86(5) \\
$\bar{d}_{14} - \bar{d}_{15}$ & --- & --1.90(6) & --12.18(12) \\
$\bar{e}_{14}$ & --- & --- & 1.18(4) \\
$\bar{e}_{17}$ & --- & --- & --0.18(6) \\
\hline
\hline
\noalign{\smallskip}
\end{tabular}
\end{table}

As obvious from figs.~\ref{fig_disp} and \ref{fig_3nf2}, there
is a close connection between the dynamics in the $\pi N$, the $NN$, and the $3N$-systems.
In the approach based upon chiral Lagrangians, the connection is made
by starting from the same Lagrangians in the evaluation of the different processes and applying the
same low-energy constants. 
Therefore, the $\pi N$ LECs as 
determined in $\pi N$ analysis are used in $NN$ (and $3N$).
In table~\ref{tab_lecs}, we show the very accurate values as extracted
in the Roy-Steiner equations analysis of ref.~\cite{Hof15,Hof16}.

\subsection{The short-range $NN$ potential}
\label{sec_short}

In conventional meson theory (sect.~\ref{sec_mes}), the short-range nuclear force
is described by the exchange of heavy mesons, notably the
$\omega(782)$. 
Qualitatively, the short-distance behavior of the $NN$ 
potential is obtained
by Fourier transform of the propagator of
a heavy meson,
\begin{equation}
\int d^3q \frac{e^{i{\vec q} \cdot {\vec r}}}{m^2_\omega
+ {\vec q}^2} \;
\sim \;
 \frac{e^{-m_\omega r}}{r} \; .
\end{equation}

ChPT is an expansion in small momenta $Q$, where $Q$ is too small
to resolve structures like a $\rho(770)$ or $\omega(782)$
meson, because $Q \ll \Lambda_\chi \approx m_{\rho,\omega}$.
But the latter relation allows us to expand the propagator 
of a heavy meson into a power series,
\begin{equation}
\frac{1}{m^2_\omega + Q^2} 
\approx 
\frac{1}{m^2_\omega} 
\left( 1 
- \frac{Q^2}{m^2_\omega}
+ \frac{Q^4}{m^4_\omega}
-+ \ldots
\right)
,
\label{eq_power}
\end{equation}
where the $\omega$ is representative
for any heavy meson of interest.
The above expansion suggests that it should be 
possible to describe the short distance part of
the nuclear force simply in terms of powers of
$Q/m_\omega$, which fits in well
with our over-all 
power expansion since $Q/m_\omega \approx Q/\Lambda_\chi$.
Since such terms act directly between nucleons, they are dubbed contact terms.

Contact terms play also an important role in renormalization.
Regularization
of the loop integrals that occur in multi-pion exchange diagrams
typically generates polynomial terms
with coefficients that are, in part, infinite or scale
dependent (cf.\ Appendix B of ref.~\cite{ME11}). Contact terms absorb infinities
and remove scale and cutoff dependences.

Thus, in EFT, the short-range $NN$ potential is described by contributions of the contact type,
which are constrained by parity, time-reversal, and the usual 
conservation laws, but not by chiral symmetry. Terms that include a factor
$\bm{\tau}_1 \cdot \bm{\tau}_2$ (owing to isospin invariance) can be left out due to Fierz ambiguity. 
Because of  parity and time-reversal only even powers of momentum
are allowed.
Thus, the expansion of the contact potential is
formally written as
\begin{equation}
V_{\rm ct} =
V_{\rm ct}^{(0)} + 
V_{\rm ct}^{(2)} + 
V_{\rm ct}^{(4)} + 
V_{\rm ct}^{(6)} 
+ \ldots \; ,
\label{eq_ct}
\end{equation}
where the superscript denotes the power or order.

The zeroth order (leading order, LO) contact potential is given by
\begin{equation}
V_{\rm ct}^{(0)}(\vec{p'},\vec{p}) =
C_S +
C_T \, \vec{\sigma}_1 \cdot \vec{\sigma}_2 \, 
\label{eq_ct0}
\end{equation}
and, in terms of partial waves, 
\be
V_{\rm ct}^{(0)}(^1 S_0)          &=&  \widetilde{C}_{^1 S_0} =
4\pi\, ( C_S - 3 \, C_T )
\nonumber
\\
V_{\rm ct}^{(0)}(^3 S_1)          &=&  \widetilde{C}_{^3 S_1} =
4\pi\, ( C_S + C_T ) \,.
\label{eq_ct0_pw}
\ee

At second order (NLO), we have
\be
V_{\rm ct}^{(2)}(\vec{p'},\vec{p}) &=&
C_1 \, q^2 +
C_2 \, k^2 
\nonumber 
\\ &+& 
\left(
C_3 \, q^2 +
C_4 \, k^2 
\right) \vec{\sigma}_1 \cdot \vec{\sigma}_2 
\nonumber 
\\
&+& C_5 \left[ -i \vec{S} \cdot (\vec{q} \times \vec{k}) \right]
\nonumber 
\\ &+& 
 C_6 \, ( \vec{\sigma}_1 \cdot \vec{q} )\,( \vec{\sigma}_2 \cdot 
\vec{q} )
\nonumber 
\\ &+& 
 C_7 \, ( \vec{\sigma}_1 \cdot \vec{k} )\,( \vec{\sigma}_2 \cdot 
\vec{k} ) \,,
\label{eq_ct2}
\ee
with $\vec k \equiv \frac12 ({\vec p}
' + \vec p)$
and  $\vec S \equiv \frac12 (\vec\sigma_1+\vec\sigma_2)$.
Partial-wave decomposition yields
\be
V_{\rm ct}^{(2)}(^1 S_0)          &=&  C_{^1 S_0} ( p^2 + {p'}^2 ) 
\nonumber \\
V_{\rm ct}^{(2)}(^3 P_0)          &=&  C_{^3 P_0} \, p p'
\nonumber \\
V_{\rm ct}^{(2)}(^1 P_1)          &=&  C_{^1 P_1} \, p p' 
\nonumber \\
V_{\rm ct}^{(2)}(^3 P_1)          &=&  C_{^3 P_1} \, p p' 
\nonumber \\
V_{\rm ct}^{(2)}(^3 S_1)          &=&  C_{^3 S_1} ( p^2 + {p'}^2 ) 
\nonumber \\
V_{\rm ct}^{(2)}(^3 S_1- ^3 D_1)  &=&  C_{^3 S_1- ^3 D_1}  p^2 
\nonumber \\
V_{\rm ct}^{(2)}(^3 D_1- ^3 S_1)  &=&  C_{^3 S_1- ^3 D_1}  {p'}^2 
\nonumber \\
V_{\rm ct}^{(2)}(^3 P_2)          &=&  C_{^3 P_2} \, p p' 
\,.
\label{eq_ct2_pw}
\ee

The fourth order (N$^3$LO) contacts are
\be
V_{\rm ct}^{(4)}(\vec{p'},\vec{p}) &=&
D_1 \, q^4 +
D_2 \, k^4 +
D_3 \, q^2 k^2 +
D_4 \, (\vec{q} \times \vec{k})^2 
\nonumber 
\\ &+& 
\left[
D_5 \, q^4 +
D_6 \, k^4 +
D_7 \, q^2 k^2 
\right.
\nonumber 
\\ &+& 
\left.
D_8 \, (\vec{q} \times \vec{k})^2 
\right] \vec{\sigma}_1 \cdot \vec{\sigma}_2 
\nonumber 
\\ &+& 
\left(
D_9 \, q^2 +
D_{10} \, k^2 
\right) \left[ -i \vec{S} \cdot (\vec{q} \times \vec{k}) \right]
\nonumber 
\\ &+& 
\left(
D_{11} \, q^2 +
D_{12} \, k^2 
\right) ( \vec{\sigma}_1 \cdot \vec{q} )\,( \vec{\sigma}_2 
\cdot \vec{q})
\nonumber 
\\ &+& 
\left(
D_{13} \, q^2 +
D_{14} \, k^2 
\right) ( \vec{\sigma}_1 \cdot \vec{k} )\,( \vec{\sigma}_2 
\cdot \vec{k})
\nonumber 
\\ &+& 
D_{15} \left[ 
\vec{\sigma}_1 \cdot (\vec{q} \times \vec{k}) \, \,
\vec{\sigma}_2 \cdot (\vec{q} \times \vec{k}) 
\right]
\,,
\label{eq_ct4}
\ee
with contributions by partial waves,
\be
V_{\rm ct}^{(4)}(^1 S_0)          &=&  \widehat{D}_{^1 S_0}          
({p'}^4 + p^4) +
                              D_{^1 S_0}          {p'}^2 p^2 
\nonumber 
\\
V_{\rm ct}^{(4)}(^3 P_0)          &=&        D_{^3 P_0}          
({p'}^3 p + p' p^3) 
\nonumber 
\\
V_{\rm ct}^{(4)}(^1 P_1)          &=&        D_{^1 P_1}          
({p'}^3 p + p' p^3) 
\nonumber 
\\
V_{\rm ct}^{(4)}(^3 P_1)          &=&        D_{^3 P_1}          
({p'}^3 p + p' p^3) 
\nonumber 
\\
V_{\rm ct}^{(4)}(^3 S_1)          &=&  \widehat{D}_{^3 S_1}          
({p'}^4 + p^4) +
                              D_{^3 S_1}          {p'}^2 p^2 
\nonumber 
\\
V_{\rm ct}^{(4)}(^3 D_1)          &=&        D_{^3 D_1}          
{p'}^2 p^2 
\nonumber 
\\
V_{\rm ct}^{(4)}(^3 S_1 - ^3 D_1) &=&  \widehat{D}_{^3 S_1 - ^3 D_1} 
p^4             +
                              D_{^3 S_1 - ^3 D_1} {p'}^2 p^2
\nonumber 
\\
V_{\rm ct}^{(4)}(^3 D_1 - ^3 S_1) &=&  \widehat{D}_{^3 S_1 - ^3 D_1} 
{p'}^4             +
                              D_{^3 S_1 - ^3 D_1} {p'}^2 p^2
\nonumber 
\\
V_{\rm ct}^{(4)}(^1 D_2)          &=&        D_{^1 D_2}          
{p'}^2 p^2 
\nonumber 
\\
V_{\rm ct}^{(4)}(^3 D_2)          &=&        D_{^3 D_2}          
{p'}^2 p^2 
\nonumber 
\\
V_{\rm ct}^{(4)}(^3 P_2)          &=&        D_{^3 P_2}          
({p'}^3 p + p' p^3) 
\nonumber 
\\
V_{\rm ct}^{(4)}(^3 P_2 - ^3 F_2) &=&        D_{^3 P_2 - ^3 F_2} {p'}p^3
\nonumber 
\\
V_{\rm ct}^{(4)}(^3 F_2 - ^3 P_2) &=&        D_{^3 P_2 - ^3 F_2} {p'}^3p
\nonumber 
\\
V_{\rm ct}^{(4)}(^3 D_3)          &=&        D_{^3 D_3}          
{p'}^2 p^2 
\,.
\label{eq_ct4_pw}
\ee

The next higher order is the sixth order (N$^5$LO) which creates contributions up to
 $F$-waves.

\subsection{The full potential and the $NN$ $T$-matrix}
\label{sec_pot1}

The full $NN$ potential is the sum of the long-, intermediate- and short-range contributions:
\beq
V=V_\pi+V_{\rm ct} \,.
\eeq
Order by order, this is given by:
\beqa
V_{\rm LO}  \equiv  V^{(0)} &=&
V_{1\pi} + V_{\rm ct}^{(0)} 
\label{eq_VVLO}
\\
V_{\rm NLO}  \equiv  V^{(2)} &=& V_{\rm LO} 
+V_{2\pi}^{(2)} 
+ V_{\rm ct}^{(2)} 
\label{eq_VVNLO}
\\
V_{\rm NNLO}  \equiv  V^{(3)} &=& V_{\rm NLO} 
+V_{2\pi}^{(3)} 
\label{eq_VVNNLO}
\\
V_{\rm N3LO}  \equiv  V^{(4)} &=& V_{\rm NNLO} +
V_{2\pi}^{(4)} +
V_{3\pi}^{(4)} 
 + V_{\rm ct}^{(4)} 
\label{eq_VVN3LO}
\\
V_{\rm N4LO}  \equiv  V^{(5)} &=& V_{\rm N3LO} + 
V_{2\pi}^{(5)} +
V_{3\pi}^{(5)} 
\label{eq_VVN4LO}
\\
V_{\rm N5LO}  \equiv  V^{(6)} &=& V_{\rm N4LO} + 
V_{2\pi}^{(6)} +
V_{3\pi}^{(6)} +
V_{4\pi}^{(6)} 
\nonumber 
\\ &&
 + V_{\rm ct}^{(6)} 
\,.
\label{eq_VVN5LO}
\eeqa

The two-nucleon system at low angular momentum, particularly
in $S$ waves, is characterized by the
presence of a shallow bound state (the deuteron)
and large scattering lengths.
Thus, perturbation theory does not apply.
In contrast to $\pi$-$\pi$ and $\pi$-$N$,
the interaction between nucleons is not suppressed
in the chiral limit ($Q\rightarrow 0$).
Weinberg~\cite{Wei91} showed that the strong enhancement of the
scattering amplitude arises from purely nucleonic intermediate
states (``infrared enhancement''). He therefore suggested to use perturbation theory to
calculate the $NN$ potential (i.e., the irreducible graphs) and to apply this potential
in a scattering equation 
to obtain the $NN$ amplitude. 

The potential $V$ is, in principal, an invariant amplitude (with relativity taken into account perturbatively) and, thus, satisfies a relativistic scattering equation, like, e.\ g., the
Blankenbeclar-Sugar (BbS) equation~\cite{BS66},
which reads,
\begin{eqnarray}
{T}({\vec p}',{\vec p})&=& {V}({\vec p}',{\vec p})+
\int \frac{d^3p''}{(2\pi)^3} \:
{V}({\vec p}',{\vec p}'') \:
\nonumber \\ &&
\times
\frac{M_N^2}{E_{p''}} \:  
\frac{1}
{{ p}^{2}-{p''}^{2}+i\epsilon} \:
{T}({\vec p}'',{\vec p}) 
\label{eq_bbs}
\end{eqnarray}
with $E_{p''}\equiv \sqrt{M_N^2 + {p''}^2}$ and $M_N$ the nucleon mass.
The advantage of using a relativistic scattering equation is that it automatically
includes relativistic kinematical corrections to all orders. Thus, in the scattering equation,
no propagator modifications are necessary when moving up to higher orders.

Defining
\begin{equation}
\widehat{V}({\vec p}',{\vec p})
\equiv 
\frac{1}{(2\pi)^3}
\sqrt{\frac{M_N}{E_{p'}}}\:  
{V}({\vec p}',{\vec p})\:
 \sqrt{\frac{M_N}{E_{p}}}
\label{eq_minrel1}
\end{equation}
and
\begin{equation}
\widehat{T}({\vec p}',{\vec p})
\equiv 
\frac{1}{(2\pi)^3}
\sqrt{\frac{M_N}{E_{p'}}}\:  
{T}({\vec p}',{\vec p})\:
 \sqrt{\frac{M_N}{E_{p}}}
\,,
\label{eq_minrel2}
\end{equation}
where the factor $1/(2\pi)^3$ is included for convenience,
the BbS equation collapses into the usual, nonrelativistic
Lippmann-Schwinger (LS) equation,
\begin{eqnarray}
 \widehat{T}({\vec p}',{\vec p})&=& \widehat{V}({\vec p}',{\vec p})+
\int d^3p''\:
\widehat{V}({\vec p}',{\vec p}'')\:
\nonumber \\ &&
\times
\frac{M_N}
{{ p}^{2}-{p''}^{2}+i\epsilon}\:
\widehat{T}({\vec p}'',{\vec p}) \, .
\label{eq_LS}
\end{eqnarray}
Since 
$\widehat V$ 
satisfies eq.~(\ref{eq_LS}), 
it may be used like a nonrelativistic potential. By the same token, 
$\widehat{T}$ 
may be considered as the nonrelativistic 
$T$-matrix.

\subsection{Regularization and non-perturbative renormalization}
\label{sec_reno}

\begin{figure*}[t]\centering
\scalebox{0.85}{\includegraphics{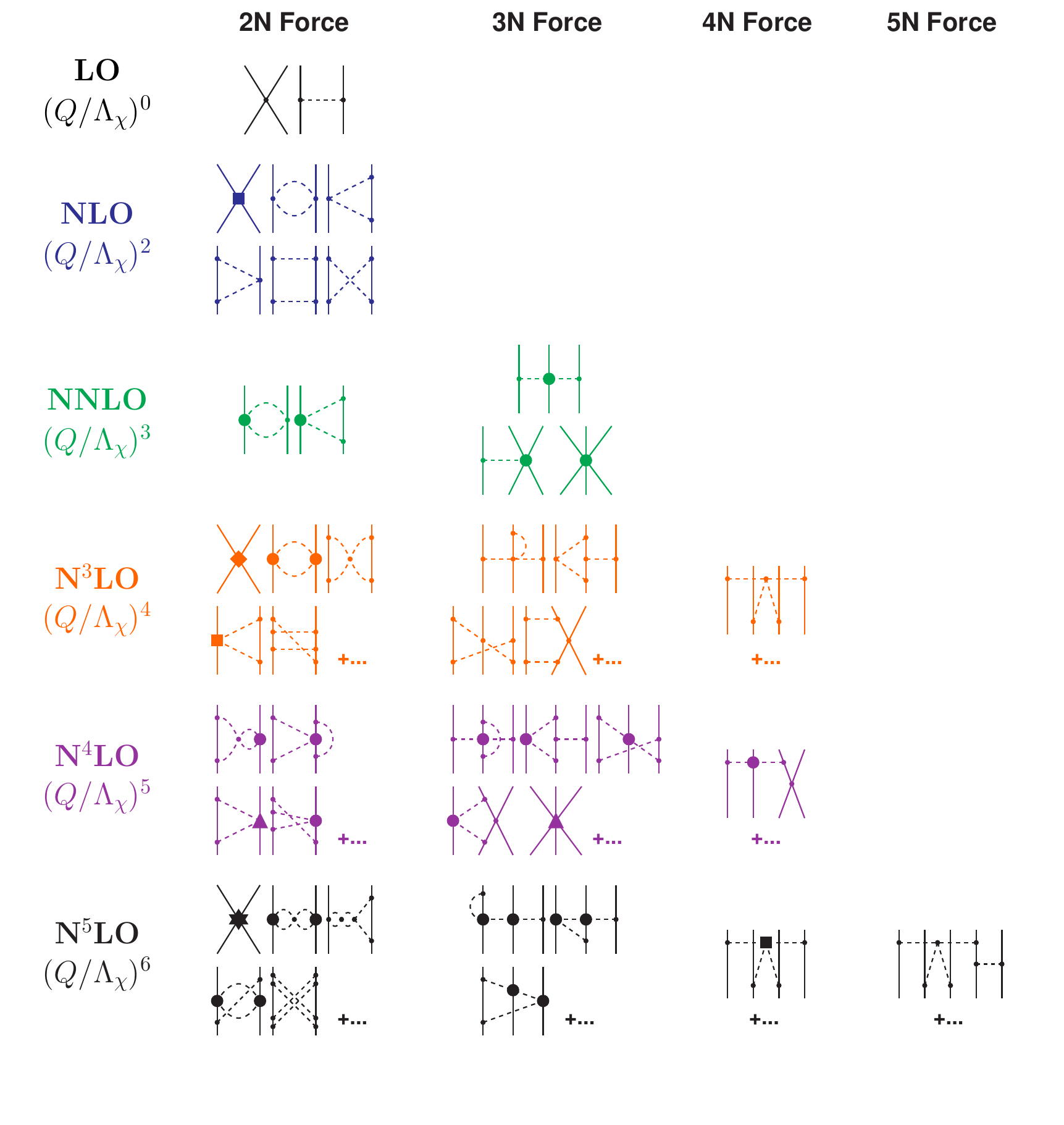}}
\vspace*{-0.75cm}
\caption{The nuclear potential in ChPT. Solid lines
represent nucleons and dashed lines pions. 
Small dots, large solid dots, solid squares, triangles, diamonds, and stars
denote vertices of index $\Delta_i = \, $ 0, 1, 2, 3, 4, and 6, respectively. 
Further explanations are
given in the text.}
\label{fig_hi}
\end{figure*}

Iteration of $\widehat V$ in the LS equation, eq.~(\ref{eq_LS}),
requires cutting $\widehat V$ off for high momenta to avoid infinities.
This is consistent with the fact that ChPT
is a low-momentum expansion which
is valid only for momenta $Q < \Lambda_\chi \approx 1$~GeV.
Therefore, it is customary to multiply the potential $\widehat V$
with a regulator function $f(p',p)$,
\begin{equation}
{\widehat V}(\vec{ p}',{\vec p})
\longmapsto
{\widehat V}(\vec{ p}',{\vec p}) \, f(p',p) \,.
\end{equation}
A frequently applied (nonlocal) regulator is 
\begin{equation}
f(p',p) = \exp[-(p'/\Lambda)^{2n}-(p/\Lambda)^{2n}] \,,
\label{eq_f}
\end{equation}
for which typically values
of the cutoff parameter $\Lambda$ around $500 \pm 100$ MeV are used.

It is pretty obvious that results for the $T$-matrix may
depend sensitively on the regulator and its cutoff parameter.
This is acceptable if one wishes to build models.
For example, the meson models of the past (cf.\ sect.~\ref{sec_mes})
always depended sensitively on the choices for the
cutoff parameters.
Within those models, this was very welcome,
because it provided additional fit parameters to improve the reproduction of the $NN$ phase-shifts and data.
However, this attitude is unacceptable for a proper EFT.

In field theories, divergent integrals are not uncommon and methods have
been developed for how to deal with them.
One regulates the integrals and then removes the dependence
on the regularization parameters (scales, cutoffs)
by renormalization. In the end, the theory and its
predictions do not depend on cutoffs
or renormalization scales.

So-called renormalizable quantum field theories, like QED,
have essentially one set of prescriptions 
that takes care of renormalization through all orders. 
In contrast, 
EFTs are renormalized order by order. 

The renormalization of {\it perturbative}
EFT calculations (that is ChPT) is not a problem; hence, the renormalization of the $NN$ {\it potential} is not a problem.

{\it The problem
is nonperturbative renormalization} [LS eq.~(\ref{eq_LS})].
This problem typically occurs in {\it nuclear} EFT because
nuclear physics is characterized by bound states which
are nonperturbative in nature.
EFT power counting may be different for nonperturbative processes as
compared to perturbative ones. Such difference may be caused by the infrared
enhancement of the reducible diagrams generated in the LS equation.

Weinberg's discussion in refs.~\cite{Wei90,Wei91} may suggest that the contact terms
introduced to renormalize the perturbatively calculated
potential, based upon naive dimensional analysis (``Weinberg counting'', cf.\ sects.~\ref{sec_power} and \ref{sec_short}),
may also be sufficient to renormalize the nonperturbative
resummation of the potential in the LS equation.

Weinberg's alleged assumption may not be correct as first pointed out by Kaplan, Savage, and Wise (KSW)~\cite{KSW96,KSW98a,KSW98b} who, therefore, suggested to treat 1PE 
perturbatively---a prescrition which, however, has convergence problems~\cite{FMS00}.
The KSW critique resulted in a flurry of publications on the renormalization of the $NN$ amplitude,
 and we refer the interested reader to 
section 4.5 of ref.~\cite{ME11} for an account of the first phase of discussion.
However, even today, no generally accepted solution to this problem has emerged and some more recent proposals can be found in 
refs.~\cite{HKK19,NTK05,Bir06,Bir07,Bir08,Bir11,LY12,Lon16,Val11,Val11a,Val16,Val16a,EGM17,Kon17,Epe18,Kol19,Val19}.

Concerning the construction of quantitative $NN$ potential
(by which we mean $NN$ potentials suitable for use in contemporary many-body nuclear methods), 
only Weinberg counting
has been used with success during the past 
25 years~\cite{ORK94,ORK96,EGM00,EM03,EGM05,Eks13,Gez14,Pia15,Pia16,Eks15,EKM15,PAA15,Car16,RKE18,Eks18,EMN17}.

In spite of the criticism, Weinberg counting may be perceived as not unreasonable by the following argument.
For a successful EFT (in its domain of validity), one must be able to claim independence of the predictions on the regulator within the theoretical error.
Also,                                         
truncation errors must decrease as we go to higher and higher orders.
These are precisely the goals of renormalization.  

Lepage~\cite{Lep97} has stressed that the cutoff independence should be examined
for cutoffs below the hard scale and not beyond. Ranges of cutoff independence within the
theoretical error are to be identified using Lepage plots~\cite{Lep97}.
A systematic investigation of this kind has been conducted in ref.~\cite{Mar13b}.
In that work, the error of the predictions was quantified by calculating the $\chi^2$/datum 
for the reproduction of the $np$ elastic scattering data
as a function of the cutoff parameter $\Lambda$ of the regulator function
eq.~(\ref{eq_f}). Predictions by chiral $np$ potentials at 
order NLO and NNLO were investigated applying Weinberg counting 
for the $NN$ contact terms.
It is found that the reproduction of the $np$ data at lab.\ energies below 200 MeV is generally poor
at NLO, while at NNLO the $\chi^2$/datum assumes acceptable values (a clear demonstration of
order-by-order improvement). Furthermore, at NNLO, 
a ``plateau'' of constant low $\chi^2$ for
cutoff parameters ranging from about 450 to 850 MeV can be identified. This may be perceived as cutoff independence
(and, thus, successful renormalization) for the relevant range of cutoff parameters.

Alternatively, one may go for a compromise between Weinberg's prescription of full resummation
of the potential
and Kaplan, Savage, and Wise's~\cite{KSW96,KSW98a,KSW98b} suggestion of perturbative pions---as discussed in
ref.~\cite{Kol19}: 1PE is resummed only in lower partial waves and all corrections are included in 
distorted-wave perturbation theory. 
However, since current {\it ab initio} calculations are tailored such that they need a potential as input,  the question is if there is a way to reconcile those (low-cutoff) potentials 
with the approch of
partially perturbative pions. A first attempt to address this issue has recently been undertaken
by Valderrama~\cite{Val19}.

\vspace*{1cm}

\subsection{Chiral nuclear forces order by order\\ ({\it vs.} phenomenological and meson models)}
\label{sec_hierarchy}

\begin{figure}[t]\centering
\scalebox{0.65}{\includegraphics{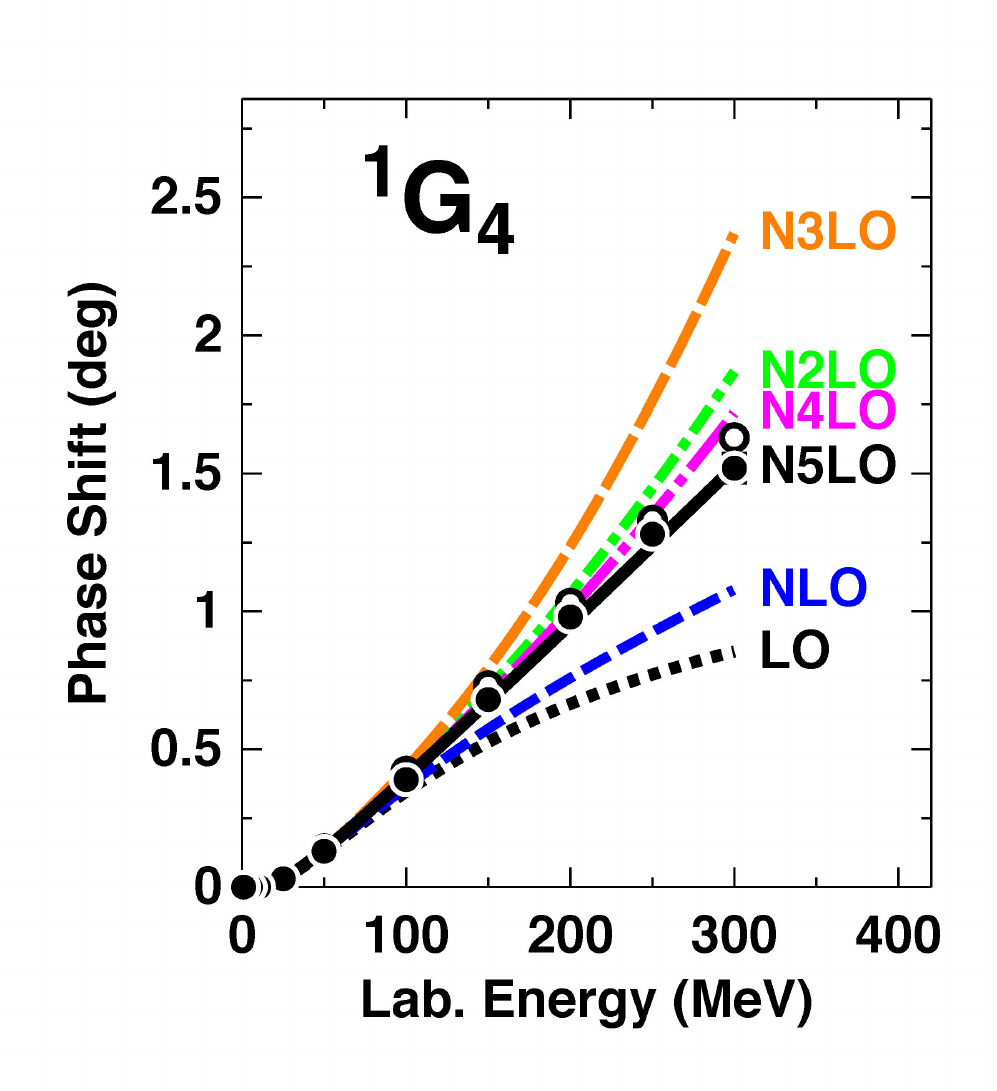}}
\caption{Phase-shifts of neutron-proton scattering in the 
singlet $G$-wave for all orders of ChPT
from LO to N$^5$LO, as denoted.
The filled and open circles represent the results from the Nijmegen multi-energy $np$ phase-shift analysis~\cite{Sto93} and the GWU single-energy $np$ analysis SP07~\cite{SP07}, respectively.}
\label{fig_ph1g4}
\end{figure}

Figure~\ref{fig_hi} provides an overview of how the nuclear potential emerges
in ChPT. We will now go through this order by order and compare to 
both phenomenology and
meson models discussed in sects.~\ref{sec_phen} and \ref{sec_mes}, respectively.

\subsubsection{Leading order (LO)}

At LO, we have the
static one-pion exchange (1PE), shown 
in the first row of fig.~\ref{fig_hi}.          
As mentioned before, 1PE became established in 1956~\cite{Sup56} and is part of any $NN$ potential model since around 1960.
So, at this order, ChPT reproduces what has been done all along.

In addition, at LO, we have two contact contributions with no momentum dependence 
($\sim Q^0$), eq.~(\ref{eq_ct0}). They are signified by the 
four-nucleon-leg diagram 
with one small-dot vertex shown in the first row of 
fig.~\ref{fig_hi}.

In spite of its simplicity, the rough LO description                      
captures some of the main attributes of the $NN$ force. 
First, through the 1PE it generates the tensor component of the force
known to be crucial for the two-nucleon bound state (deuteron quadrupole moment). Second, 
it predicts correctly 
$NN$ phase parameters for partial waves of very high orbital angular momentum.               
The two terms, eq.~(\ref{eq_ct0_pw}), which result from a partial-wave expansion of the contact term
impact states of zero orbital angular momentum and 
allow to fit the $S$-wave scattering lengths and the deuteron binding energy.

In fig.~\ref{fig_ph1g4} we show the contributions to the phase shifts
in peripheral $NN$ scattering.
Note that $NN$ scattering in peripheral partial waves
is of special interest---for several reasons.
First, these partial waves probe the long- and 
intermediate-range of the nuclear force. Due to the 
high angular momentum `barrier', there is only small sensitivity to short-range
contributions and, in fact, the contact terms up to and including 6th order (N$^5$LO)  
make no contributions for orbital angular momenta $L\geq 4$.
Thus, for $G$ and higher waves and energies below the pion-production
threshold, we have a window in which the $NN$ interaction
is governed by chiral symmetry alone (chiral one- and multi-pion exchanges), and we can 
conduct a clean test of how well the theory works.
Due to the smallness of the phase shifts in peripheral waves, 
the calculation is conducted perturbatively and no regulator is applied,
which is another advantage.

In the $^1G_4$ state shown
in fig.~\ref{fig_ph1g4}, 
the 1PE (LO) is obviously insufficient to describe the data.
The difference between the 1PE prediction and the data is to
be provided by two- and three-pion exchanges, i.e. the intermediate-range part
of the nuclear force. How well that is accomplished is a crucial test for
any theory of nuclear forces, and it will be interesting to see if and how the higher orders
of ChPT fill that gap.

\subsubsection{Next-to-leading order (NLO)}

Note that there are no terms with power 
$\nu=1$, as they would violate parity conservation 
and time-reversal invariance.
Thus, NLO is $\nu=2$.
Two-pion exchange makes its first appearance at this order, which is why
it is referred to as ``leading 2PE''.
However, this leading 2PE is not ``leading'' at all, because it is very weak
as clearly seen in fig.~\ref{fig_ph1g4} (blue dashed curve labeled NLO).
The reasons are as follows:
 Loops carry already the power $\nu=2$ [cf.\ eq.~(\ref{eq_nunn})],
and so only                                  
$\pi NN$ and $\pi \pi NN$ vertices with $\Delta_i = 0$ can contribute at this order. 
These vertices are known to be weak.
The NLO box and crossed box 2PE diagrams were included already in some meson models of the 1970's (cf.\ first row of fig.~\ref{fig_2pi}) and found to be insufficient
to describe the intermediate-range attraction of the nuclear force.

At NLO, seven new contacts appear, eq.~(\ref{eq_ct2}), which 
impact $L = 0$ and $L = 1$ states, eq.~(\ref{eq_ct2_pw}). (As always in 
fig.~\ref{fig_hi}, 
two-nucleon contact terms are indicated 
by four-nucleon-leg graphs and a single vertex of appropriate shape, in this case a solid square.) 
At this power, the contact operators                                  
include central, spin-spin, spin-orbit, and tensor terms; that is,
all the spin structures needed for a realistic description of the 2NF. 
However, the 2NF is not realistic at all at this order, because the medium-range attraction lacks strength.

\subsubsection{Next-to-next-to-leading order (NNLO)}

\begin{figure}[t]\centering
\scalebox{0.6}{\includegraphics{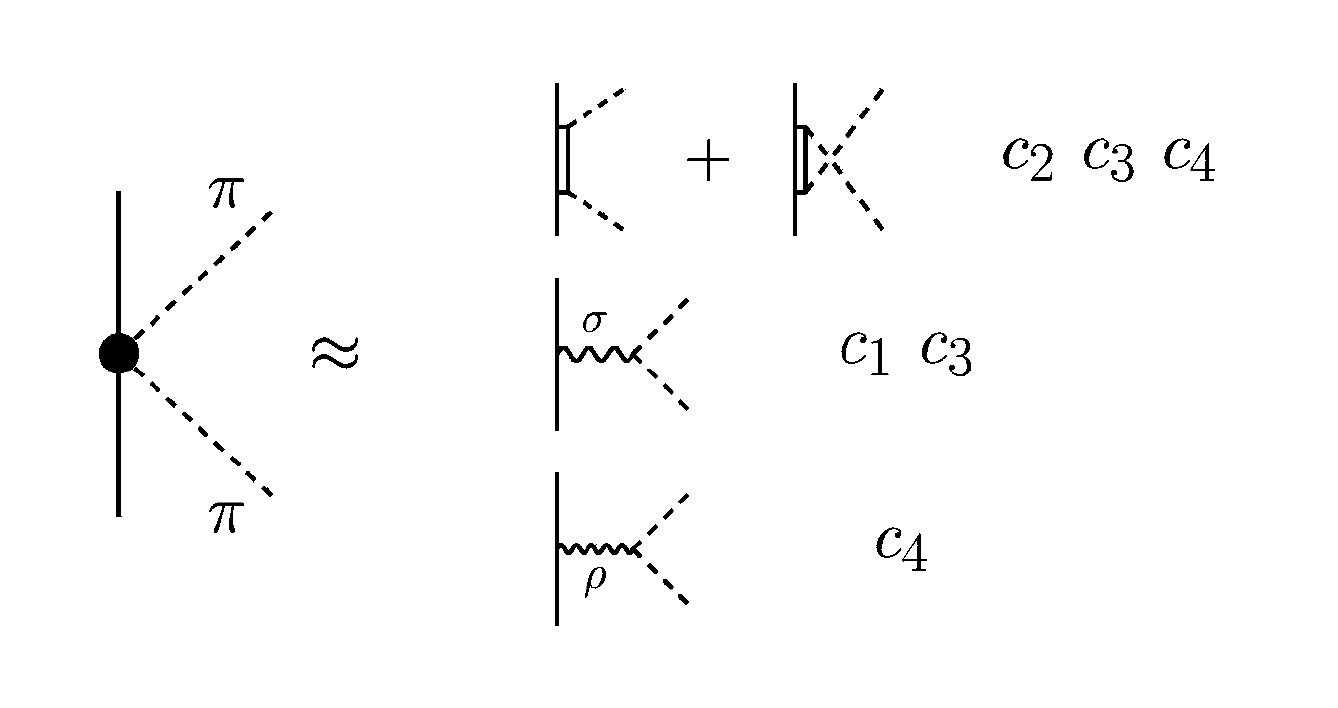}}
\caption{Interpretation of the second-order seagull graph (large solid dot)
in terms of resonance exchanges. The $c_i$ refer to the LECs used 
in the ${\cal L}_{\pi N}^{(2)}$ Lagrangian.}
\label{fig_ressat}
\end{figure}

\begin{figure}[t]\centering
\scalebox{0.5}{\includegraphics{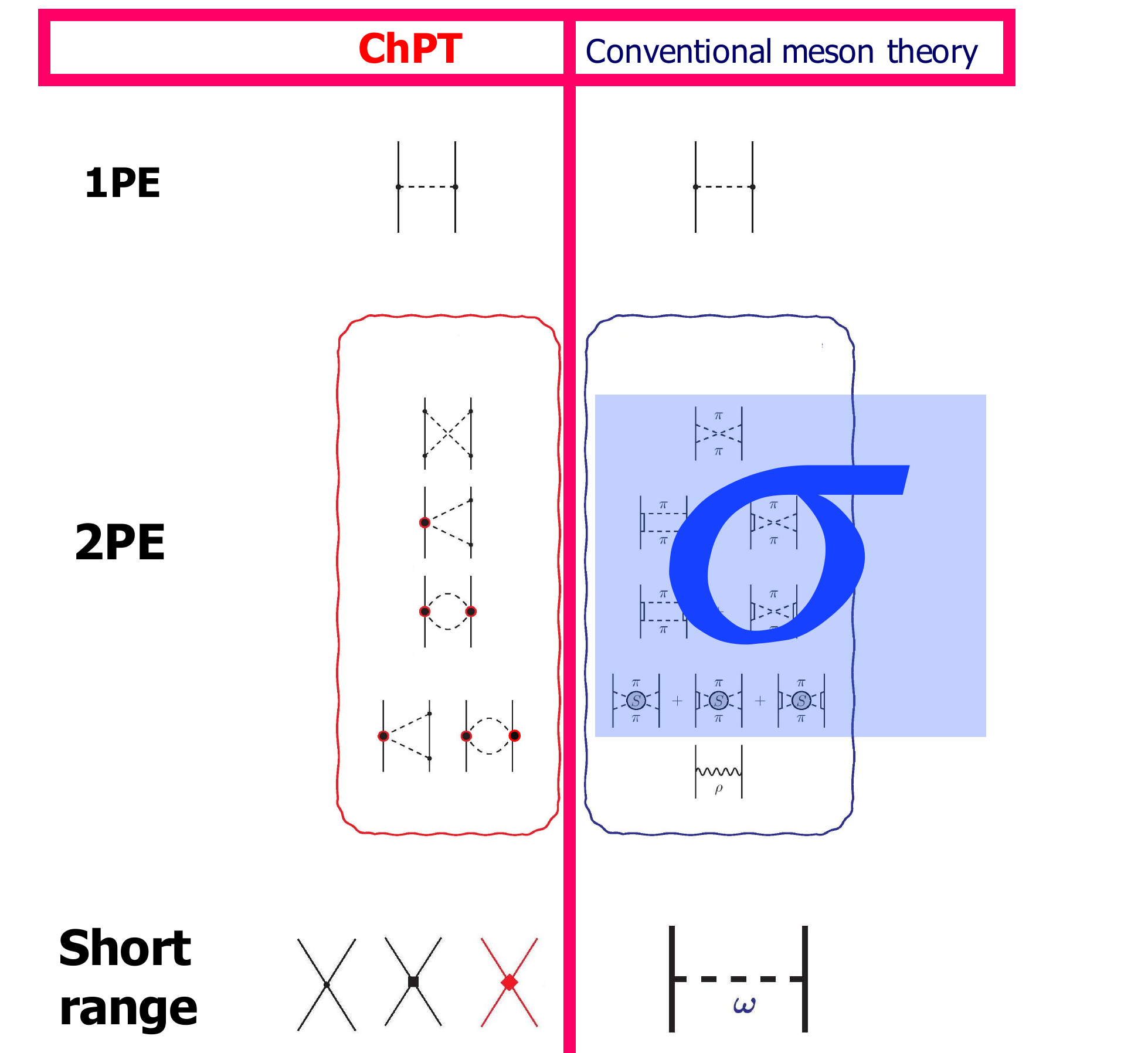}}
\caption{ChPT {\it versus} conventional meson theory for the $NN$ interaction.}
\label{fig_chpt_mes}
\end{figure}

At NNLO, 
the 2PE contains one     
$\pi\pi NN$ seagull vertex with two derivatives 
(denoted by a large solid dot in fig.~\ref{fig_hi}),
which is part of the ${\cal L}_{\pi N}^{(2)}$ Lagrangian and proportional
to the $c_i$ LECs~\cite{ME11,KGE12}.         
In terms of resonance saturation~\cite{BKM97}, this vertex 
is equivalent to correlated/resonant 2PE
and intermediate $\Delta(1232)$-isobar excitation as illustrated in fig.~\ref{fig_ressat}.
Note that the NNLO football diagram in fig.~\ref{fig_hi} vanishes (for purely mathematical reasons), and only the triangle diagram with one large solid dot contributes.
Interpreting this vertex according to fig.~\ref{fig_ressat} suggests that the NNLO triangle 
is equivalent to the $2\pi N \Delta$, $\pi\pi$-$S_{\rm CORR}$ ($\simeq \sigma$)
and $\rho$ diagrams of fig.~\ref{fig_2pi}.
Thus, there are strong parallels between conventional meson theory and chiral EFT,
which is further elucidated in fig.~\ref{fig_chpt_mes}.

As was already well known from conventional meson theory~\cite{Mac89,MHE87}, a 2PE which
includes resonance exchange generates the strong  intermediate-range     
attraction needed for a realistic $NN$ force.
 The N2LO (or NNLO) phase shifts in fig.~\ref{fig_ph1g4} clearly confirm it.

Besides 2NF contributions, the diagramatic display in fig.~\ref{fig_hi} includes also many-nucleon forces.                      
Three-nucleon forces appear at NLO,                          
but their net contribution vanishes at that order~\cite{Wei92}.
The first non-zero 3NF contribution is found 
at NNLO~\cite{Kol94,Epe02b}. It is therefore easy to understand why  
3NF are very weak as compared to the 2NF which contributes already at 
$(Q/\Lambda_\chi)^0$.

\subsubsection{Next-to-next-to-next-to-leading order (N$^3$LO)}

Starting at N$^3$LO, 
the number of diagrams grows out of proportion, such that, in fig.~\ref{fig_hi}, we can display only 
a few representative samples of them. There is a large attractive one-loop 2PE contribution (the bubble diagram with two large solid dots), which 
leads to an over-estimation of the medium-range attraction (cf.\  fig.~\ref{fig_ph1g4}).
The equivalent diagram from conventional meson theory is clearly the double
$\Delta$-exitation, $2\pi \Delta \Delta$, shown in fig.~\ref{fig_2pi}, also known to be very attractive~\cite{MHE87}.

{\it Two-loop} 2PE makes its first appearance (but it is small at this order).
Finally, 3PE occurs for the first time, but has negligible size~\cite{Kai00a,Kai00b}.

The most important feature is the presence 
of 15 additional contacts $\sim Q^4$, eq.~(\ref{eq_ct4}), signified 
by the four-nucleon-leg diagram with the diamond-shaped vertex. 
These contacts impact states with orbital angular momentum up to $L = 2$
[cf.\ eq.~(\ref{eq_ct4_pw})], 
and are the reason for the                            
 quantitative description of the
two-nucleon force (up to approximately 300 MeV
in terms of laboratory energy) 
at this order~\cite{ME11,EM03}.

More 3NF diagrams show up 
at N$^3$LO~\cite{Ber08,Ber11}, as well as the first contributions to 
four-nucleon forces (4NF)~\cite{Epe07}     
We then see that forces involving more and more nucleons appear for the
first time at higher and higher orders, which 
gives theoretical support to the fact that          
2NF $\gg$ 3NF $\gg$ 4NF, etc..

\subsubsection{N$^4$LO}

Further (two-loop) 2PE and 3PE occur at N$^4$LO (fifth order)~\cite{Ent15a}.
 They turn out to be moderately repulsive (cf.\ fig.~\ref{fig_ph1g4}), thus
compensating for the surplus attraction generated at N$^3$LO by the bubble diagram with two solid dots. 

In this context, it is worth noting that also in conventional meson 
theory~\cite{Mac89,MHE87} (sect.~\ref{sec_2pi}) the one-loop models for the 2PE contribution always show 
some excess attraction (cf. fig.~10 of ref.~\cite{ME11}). The same is 
true for the dispersion theoretic approach pursued by the Paris 
group (see, e.~g., the predictions for $^1D_2$, $^3D_2$, and $^3D_3$
in fig.~8 of ref.~\cite{Vin79} which are all too attractive). 
In conventional meson theory~\cite{Mac89,MHE87}, this surplus
attraction is compensated by heavy-meson exchanges ($\rho$-, $\omega$-, and $\pi\rho$-exchange) 
which, however, have no place in chiral effective field theory. 
Instead, in the latter approach, two-loop 
$2\pi$- and $3\pi$-exchanges provide the corrective action.

The long- and intermediate-range 3NF contributions at this order have been evaluated~\cite{KGE12,KGE13}, but not yet applied in nuclear structure calculations. They are
expected to be sizeable. Moreover, a new set of 3NF contact terms appears~\cite{GKV11}.
The N$^4$LO 4NF has not been derived yet. Due to the subleading $\pi\pi N N$ seagull vertex (large solid dot), this 4NF could be sizeable.

\subsubsection{N$^5$LO}

We, finally, turn to N$^5$LO (sixth order). The dominant 2PE and 3PE contributions to the 2NF have been derived by Entem {\it et al.} in ref.~\cite{Ent15b}. The effects are small indicating the desired trend towards convergence of the chiral expansion for the 2NF.
The final result is right on the data, see fig.~\ref{fig_ph1g4}. 
In addition, a new set of 26 $NN$ contact terms $\sim Q^6$ occurs that contributes up to $F$-waves (represented by the $NN$ diagram with a star in fig.~\ref{fig_hi}),
bringing the total number of $NN$ contacts to 50~\cite{EM03a}.
The three-, four-, and five-nucleon forces of this order have not yet been derived.

\subsection{Quantitative chiral $NN$ potentials
\label{sec_pot2}}

\begin{figure*}[t]\centering
\vspace*{-0.5cm}
\hspace*{-0.4cm}
\scalebox{0.40}{\includegraphics{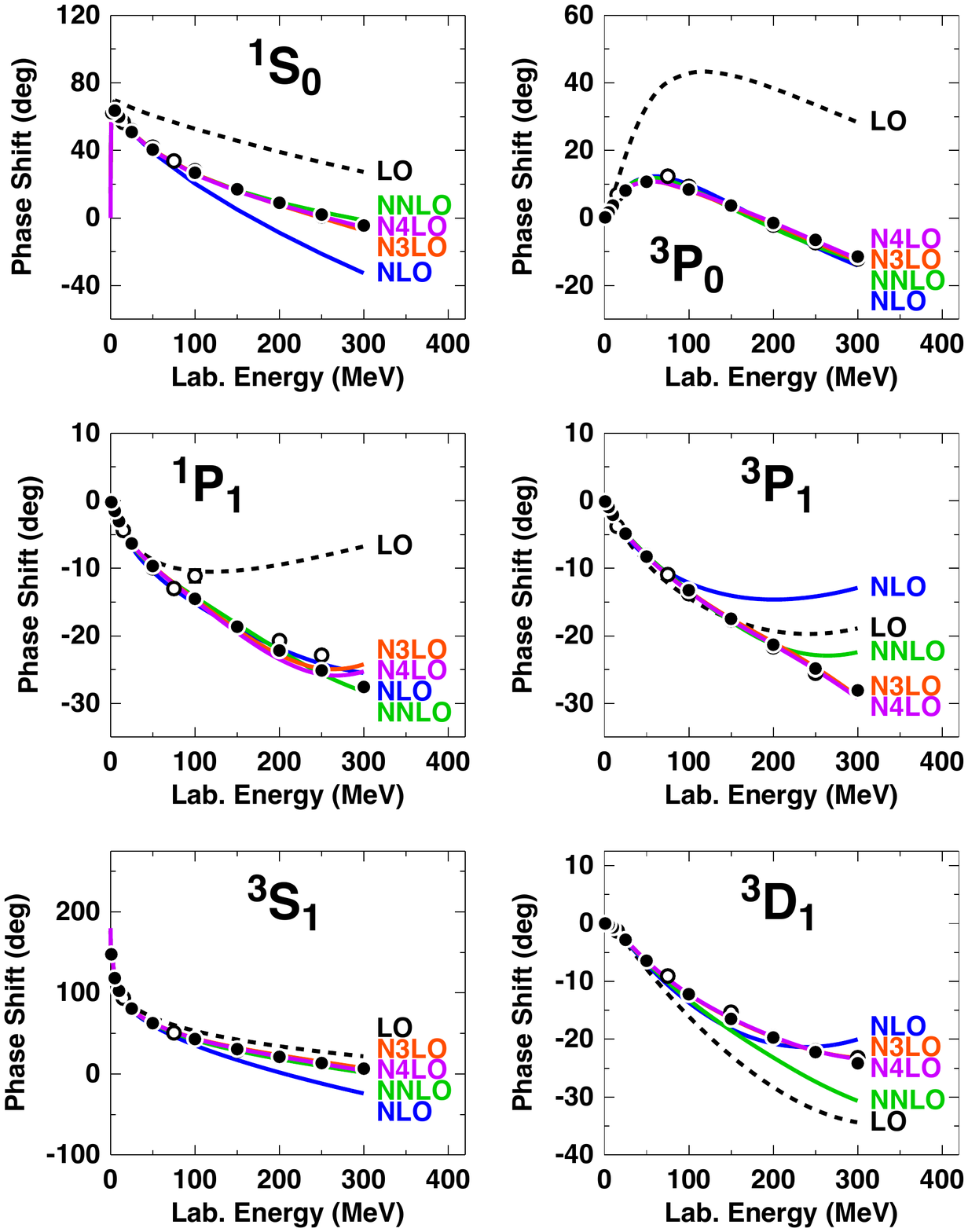}}
\hspace*{-0.5cm}
\scalebox{0.40}{\includegraphics{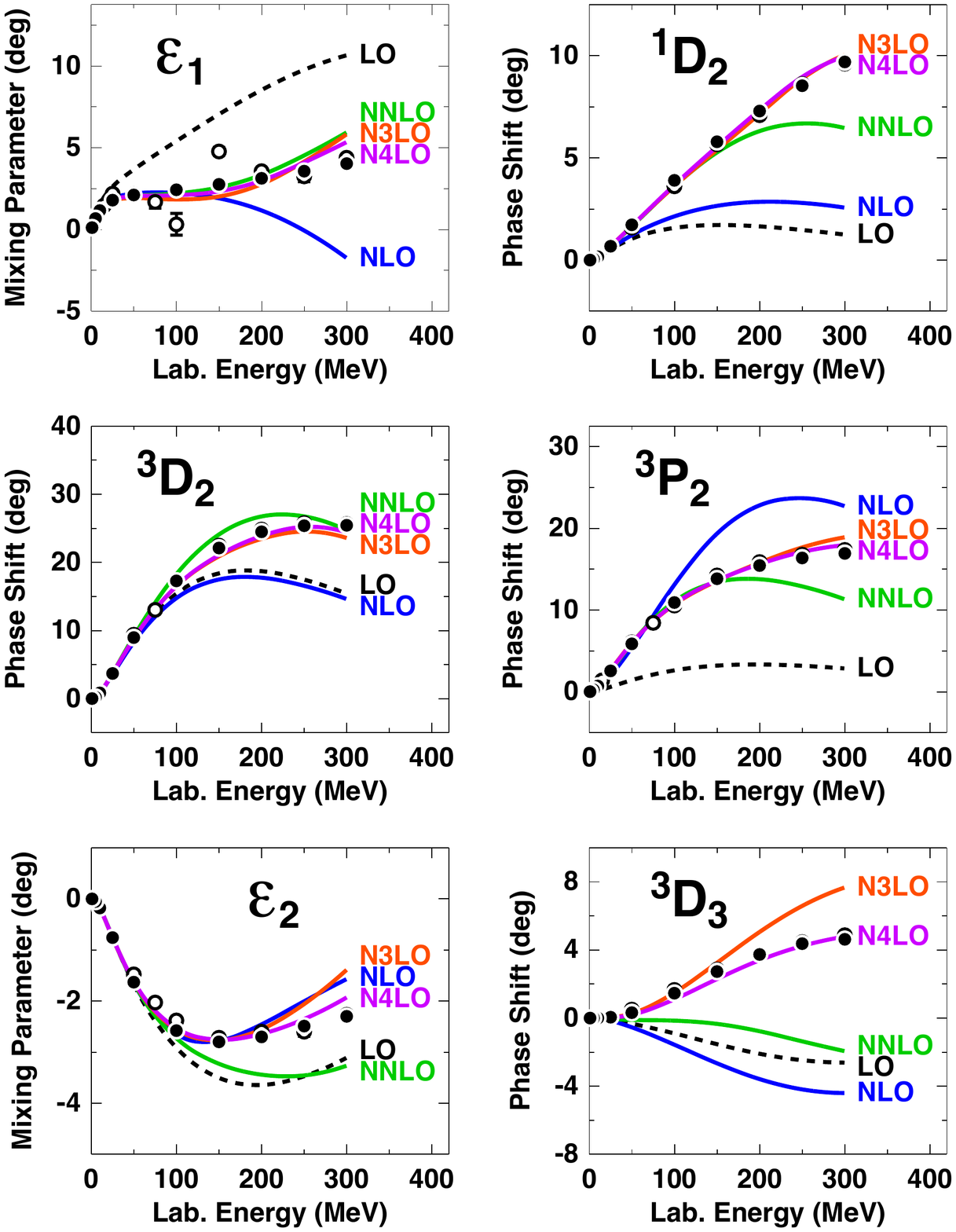}}
\vspace*{-0.5cm}
\caption{Chiral expansion of neutron-proton scattering as represented by the phase shifts 
in $S$, $P$, and $D$ waves and mixing parameters $\epsilon_1$ and $\epsilon_2$.
 Five orders ranging from LO to N$^4$LO are shown as 
denoted.
Filled and open circles as in fig.~\ref{fig_ph1g4}.
(From ref.~\cite{EMN17}.)
\label{fig_ph1a}}
\end{figure*}

\begin{table*}[t]
\caption{$\chi^2/$datum for the fit of the 2016 $NN$ data base by $NN$ potentials at various orders of chiral EFT. (From ref.~\cite{EMN17}.)
\label{tab_chi}}
\smallskip
\begin{tabular*}{\textwidth}{@{\extracolsep{\fill}}ccccccc}
\hline 
\hline 
\noalign{\smallskip}
 $T_{\rm lab}$ bin (MeV) & No.\ of data & LO & NLO & NNLO & N$^3$LO & N$^4$LO \\
\hline
\noalign{\smallskip}
\multicolumn{7}{c}{\bf proton-proton} \\
0--100 & 795 & 520 & 18.9  & 2.28   &  1.18 & 1.09 \\
0--190 & 1206  & 430 & 43.6  &  4.64 & 1.69 & 1.12 \\
0--290 & 2132 & 360 & 70.8  & 7.60  &  2.09  & 1.21 \\
\hline
\noalign{\smallskip}
\multicolumn{7}{c}{\bf neutron-proton} \\
0--100 & 1180 & 114 & 7.2  &  1.38  & 0.93  & 0.94 \\
0--190 & 1697 &  96 & 23.1  &  2.29 &  1.10  & 1.06 \\
0--290 & 2721 &  94 &  36.7 & 5.28  &  1.27 & 1.10 \\
\hline
\noalign{\smallskip}
\multicolumn{7}{c}{\boldmath $pp$ plus $np$} \\
0--100 & 1975 & 283 &  11.9 &  1.74  & 1.03   & 1.00  \\
0--190 & 2903 & 235 &  31.6 &  3.27 & 1.35   &  1.08 \\
0--290 & 4853 & 206 &  51.5 & 6.30  & 1.63   &  1.15 \\
\hline
\hline
\noalign{\smallskip}
\end{tabular*}
\end{table*}

The previous section was mainly focused on the (long- and intermediate-ranged) pion-exchange contributions to the $NN$ interaction, which are governed by chiral symmetry and control the higher partial waves. However, a complete $NN$ potential must include also the lower partial waves, where the short-range force represented by contact terms
plays a crucial role.
Thus, complete $NN$ potentials depend on two sets of parameters, the $\pi N$ and the $NN$ LECs.
The $\pi N$ LECs are the coefficients that appear in the $\pi N$ Langrangians and
 are determined in $\pi N$ analysis~\cite{Hof15,Hof16} (cf.\ table~\ref{tab_lecs}).
The $NN$ LECs are the coefficients of the $NN$ contact terms,
 eqs.~(\ref{eq_ct0})-(\ref{eq_ct4_pw}).
They are fixed by an optimal fit to the $NN$ data below pion-production threshold, see
ref.~\cite{EMN17} for details.

$NN$ potentials are then constructed order by order
and the accuracy improves as the order increases.
How well the chiral expansion converges in 
important lower partial waves
is demonstrated in fig.~\ref{fig_ph1a},
where we show phase parameters for potentials developed
through all orders from LO to N$^4$LO~\cite{EMN17}.~\footnote{Alternative chiral $NN$ potentials can be found in refs.~\cite{EGM00,EM03,EGM05,Eks13,Gez14,Pia15,Pia16,Eks15,EKM15,PAA15,Car16,RKE18,Eks18}.}
These figures clearly reveal
substantial improvements in the reproduction of the empirical
phase shifts with increasing order.

More indicative for the quality of a theory is the ability to reproduce original data. 
Therefore, we show in table~\ref{tab_chi}
the $\chi^2$/datum for the reproduction of the $NN$ data at various orders of chiral EFT. 
The bottom line of table~\ref{tab_chi} summarizes the essential point.
For the close to 5000 $pp$ plus $np$ data below 290 MeV (pion-production threshold),
the $\chi^2$/datum 
is 51.4 at NLO and 6.3 at NNLO. Note that the number of $NN$ contact terms is the same for both orders. The improvement is entirely due to an improved description of the 2PE contribution
at NNLO as discussed in the previous section.
Continuing on the bottom line of table~\ref{tab_chi}, after NNLO,
the $\chi^2$/datum further improves to 1.63 at N$^3$LO, which is largely due to an increase in the number of contact terms.
 Finally at N$^4$LO, the almost perfect value of 1.15 is achieved---great convergence.

\begin{table*}[t]
\small
\caption{Two- and three-nucleon bound-state properties as predicted by
  $NN$ potentials at various orders of chiral EFT ($\Lambda = 500$ MeV in all cases).
(Deuteron: Binding energy $B_d$, asymptotic $S$ state $A_S$,
asymptotic $D/S$ state $\eta$, structure radius $r_{\rm str}$,
quadrupole moment $Q$, $D$-state probability $P_D$; the predicted
$r_{\rm str}$ and $Q$ are without meson-exchange current contributions
and relativistic corrections. Triton: Binding energy $B_t$.)
$B_d$ is fitted, all other quantities are predictions. (From ref.~\cite{EMN17}.)
\label{tab_deu}}
\smallskip
\begin{tabular*}{\textwidth}{@{\extracolsep{\fill}}lllllll}
\hline 
\hline 
\noalign{\smallskip}
 & LO & NLO & NNLO & N$^3$LO & N$^4$LO & Empirical$^a$ \\
\hline
\noalign{\smallskip}
{\bf Deuteron} \\
$B_d$ (MeV) &
 2.224575& 2.224575 &
 2.224575 & 2.224575 & 2.224575 & 2.224575(9) \\
$A_S$ (fm$^{-1/2}$) &
 0.8526& 0.8828 &
0.8844 & 0.8853 & 0.8852 & 0.8846(9)  \\
$\eta$         & 
 0.0302& 0.0262 &
0.0257& 0.0257 & 0.0258 & 0.0256(4) \\
$r_{\rm str}$ (fm)   & 1.911
      & 1.971 & 1.968
       & 1.970
       & 1.973 &
 1.97507(78) \\
$Q$ (fm$^2$) &
 0.310& 0.273&
 0.273 & 
 0.271 & 0.273 &
 0.2859(3)  \\
$P_D$ (\%)    & 
 7.29& 3.40&
4.49 & 4.15 & 4.10 & --- \\
\hline
\noalign{\smallskip}
{\bf Triton} \\
$B_t$ (MeV) & 11.09  & 8.31  & 8.21 & 8.09  & 8.08 & 8.48 \\
\hline
\hline
\noalign{\smallskip}
\end{tabular*}
\footnotesize
$^a$See table XVIII of ref.~\cite{Mac01} for references;
the empirical value for $r_{\rm str}$ is from ref.~\cite{Jen11}.\\
\end{table*}

The evolution of the deuteron properties from LO to N$^4$LO  of chiral EFT is shown in table~\ref{tab_deu}.
In all cases, the deuteron binding energy is fit to its empirical value of 2.224575 MeV
using the non-derivative $^3S_1$ contact. All other deuteron properties are predictions.
Already at NNLO, the deuteron has converged to its empirical properties and stays there
through the higher orders.

At the bottom of table~\ref{tab_deu}, we also show the predictions for the triton binding
as obtained in 34-channel charge-dependent Faddeev calculations using only 2NFs. The results show smooth and steady convergence, order by order, towards a value around 8.1 MeV that is reached at the highest orders shown. This contribution from the 2NF will require only a moderate 3NF. 
The relatively low deuteron $D$-state probabilities ($\approx 4.1$\% at N$^3$LO and N$^4$LO) and the concomitant generous triton binding energy predictions are
a reflection of the fact that the $NN$ potentials of ref.~\cite{EMN17} are soft (which is, at least in part, due to their non-local character).

\subsection{Chiral many-body forces}
\label{sec_manyNF}

Two-nucleon forces derived from chiral EFT 
have been applied, often successfully, in the many-body system.                                  
On the other hand, over the past several years we have also learnt that, for some few-nucleon
reactions and nuclear structure issues, 3NFs are indispensable.              
The most well-known cases are the so-called $A_y$ puzzle of $N$-$d$ scattering~\cite{EMW02},
the ground state of $^{10}$B~\cite{Cau02}, and the saturation of nuclear matter~\cite{Heb11,Sam12,Cor14,Sam15,MS16,Sam18}.
As we observed previously, 
the EFT approach generates          
consistent two- and many-nucleon forces in a natural way 
(cf.\ the overview given in fig.~\ref{fig_hi}).
We now shift our focus to chiral three- and four-nucleon forces.

\subsubsection{Three-nucleon forces}
\label{sec_3nfs}

Weinberg~\cite{Wei92} was the first to discuss     
nuclear three-body forces in the context of ChPT. Not long after that, 
the first 3NF at NNLO was derived by van Kolck~\cite{Kol94}.

For a 3NF, we have $A=3$ and $C=1$ and, thus, eq.~(\ref{eq_nu})
implies
\begin{equation}
\nu = 2 + 2L + 
\sum_i \Delta_i \,.
\label{eq_nu3nf}
\end{equation}
This allows to analyze 3NF contributions in a systematic way.

\paragraph{Next-to-leading order}

The lowest possible power is obviously $\nu=2$ (NLO), which
occurs in the absence of loops ($L=0$) and with 
only leading vertices
($\sum_i \Delta_i = 0$). 
As discussed by Weinberg~\cite{Wei92} and van Kolck~\cite{Kol94}, 
the contributions from these diagrams
vanish at NLO. So, the bottom line is that there is no genuine 3NF contribution at NLO.
The first non-vanishing 3NF appears at NNLO.

\paragraph{Next-to-next-to-leading order}

\begin{figure}[t]\centering
\scalebox{0.5}{\includegraphics{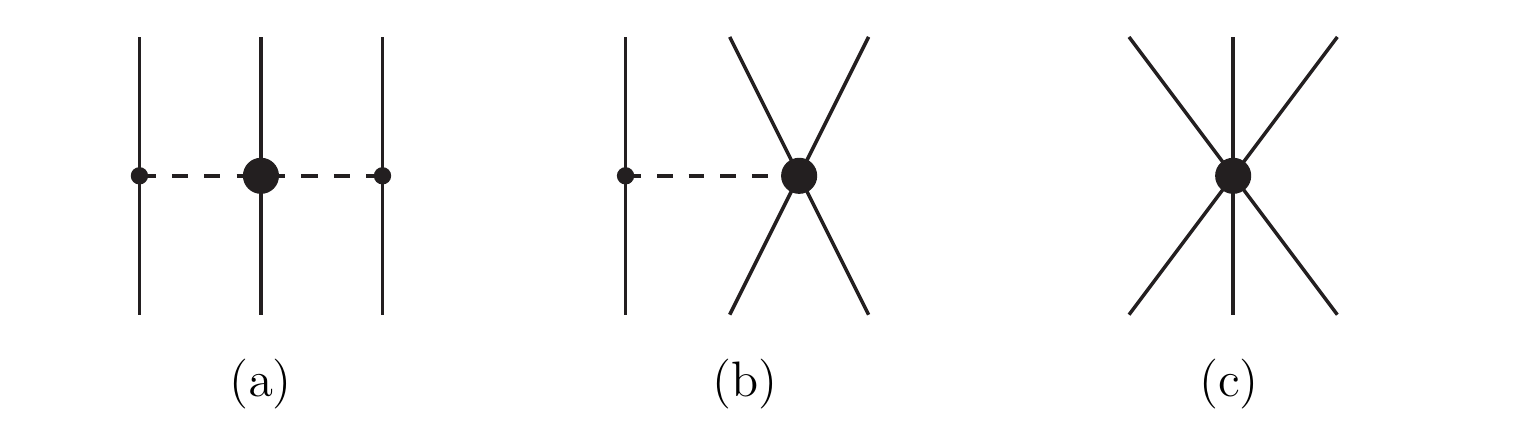}}
\caption{The three-nucleon force at NNLO
with (a) 2PE, (b) 1PE, and (c) contact diagrams.
Notation as in fig.~\ref{fig_hi}.}
\label{fig_3nf_nnlo}
\end{figure}

The power $\nu=3$ (NNLO) is obtained when
there are no loops ($L=0$) and 
$\sum_i \Delta_i = 1$, i.e., 
$\Delta_i=1$ for one vertex 
while $\Delta_i=0$ for all other vertices.
There are three topologies which fulfill this condition,
known as the 2PE, 1PE,
and contact graphs~\cite{Kol94,Epe02b}
(fig.~\ref{fig_3nf_nnlo}).

The 2PE 3N-potential is derived to be
\begin{equation}
V^{\rm 3NF}_{\rm 2PE} = 
\left( \frac{g_A}{2f_\pi} \right)^2
\frac12 
\sum_{i \neq j \neq k}
\frac{
( \vec \sigma_i \cdot \vec q_i ) 
( \vec \sigma_j \cdot \vec q_j ) }{
( q^2_i + m^2_\pi )
( q^2_j + m^2_\pi ) } \;
F^{ab}_{ijk} \;
\tau^a_i \tau^b_j
\label{eq_3nf_nnloa}
\end{equation}
with $\vec q_i \equiv \vec{p_i}' - \vec p_i$, 
where 
$\vec p_i$ and $\vec{p_i}'$ are the initial
and final momenta of nucleon $i$, respectively, and
\begin{eqnarray}
F^{ab}_{ijk} & = & \delta^{ab}
\left[ - \frac{4c_1 m^2_\pi}{f^2_\pi}
+ \frac{2c_3}{f^2_\pi} \; \vec q_i \cdot \vec q_j \right]
\nonumber \\ &&
+ 
\frac{c_4}{f^2_\pi}  
\sum_{c} 
\epsilon^{abc} \;
\tau^c_k \; \vec \sigma_k \cdot [ \vec q_i \times \vec q_j] \; .
\label{eq_3nf_nnlob}
\end{eqnarray}  

It is interesting to observe that there are clear analogies between this force and earlier          
2PE 3NFs already proposed decades ago, particularly the Fujita-Miyazawa~\cite{FM57} and
the Tucson-Melbourne (TM)~\cite{Coo79} forces (cf.\ sect.~\ref{sec_mes3nf}).
In fact, based upon the chiral 3NF at NNLO, the TM force was corrected~\cite{FHK99}
leading to what became known as the TM' or TM99 force~\cite{CH01}.

The 2PE 3NF does not introduce additional fitting constants, 
since the LECs $c_1$, $c_3$, and $c_4$ are already present in the 2PE 2NF.
Besides, since $c_i$'s determined in $\pi N$ analysis~\cite{Hof15,Hof16} are used,
the consistency of the chiral 2PE 3NF with the empirical $\pi N$ amplitude
is automatic and guaranteed (cf.\ discussion of this issue in sect.~\ref{sec_mes3nf}).

The other two 3NF contributions shown in fig.~\ref{fig_3nf_nnlo}
are given by
\begin{equation}
V^{\rm 3NF}_{\rm 1PE} = 
-D \; \frac{g_A}{8f^2_\pi} 
\sum_{i \neq j \neq k}
\frac{\vec \sigma_j \cdot \vec q_j}{
 q^2_j + m^2_\pi }
( \mbox{\boldmath $\tau$}_i \cdot \mbox{\boldmath $\tau$}_j ) 
( \vec \sigma_i \cdot \vec q_j ) 
\label{eq_3nf_nnloc}
\end{equation}
and 
\begin{equation}
V^{\rm 3NF}_{\rm ct} = E \; \frac12
\sum_{i \neq j \neq k}
 \mbox{\boldmath $\tau$}_i \cdot \mbox{\boldmath $\tau$}_j  \; .
\label{eq_3nf_nnlod}
\end{equation}
These 3NF potentials introduce 
two additional constants, $D$ and $E$, which can be constrained in             
 more than one way.                                    
One may use 
the triton binding energy and the $nd$ doublet scattering
length $^2a_{nd}$~\cite{Epe02b}
or an optimal global fit of the properties of light nuclei~\cite{Nav07}.
 Alternative choices include 
the binding energies of $^3$H and $^4$He~\cite{Nog06} or
the binding energy of $^3$H and the point charge radius of $^4$He~\cite{Heb11}.
Another method makes use of
the triton binding energy and the Gamow-Teller matrix element of tritium $\beta$-decay~\cite{Mar12}.
When the values of $D$ and $E$ are determined, the results for other
observables involving three or more nucleons are true theoretical predictions.

Applications of the leading 3NF include few-nucleon 
reactions~\cite{Epe02b,NRQ10,Viv13}, structure of light- and medium-mass nuclei~\cite{Eks15,Nav07a,Rot11,Rot12,Hag12a,Hag12b,BNV13,Her13,Hag14a,Bin14,HJP16,Sim16,Sim17,Mor18,Som19}, and infinite matter~\cite{Sam12,Cor14,Sam15,MS16,Sam18,Heb11,HS10,Hag14b,Cor13},
often with 
satisfactory results. Some problems, though, remain unresolved, such as 
the well-known `$A_y$ puzzle'~\cite{Epe02b,EMW02,Viv13}.

In summary, the leading 3NF of ChPT is a remarkable contribution. It gives validation to, 
and provides a better framework for, 
3NFs which were proposed already five decades ago; it alleviates existing problems            
in few-nucleon reactions and the spectra of
light nuclei.
Nevertheless, we still face several challenges.                   
With regard to the 2NF, we have discussed earlier that it is necessary 
to go to order four or even five for convergence and high-precison predictions. 
Thus, the 3NF at N$^3$LO must be considered simply as a matter of consistency with the 2NF sector. 
At the same time, one hopes that its 
inclusion may result in further improvements with the aforementioned unresolved problems.

\begin{figure}[t]\centering
\scalebox{0.63}{\includegraphics{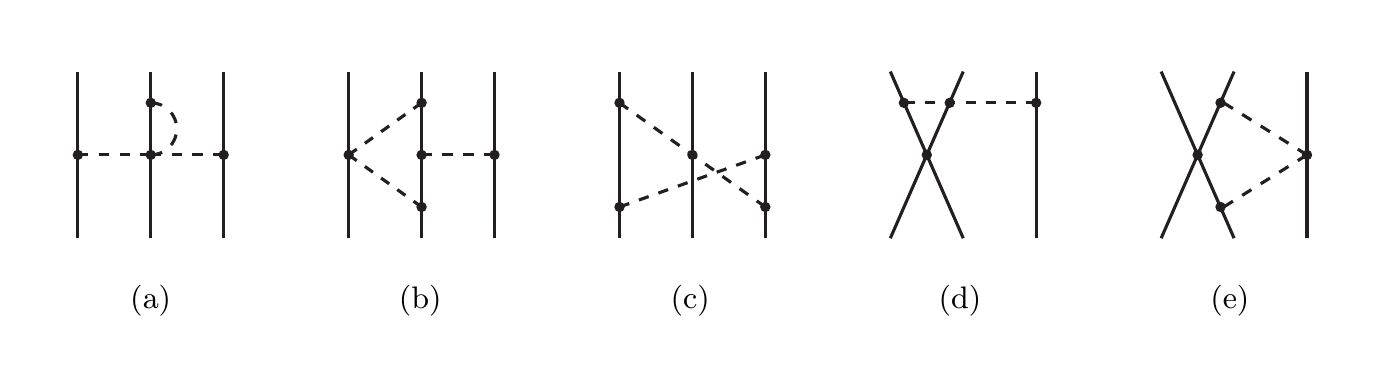}}
\vspace*{-0.75cm}
\caption{Leading one-loop 3NF diagrams at N$^3$LO.
We show one representative example for each of five topologies,
which are: (a) 2PE, (b) 1PE-2PE, (c) ring, (d) contact-1PE, (e) contact-2PE.
Notation as in fig.~\ref{fig_hi}.}
\label{fig_3nf_n3lo}
\end{figure}

\paragraph{Next-to-next-to-next-to-leading order}
\label{sec_3nfn3lo}

At N$^3$LO, the loop diagrams shown in fig.~\ref{fig_3nf_n3lo} occur.
Since a loop carries
$L=1$, all $\Delta_i$ have to be zero
to ensure $\nu=4$ [cf.\ eq.~(\ref{eq_nu3nf})].
Thus, these one-loop 3NF diagrams can include
only leading order vertices, the parameters of which
are fixed from $\pi N$ and $NN$ analysis.
The diagrams
have been evaluated by the Bochum-Bonn group~\cite{Ber08,Ber11}.
The long-range part of the chiral N$^3$LO 3NF has been
tested in the triton and in three-nucleon scattering~\cite{Gol14}
leaving the $N$-$d$ $A_y$ puzzle unresolved.
The long- and short-range parts of this
force have been applied in 
nuclear and neutron matter calculations~\cite{Kru13,Dri16,Heb15,DHS19}
as well as
in the structure of medium-mass nuclei~\cite{Hop19,Rot19}.

\paragraph{The 3NF at N$^4$LO}
\label{sec_3nfn4lo}

\begin{figure}[t]\centering
\scalebox{0.63}{\includegraphics{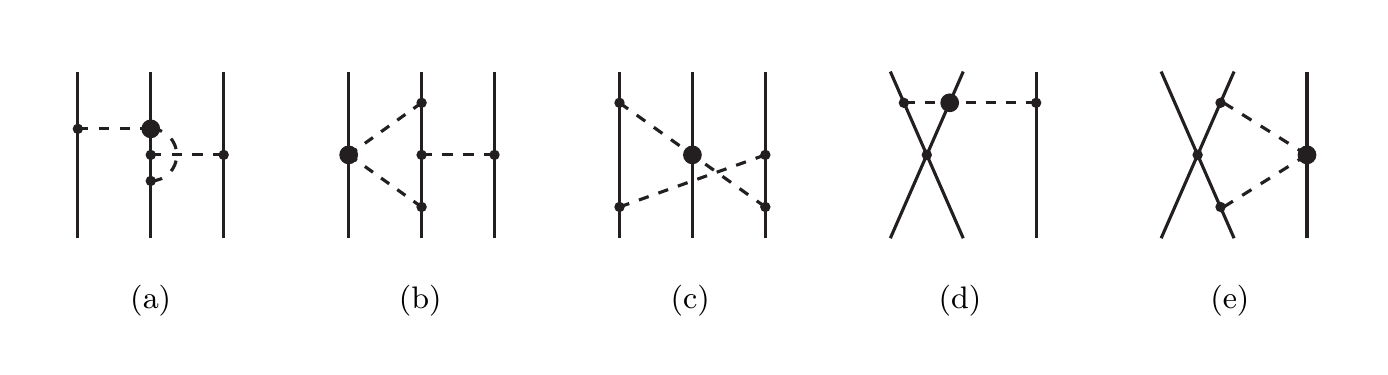}}
\vspace*{-0.75cm}
\caption{Sub-leading one-loop 3NF diagrams which appear at N$^4$LO
with topologies similar to fig.~\ref{fig_3nf_n3lo}.
Notation as in fig.~\ref{fig_hi}.}
\label{fig_3nf_n4loloops}
\end{figure}

\begin{figure}[t]\centering
\scalebox{0.8}{\includegraphics{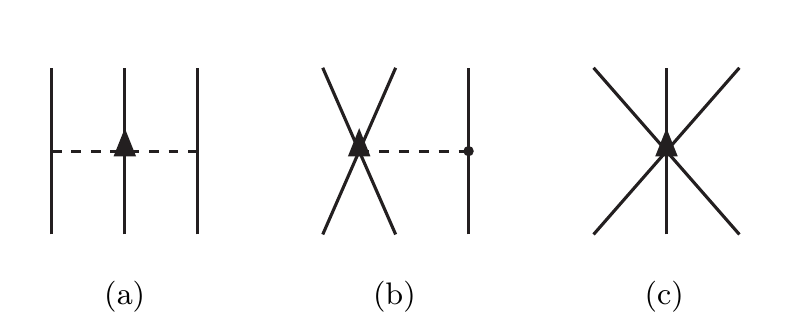}}
\vspace*{-0.2cm}
\caption{3NF tree graphs at N$^4$LO ($\nu=5$) denoted by: (a) 2PE, (b) 1PE-contact, and (c) contact. Notation as in fig.~\ref{fig_hi}.}
\label{fig_3nf_n4lotrees}
\end{figure}

\begin{figure*}[t]\centering
\hspace*{-0.2cm}
\scalebox{1.2}{\includegraphics{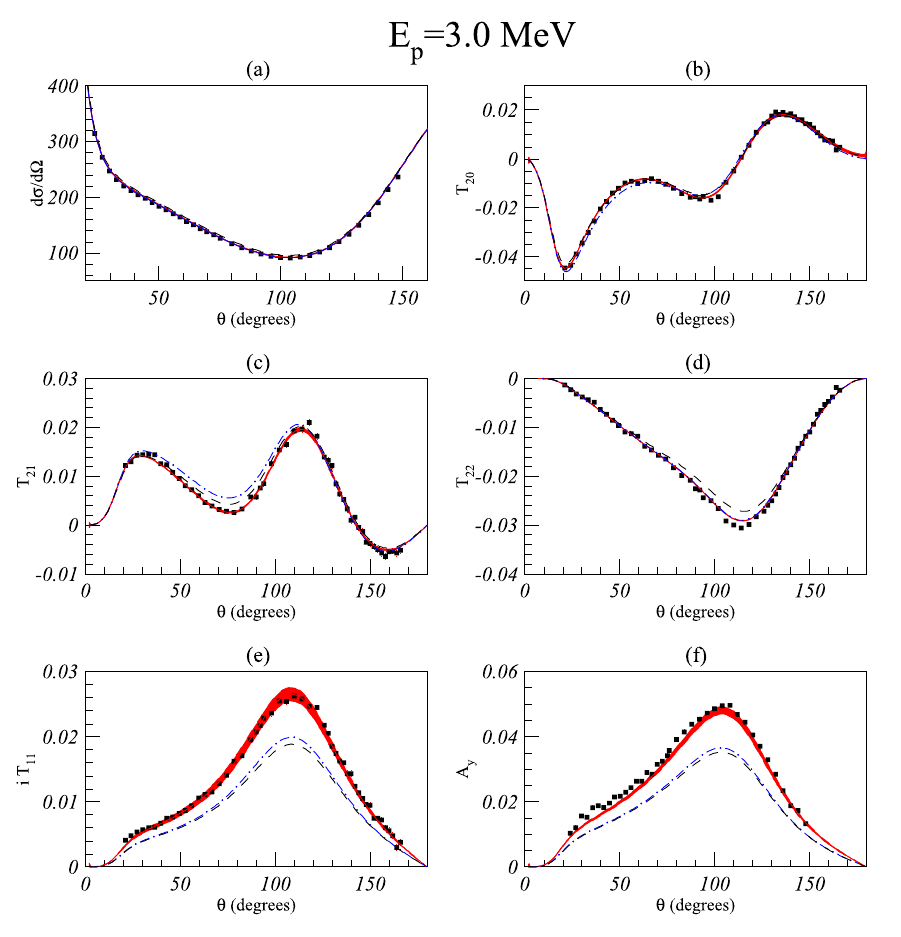}}
\caption{Cross section and polarization observables of $p$-$d$ scattering at 3 MeV
proton energy as predicted by the AV18 2NF (black dashed lines), the AV18 2NF + UIX 3NF
(blue dash-dotted lines),
and calculations that include chiral N$^4$LO 3NF contact terms (red bands). 
The width of the red bands reflect the perceived uncertainty of the predictions.
Data from ref.~\cite{Shi95}. (Figure reproduced 
from ref.~\cite{Gir19} with permission.)}
\label{fig_ay}
\end{figure*}

In regard to some unresolved issues, one may go ahead and look
at the next order of 3NFs, which is N$^4$LO or $\nu=5$.
The loop contributions that occur at this order
are obtained by replacing in the N$^3$LO loops
one vertex by a $\Delta_i=1$ vertex (with LEC $c_i$), fig.~\ref{fig_3nf_n4loloops},
which is why these loops may be more sizable than the N$^3$LO loops.
The 2PE, 1PE-2PE, and ring topologies have been evaluated~\cite{KGE12,KGE13} so far.
In addition, we have three `tree' topologies (fig.~\ref{fig_3nf_n4lotrees}), which include
a new set of 3N contact interactions that has been derived
by the Pisa group~\cite{GKV11}.
{\it The N$^4$LO 3NF contacts have been applied with success in calculations of few-body reactions at low energy solving the $p$-$d$ $A_y$ puzzle~\cite{Gir19}, fig.~\ref{fig_ay}.}

\subsubsection{Four-nucleon forces}

\begin{figure}[t]\centering
\scalebox{0.5}{\includegraphics{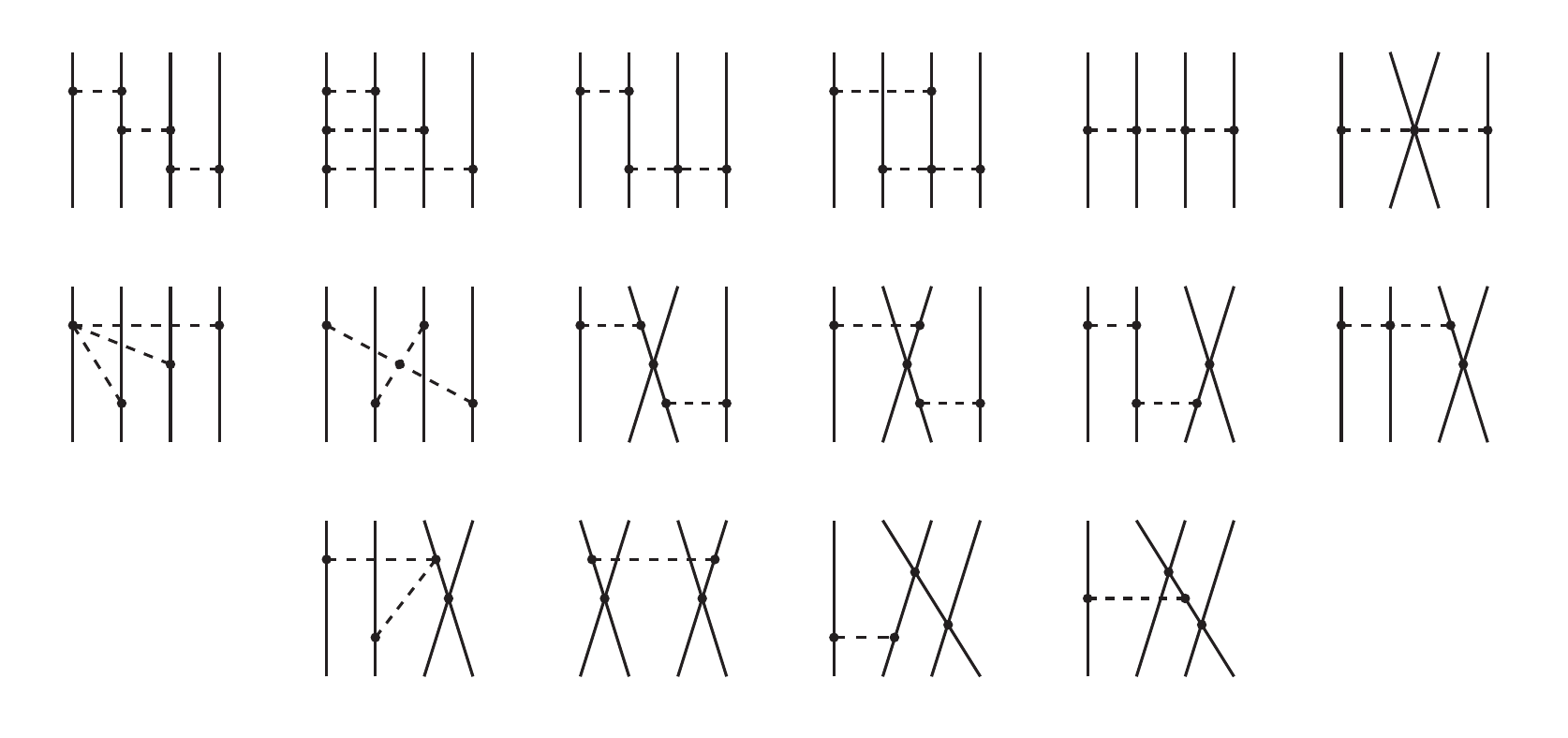}}
\caption{Leading four-nucleon force at N$^3$LO.}
\label{fig_4nf_n3lo}
\end{figure}

For connected ($C=1$) $A=4$ diagrams, eq.~(\ref{eq_nu}) yields
\begin{equation}
\nu = 4 + 2L + 
\sum_i \Delta_i \,.
\label{eq_nu4nf}
\end{equation}
We then see that the first (connected) non-vanishing 4NF is generated at $\nu = 4$ (N$^3$LO), with                   
all vertices of leading type, fig.~\ref{fig_4nf_n3lo}. 
This 4NF has no loops and introduces no novel parameters~\cite{Epe07}.

For a reasonably convergent series, terms                      
of order $(Q/\Lambda_\chi)^4$ should be small and, therefore,              
chiral 4NF contributions are expected to be very weak.    
This has been confirmed in calculations of the energy of              
$^4$He~\cite{Roz06} as well as
neutron matter and symmetric nuclear matter~\cite{Kru13}.

The effects of the leading chiral 4NF in symmetric nuclear matter and pure neutron matter have been
worked out by Kaiser {\it et al.}~\cite{Kai12,KM16}.

\subsection{Uncertainty quantification}
\label{sec_uncert}

When applying chiral two- and many-body forces in {\it ab initio} calculations producing predictions for observables of nuclear structure and reactions, 
major sources of uncertainties are~\cite{FPW15}:
\begin{enumerate}
\item
Experimental errors of the input $NN$ data that the 2NFs are based upon and the input
few-nucleon data to which the 3NFs are adjusted.
\item
Uncertainties in the Hamiltonian due to 
       \begin{enumerate}
       \item
       uncertainties in the determination of the $NN$ and $3N$ contact LECs,
       \item
       uncertainties in the $\pi N$ LECs, 
       \item
       regulator dependence, 
       \item
       EFT truncation error.
       \end{enumerate}
\item
Uncertainties associated with the few- and many-body methods applied.
\end{enumerate}

For a thorough discussion of all aspects, see ref.~\cite{EMN17}, where it was concluded that regulator dependence and EFT truncation error are the major source of uncertainty.

The choice of the regulator function and its cutoff parameter creates uncertainty.  
Originally, cutoff variations were perceived as a demonstration of the uncertainty at a given order
(equivalent to the truncation error).
However, in various investigations~\cite{Sam15,EKM15} it has been shown that this is not correct and that cutoff variations,
in general, underestimate this uncertainty. 
Therefore, the truncation error is better determined by sticking literally to what
 `truncation error' means, namely, the error due to
 ignoring contributions from orders beyond the given order $\nu$. 
 The largest such contribution is the one of order $(\nu + 1)$,
 which one may, therefore, consider as representative for the magnitude of what is left out.
This suggests that the truncation error at order $\nu$ can reasonably be defined as
\begin{equation}
\Delta X_\nu = |X_\nu - X_{\nu+1}| \,, 
\end{equation}
where $X_\nu$ denotes the prediction for observable $X$ at order $\nu$. If $X_{\nu+1}$ is not available, then one may use, 
\begin{equation}
\Delta X_\nu = |X_{\nu-1} - X_\nu|Q/\Lambda \,,
\end{equation}
 choosing a typical value for the momentum $Q$, or $Q=m_\pi$.
Alternatively, one may also apply more elaborate definitions, like the one given in ref.~\cite{EKM15}.
Note that one should not add up (in quadrature) the uncertainties due to regulator dependence and the truncation error, because they are not independent. In fact, it is appropriate to 
leave out the uncertainty due to regulator dependence entirely and just focus on the truncation error~\cite{EKM15}. The latter should be estimated using the same cutoff (e.~g., $\Lambda = 500$ MeV)
in all orders considered.

The bottom line is that the most substantial uncertainty is the truncation error. This is the dominant source of (systematic) error that can be reliably estimated in the EFT approach.

\section{Conclusions}
\label{sec_concl}

\begin{figure}[t]\centering
\vspace*{0.1cm}
\hspace*{-0.2cm}
\scalebox{0.43}{\includegraphics{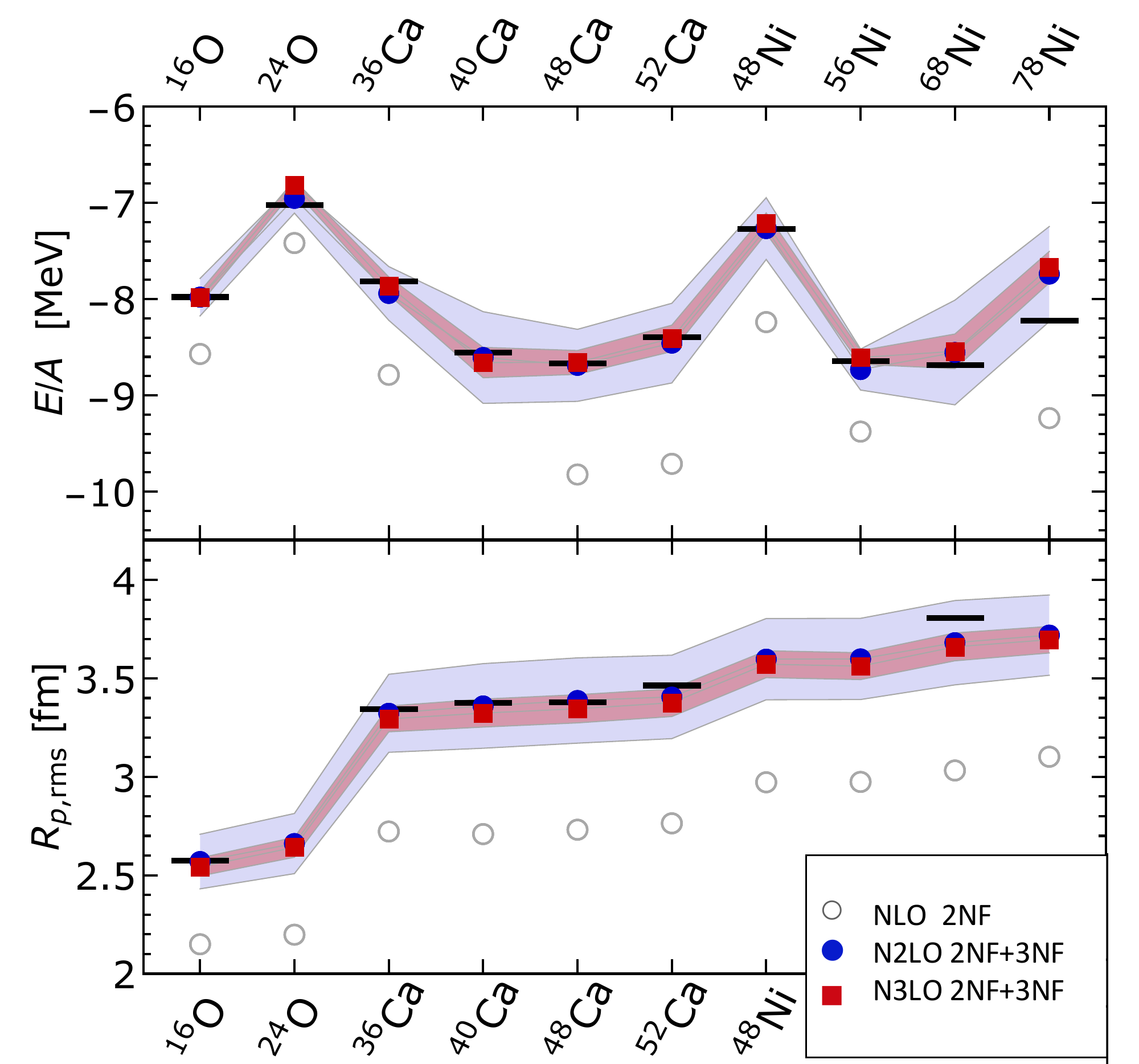}}
\vspace*{0.2cm}
\caption{Ground-state energies (top panel) and point-proton rms radii (bottom panel) 
for selected medium-mass isotopes as obtained in the {\it ab initio} 
many-body calculations by the group of Roth~\cite{Rot19}
at various orders of chiral EFT as explained in the legend. The light blue and pink bands represent 
the uncertainties at N$^2$LO and N$^3$LO, respectively, due to truncation error.
Black bars indicate the experimental data. (Figure courtesy of R. Roth)}
\label{fig_roth}
\end{figure}

In this article, we have contrasted the traditional models for nuclear forces
with the chiral EFT approach. 
The principal superiority of chiral EFT lies in the fact that---via symmetries---it is 
much more closely related to low-energy QCD than any of the earlier phenomenologies.

Moreover, while in the traditional (meson-theoretic) approach nuclear forces are 
expressed as an expansion in terms
of decreasing ranges (or increasing meson masses), chiral EFT
is an expansion in powers of (low) nucleon momenta. Such expansion allows
to estimate the uncertainty of the predictions.
Via fig.~\ref{fig_ressat}, we showed that
there is a close relationship between both schemes, 
as it should be, since,
afterall, both deal with the
same basic content (cf.\ also fig.~\ref{fig_chpt_mes}).
However, while the contributions represented
 in figs.~\ref{fig_obep}-\ref{fig_pirho} offer no formal guidance
concerning what is more or less important,
the order-by-order hierarchy of fig.~\ref{fig_hi} introduces a superior scheme.

The advantage of the chiral EFT approach becomes even more obvious for 
multi-nucleon forces. The suggested 3NFs
that were advanced during Era II, 
figs.~\ref{fig_3nf1}-\ref{fig_3nf4}, 
show a jumble of contributions
whose individual relevance and size are hard to estimate. 
Diagrams such as those shown in 
figs.~\ref{fig_3nf1}-\ref{fig_3nf4}
have been included in microscopic
calculations in a more or less arbitrary way
and with questionable success as discussed in sect.~\ref{sec_mes3nf}.
Thus, the many-body force sector, in particular,
calls for more and better systematics. The order-by-order arrangement in chiral EFT
as shown in the overview of fig.~\ref{fig_hi} and detailed in 
figs.~\ref{fig_3nf_nnlo}-\ref{fig_4nf_n3lo} introduces this much needed
organization.

To summarize,
the strong formal points which render chiral EFT superior to the traditional phenomenlogical or meson-theoretic approaches are: Chiral EFT 
\begin{itemize}
\item
is rooted in low-energy QCD (sect.~\ref{sec_eraiii});
\item
is, in principal, model-independent (sect.~\ref{sec_reno});
\item
occurs with an organizational scheme (`power counting', sect.~\ref{sec_power}) that allows to quantify the uncertainty of  predictions (sect.~\ref{sec_uncert});
\item
generates two- and many-body forces on an equal footing 
(cf.\ fig.~\ref{fig_hi} and sect.~\ref{sec_hierarchy}).
\end{itemize}

But the strength of chiral EFT extends beyond a purely formal level.
{\it Ab inito} calculations applying chiral nuclear forces are much more successful and
quantitative than calculations applying the traditional two- and three-body forces.
We recall here only a few although outstanding examples.
An old problem in microscopic nuclear theory has been the proper 
explanation of nuclear matter saturation. Though it was speculated early on
that the addition of a 3NF might solve the problem, nonrelativistic calculations with phenomenological
3NFs failed to do so~\cite{Day83,CPW83,APR98} as demonstrated in fig.~\ref{fig_snm}.
In contrast, the chiral 3NF at only leading order (NNLO) has the ability to solve that problem~\cite{Heb11,Sam12,Cor14,Sam15,MS16,Sam18} (cf.\ fig.~\ref{fig_snm}).
Similar observations can be made about intermediate-mass nuclei:
While the traditional approach fails badly in the intermediate-mass region~\cite{Lon17},
chiral EFT based two- and three-body forces generate excellent
 predictions~\cite{Eks15,HJP16,Sim16,Sim17,Mor18,Som19,Rot19}
 (fig.~\ref{fig_roth}).
The $N$-$d$ $A_y$ puzzle, which could never be resolved with phenomenological
3NFs~\cite{Kie10},
is another example for the success of
EFT based 3NFs (this time, of higher order)~\cite{Gir19}, see fig.~\ref{fig_ay}.

Thus, it is fair to say that chiral EFT represents substantial progress 
and improvement as compared to 
the traditional approaches. But since EFT is a {\it field theory},
the standards to which it must measure up are higher
than for a model. A sound EFT must be renormalizable and 
allow for a proper power counting (order-by-order arrangement).
The presently used chiral nuclear potentials are based on naive dimensional analysis 
(`Weinberg counting', sect.~\ref{sec_power}) and apply a
cutoff regularization scheme (sect.~\ref{sec_reno}).
In that scheme, one wishes to see cutoff independence of the results.
Such independence is seen to a good degree below the 
breakdown scale~\cite{Mar13b,Cor13,Cor14},
but to which extent that is satisfactory is controversial.
The problem is due to the nonperturbative
resummation necessary for typical nuclear physics problems (bound states).
However, there is hope that from the current 
discussion~\cite{HKK19,NTK05,Bir06,Bir07,Bir08,Bir11,LY12,Lon16,Val11,Val11a,Val16,Val16a,EGM17,Kon17,Epe18}
constructive solutions may emerge~\cite{Kol19,Val19}.

In conclusion, considering both formal aspects and evidence of successful applications, 
one may say that chiral EFT has brought a considerable degree of satisfaction
to the field of nuclear forces.
It is unclear, though, whether the remaining unresolved issues will be settled in a 
satisfactory manner in the near future.

\section*{Acknowledgments}
This work was supported in part by the U.S. Department of Energy
under Grant No.~DE-FG02-03ER41270.

%

\begin{thebibliography}{999}
%
%
\bibitem{Wei67}
S. Weinberg, Phys. Rev. Lett. {\bf 18}, 188 (1967).
\bibitem{Wei68}
S. Weinberg, Phys. Rev. {\bf 166}, 1568 (1968).
\bibitem{Wei79} S. Weinberg, {Physica} {\bf 96A}, 327 (1979).
\bibitem{Wei97} 
  S.~Weinberg, `` What is quantum field theory, and what did we think it was?'',
  in: {\it Conceptual foundations of quantum field theory},
  T. Cao, ed. (Cambridge University Press, Cambridge, 1999)  pp.\ 241-251 [hep-th/9702027].
\bibitem{Wei09} 
  S.~Weinberg,
  ``Effective Field Theory, Past and Future,''
  PoS CD {\bf 09}, 001 (2009)
  [arXiv:0908.1964 [hep-th]].
\bibitem{Yuk35} 
H. Yukawa, {Proc. Phys. Math. Soc. Japan} {\bf 17}, 48 (1935).
\bibitem{Fer34}
E. Fermi, {Z. Phys.} {\bf 88}, 161 (1934).
\bibitem{NA37}
S. H. Neddermeyer and C. D. Anderson, {Phys. Rev.} {\bf 51}, 884 (1937).
\bibitem{YS37} 
H. Yukawa and S. Sakata, {Proc. Phys. Math. Soc. Japan} {\bf 19}, 1084 (1937).
\bibitem{Kem38}
N. Kemmer,  {Proc. Roy. Soc. (London)} {\bf A166}, 127 (1938).
\bibitem{LOP47}
C. M. G. Lattes, G. P. S. Occhialini, and C. F. Powell, {Nature} {\bf 160}, 453, 486 (1947).
\bibitem{GL48}
E. Gardner and C. M. G. Lattes, {Science} {\bf 107}, 270 (1948).
\bibitem{TMO52} 
M. Taketani, S. Machida, and S. Onuma,
{Prog. Theor. Phys. (Kyoto)} {\bf 7}, 45 (1952).
\bibitem{BW53} K. A. Brueckner and K. M. Watson, 
{Phys. Rev.}  {\bf 90}, 699; {\bf 92}, 1023 (1953).
\bibitem{Mar52}
R. E. Marshak, {\it Meson Physics} (Dover Publications, New York, 1952).
\bibitem{SBH55}
S. S. Schweber, H. A. Bethe, and F. de Hoffmann, {\it Mesons and Fields, Volume I Fields}
(Row, Peterson and Company, Evanston, IL, 1955).
\bibitem{BH55}
H. A. Bethe, and F. de Hoffmann, {\it Mesons and Fields, Volume II Mesons}
(Row, Peterson and Company, Evanston, IL, 1955).
\bibitem{Mor63}
M. J. Moravcsik, {\it The Two-Nucleon Interaction}
(Clarendon Press, Oxford, 1963).
\bibitem{GL84} J. Gasser and H. Leutwyler,
{Ann.\ Phys.} {\bf 158}, 142 (1984); 
{Nucl. Phys.} {\bf B250}, 465 (1985).
\bibitem{GSS88} J. Gasser, M. E. Sainio, and A. \v{S}varc,
{Nucl.\ Phys.} {\bf B307}, 779 (1986).
\bibitem{Wei90} S. Weinberg, {Phys.\ Lett.} {\bf B251}, 288 (1990).
\bibitem{Wei91} S. Weinberg, {Nucl.\ Phys.} {\bf B363}, 3 (1991).
\bibitem{Wei92} S. Weinberg, {Phys.\ Lett.\ B} {\bf 295}, 114 (1992).
\bibitem{ORK94}
C. Ord\'o\~nez, L. Ray, and U. van Kolck,
{Phys.\ Rev.\ Lett.} {\bf 72}, 1982 (1994).
\bibitem{ORK96}
C. Ord\'o\~nez, L. Ray, and U. van Kolck,
{Phys.\ Rev.\ C} {\bf 53}, 2086 (1996).
\bibitem{Sup56}
Proc. Theor. Phys.  (Kyoto), Suppl. {\bf 3}, 1-174 (1956).
\bibitem{GT57}
 J.~L.~Gammel and R.~M.~Thaler,
Phys.\ Rev.\  {\bf 107}, 291,
 1337 (1957).
\bibitem{HJ62}
T.~Hamada and I.~D.~Johnston,
  Nucl.\ Phys.\  {\bf 34}, 382 (1962).
\bibitem{Rei68}
R.~V.~Reid, Jr.,
  Annals Phys.\  {\bf 50}, 411 (1968).
\bibitem{LP81} 
  I.~E.~Lagaris and V.~R.~Pandharipande,
  Nucl.\ Phys.\ A {\bf 359}, 331 (1981).
\bibitem{WSA84}
  R.~B.~Wiringa, R.~A.~Smith and T.~L.~Ainsworth,
  Phys.\ Rev.\ C {\bf 29}, 1207 (1984).
\bibitem{WSS95} 
 R.~B.~Wiringa, V.~G.~J.~Stoks and R.~Schiavilla,
  Phys.\ Rev.\ C {\bf 51}, 38 (1995)
\bibitem{Cha57} 
  O.~Chamberlain, E.~Segre, R.~D.~Tripp, C.~Wiegand and T.~Ypsilantis,
{Phys.\ Rev.}  {\bf 105}, 288 (1957).
\bibitem{SYM57} 
  H.~P.~Stapp, T.~J.~Ypsilantis and N.~Metropolis,
{Phys.\ Rev.}  {\bf 105}, 302 (1957).
\bibitem{Sak60}
J. J. Sakurai, {Phys. Rev.} {\bf 119}, 1784 (1960).
\bibitem{Bre60}
G. Breit, {Phys. Rev.} {\bf 120}, 287 (1960).
\bibitem{Nam57}
Y. Nambu, {Phys. Rev.} {\bf 106}, 1366 (1957).
\bibitem{FF59}
W. R. Frazer and J. R. Fulco, {Phys. Rev. Lett.} {\bf 2}, 365 (1959).
\bibitem{FF60}
W. R. Frazer and J. R. Fulco, {Phys. Rev.} {\bf 117}, 1609 (1960).
\bibitem{Mag61}
B. C. Maglic {\it et al.}, {Phys. Rev. Lett.} {\bf 7}, 178 (1961).
\bibitem{Erw61}
A. R. Erwin {\it et al.}, {Phys. Rev. Lett.} {\bf 6}, 628 (1961).
\bibitem{BS64} R. A. Bryan and B. L. Scott, Phys. Rev. {\bf 135}, B434 (1964), and references therein.
\bibitem{BS67} R. A. Bryan and B. L. Scott, Phys Rev. {\bf 164}, 1215 (1967).
\bibitem{BS69} R. A. Bryan and B. L. Scott, Phys. Rev. {\bf 177}, 1435, (1969).
\bibitem{Mor72}
M. J. Moravcsik, {Rep. Prog. Phys.} {\bf 35}, 587 (1972).
\bibitem{Sto94} 
V.\ G.\ J.\ Stoks, R.\ A.\ M.\ Klomp, 
C.\ P.\ F.\ Terheggen, and J.\ J.\ de Swart, 
{Phys.\ Rev.\ C} {\bf 49}, 2950 (1994).
\bibitem{MSS96}
R.~Machleidt, F.~Sammarruca and Y.~Song,
  Phys.\ Rev.\ C {\bf 53}, R1483 (1996).
\bibitem{Mac01} 
 R. Machleidt, {Phys. Rev. C} {\bf 63} 024001 (2001).
\bibitem{PDG} 
M.~Tanabashi {\it et al.} [Particle Data Group],
  Phys.\ Rev.\ D {\bf 98}, 030001 (2018).
\bibitem{Mac89} 
R. Machleidt, {Adv. Nucl. Phys.} {\bf 19}, 189 (1989).
\bibitem{Sup67}
Proc. Theor. Phys.  (Kyoto), Suppl. {\bf 39}, 1-346 (1967).
\bibitem{RMP67} 
A. E. S. Green, M. H. MacGregor and R. Wilson, {Rev. Mod. Phys.} {\bf 39} (1967) 495.
\bibitem{NRS78} 
M. M. Nagels, T. A. Rijken, and J. J. de Swart, {Phys. Rev. D} 
{\bf 17}, 768 (1978).
\bibitem{GTG71} 
A. Gersten, R. Thompson, and A. E. S. Green,
{Phys. Rev. D} {\bf 3}, 2076 (1971).
\bibitem{Sch72}
G. Schierholz, {Nucl. Phys.} {\bf B40}, 335 (1972).
\bibitem{FT75}
J. Fleischer and J. A. Tjon, {Nucl. Phys.} {\bf B84}, 375 (1975).
\bibitem{FT80}
J. Fleischer and J. A. Tjon, {Phys. Rev. D} {\bf 24}, 87 (1980).
\bibitem{Erk74}
K. Erkelenz, {Phys. Reports} {\bf 13C}, 191 (1974).
\bibitem{HM75}
K. Holinde and R. Machleidt, {Nucl. Phys.} {\bf A247}, 495 (1975).
\bibitem{HM76}
K. Holinde and R. Machleidt, {Nucl. Phys.} {\bf A256}, 479 (1976).
\bibitem{BG79}
W. W. Buck and F. Gross, 
Phys. Rev. D {\bf 20}, 2361 (1979).
\bibitem{GVH92}
F. Gross, J. W. Van Orden, and K. Holinde,
Phys. Rev. C {\bf 45}, 2094 (1992).
 \bibitem{GS08}
F. Gross and A. Stadler,
Phys. Rev. C {\bf 78}, 014005 (2008).
\bibitem{CDR72}
M. Chemtob, J. W. Durso, and D. O. Riska, {Nucl. Phys.} {\bf B38}, 141 (1972).
\bibitem{JRV75}
 A. D. Jackson, D. O. Riska. and B. Verwest, {Nucl. Phys.} {\bf A249}, 397 (1975).
\bibitem{BJ76}
G. E. Brown and A. D. Jackson, {\it The Nucleon-Nucleon Interaction}
(North-Holland, Amsterdam, 1976).
\bibitem{CV63}
W. N. Cottingham and R. Vinh Mau, {Phys. Rev.} {\bf 130}, 735 (1963).
\bibitem{Cot73} 
W. N. Cottingham, M. Lacombe, B. Loiseau, J. M. Richard, and R. Vinh Mau,  
{Phys. Rev. D} {\bf 8}, 800 (1973).
\bibitem{Vin73} 
R. Vinh Mau,  J. M. Richard, B. Loiseau, M. Lacombe, and W. N. Cottingham,
{Phys. Lett. B} {\bf 44}, 1 (1973).
\bibitem{Lac80} M. Lacombe, B. Loiseau, J. M. Richard, R. Vinh Mau, J. 
C\^{o}t\'{e}, P. Pir\`{e}s, and R. de Tourreil, {Phys. Rev. C} {\bf 21}, 861 
(1980).
\bibitem{Vin79} R. Vinh Mau, 
``The Paris Nucleon-Nucleon Interaction'',
in: {\it Mesons in Nuclei}, Vol. I, eds. M. Rho and D. H. Wilkinson (North-Holland,
Amsterdam, 1979) pp.~151-196.
\bibitem{PL70} 
  M.~H.~Partovi and E.~L.~Lomon,
{Phys.\ Rev.\ D} {\bf 2}, 1999 (1970).
\bibitem{MHE87} 
R. Machleidt, K. Holinde, and Ch. Elster, {Phys. Reports} {\bf 
149}, 1 (1987).
\bibitem{Neg70}
J. W. Negele, Phys. Rec. C {\bf 1}, 1260 (1970).
\bibitem{Coe70} 
  F.~Coester, S.~Cohen, B.~Day and C.~M.~Vincent,
  Phys.\ Rev.\ C {\bf 1}, 769 (1970).
  \bibitem{CDG72} 
  F.~Coester, B.~Day and A.~Goodman,
  Phys.\ Rev.\ C {\bf 5}, 1135 (1972).
  \bibitem{Day83} 
  B.~D.~Day,
  Comments Nucl.\ Part.\ Phys.\  {\bf 11}, 115 (1983).
  \bibitem{BKT74} 
  R.~A.~Brandenburg, Y.~E.~Kim and A.~Tubis,
  Phys.\ Lett.\  {\bf 49B}, 205 (1974).
  \bibitem{BSM77}
  R. A. Brandenburg, P. U. Sauer, and R. Machleidt, 
Z. Physik {\bf A280}, 93 (1977).
\bibitem{PH39} 
  H.~Primakoff and T.~Holstein,
  Phys.\ Rev.\  {\bf 55}, 1218 (1939).
  \bibitem{BM84} 
R.~Brockmann and R.~Machleidt,
  Phys.\ Lett.\  {\bf 149B}, 283 (1984).
  \bibitem{BM90} 
R.~Brockmann and R.~Machleidt,
  Phys.\ Rev.\ C {\bf 42}, 1965 (1990).
  \bibitem{HM87} 
  B.~Ter Haar and R.~Malfliet,
  Phys.\ Rept.\  {\bf 149}, 207 (1987).
  \bibitem{Bro87}
  G. E. Brown, W. Weise, G. Baym, and J. Speth,
  Comments Nucl. Part. Phys. {\bf 17}, 39 (1987).
 \bibitem{AT92} 
  A.~Amorim and J.~A.~Tjon,
  Phys.\ Rev.\ Lett.\  {\bf 68}, 772 (1992).
\bibitem{AS03} 
  D.~Alonso and F.~Sammarruca,
  Phys.\ Rev.\ C {\bf 67}, 054301 (2003).
  \bibitem{MSM17} 
H. Muether, F. Sammarruca, and Z. Ma, 
Int. J. Mod. Phys. E {\bf 26}, 173001 (2017).
  \bibitem{FM57}
J.-I. Fujita and H. Miyazawa, {Prog. Theor. Phys.} {\bf 17}, 360 (1957).
\bibitem{Li08}
Z.~H.~Li, U.~Lombardo, H.-J.~Schulze and W.~Zuo,
  Phys.\ Rev.\ C {\bf 77}, 034316 (2008).
 \bibitem{BB75} 
  S.~Barshay and G.~E.~Brown,
  Phys.\ Rev.\ Lett.\  {\bf 34}, 1106 (1975).
\bibitem{BGG68} 
  G.~E.~Brown, A.~M.~Green and W.~J.~Gerace,
  Nucl.\ Phys.\ A {\bf 115}, 435 (1968).
  \bibitem{Bro72}
  G. E. Brown, 
  Comments Nucl. Part. Phys. {\bf 5}, 6 (1972).
\bibitem{CSB75} 
  S.~A.~Coon, M.~D.~Scadron and B.~R.~Barrett,
  Nucl.\ Phys.\ A {\bf 242}, 467 (1975)
 \bibitem{Coo79} 
  S.~A.~Coon, M.~D.~Scadron, P.~C.~McNamee, B.~R.~Barrett, D.~W.~E.~Blatt and B.~H.~J.~McKellar,
  Nucl.\ Phys.\ A {\bf 317}, 242 (1979).
    \bibitem{CG81} 
  S.~A.~Coon and W.~Gloeckle,
  Phys.\ Rev.\ C {\bf 23}, 1790 (1981).
  \bibitem{CDR83} 
  H.~T.~Coelho, T.~K.~Das and M.~R.~Robilotta,
  Phys.\ Rev.\ C {\bf 28}, 1812 (1983).
  \bibitem{RI84} 
  M.~R.~Robilotta and M.~P.~Isidro Filho,
  Nucl.\ Phys.\ A {\bf 414}, 394 (1984).
  \bibitem{ECM85} 
  R.~G.~Ellis, S.~A.~Coon and B.~H.~J.~McKellar,
  Nucl.\ Phys.\ A {\bf 438}, 631 (1985).
  \bibitem{CP93} 
  S.~A.~Coon and M.~T.~Pena,
  Phys.\ Rev.\ C {\bf 48}, 2559 (1993).
  \bibitem{CPW83} 
  J.~Carlson, V.~R.~Pandharipande and R.~B.~Wiringa,
  Nucl.\ Phys.\ A {\bf 401}, 59 (1983).
  \bibitem{Pud95} 
  B.~S.~Pudliner, V.~R.~Pandharipande, J.~Carlson and R.~B.~Wiringa,
  Phys.\ Rev.\ Lett.\  {\bf 74}, 4396 (1995).
    \bibitem{EMN17}
D. R. Entem, R. Machleidt, and Y. Nosyk, 
{Phys. Rev. C} {\bf 96}, 024004 (2017).
  \bibitem{Sam18} 
  F.~Sammarruca, L.~E.~Marcucci, L.~Coraggio, J.~W.~Holt, N.~Itaco and R.~Machleidt,
  arXiv:1807.06640 [nucl-th].
  \bibitem{Pud97} 
  B.~S.~Pudliner, V.~R.~Pandharipande, J.~Carlson, S.~C.~Pieper and R.~B.~Wiringa,
  Phys.\ Rev.\ C {\bf 56}, 1720 (1997).
  \bibitem{APR98} 
  A.~Akmal, V.~R.~Pandharipande and D.~G.~Ravenhall,
  Phys.\ Rev.\ C {\bf 58}, 1804 (1998).
  \bibitem{Pie01} 
  S.~C.~Pieper, V.~R.~Pandharipande, R.~B.~Wiringa and J.~Carlson,
  Phys.\ Rev.\ C {\bf 64}, 014001 (2001)
  \bibitem{Fri88} 
  J.~L.~Friar, B.~F.~Gibson, G.~L.~Payne and S.~A.~Coon,
  Few-Body Systems {\bf 5}, 13 (1988).
  \bibitem{Pie08}
  S. C. Pieper, AIP Conf. Proc. {\bf 1011}, 143 (2008).
  \bibitem{SW84} 
  B.~D.~Serot and J.~D.~Walecka,
  Adv.\ Nucl.\ Phys.\  {\bf 16}, 1 (1986).
  \bibitem{Rei89} 
  P.~G.~Reinhard,
  Rept.\ Prog.\ Phys.\  {\bf 52}, 439 (1989).
  \bibitem{Bro98}
  B. A. Brown, Phys. Rev. C {\bf 58}, 220 (1998).
 \bibitem{Lon17} 
  D.~Lonardoni, A.~Lovato, S.~C.~Pieper and R.~B.~Wiringa,
  Phys.\ Rev.\ C {\bf 96}, no. 2, 024326 (2017)
  \bibitem{Mar13a} 
  P.~Maris, J.~P.~Vary, S.~Gandolfi, J.~Carlson and S.~C.~Pieper,
  Phys.\ Rev.\ C {\bf 87}, no. 5, 054318 (2013)
  \bibitem{EMW02}
D. R. Entem, R. Machleidt, and H. Witala,
{Phys. Rev. C} {\bf 65}, 064005 (2002).
  \bibitem{Kie10} 
  A.~Kievsky, M.~Viviani, L.~Girlanda and L.~E.~Marcucci,
  Phys.\ Rev.\ C {\bf 81}, 044003 (2010).
\bibitem{HSS83} 
  C.~Hajduk, P.~U.~Sauer and W.~Struve,
  Nucl.\ Phys.\ A {\bf 405}, 581 (1983).
  \bibitem{HSY83} 
  C.~Hajduk, P.~U.~Sauer and S.~N.~Yang,
  Nucl.\ Phys.\ A {\bf 405}, 605 (1983).
  \bibitem{SHS83} 
  W.~Struve, C.~Hajduk and P.~U.~Sauer,
  Nucl.\ Phys.\ A {\bf 405}, 620 (1983).
\bibitem{DMS03} 
  A.~Deltuva, R.~Machleidt and P.~U.~Sauer,
  Phys.\ Rev.\ C {\bf 68}, 024005 (2003).
  \bibitem{PRB92} 
  A.~Picklesimer, R.~A.~Rice and R.~Brandenburg,
  Phys.\ Rev.\ C {\bf 46}, 1178 (1992); and references therein.
\bibitem{PRB95} 
  A.~Picklesimer, R.~A.~Rice and R.~Brandenburg,
  Few Body Syst.\  {\bf 19}, 47 (1995).
  \bibitem{HM77} 
  K.~Holinde and R.~Machleidt,
  Nucl.\ Phys.\ A {\bf 280}, 429 (1977).
\bibitem{MH80} 
  R.~Machleidt and K.~Holinde,
  Nucl.\ Phys.\ A {\bf 350}, 396 (1980).
\bibitem{ME11}
R. Machleidt and D. R. Entem, {Phys Reports} {\bf 503}, 1 (2011).
\bibitem{EHM09}
E.~Epelbaum, H.~W.~Hammer and U.~G.~Mei\ss ner,
  Rev.\ Mod.\ Phys.\  {\bf 81}, 1773 (2009).
    \bibitem{HKK19}
   H.-W.~Hammer, S.~K\H{o}nig and U.~van Kolck,
  ``Nuclear effective field theory: status and perspectives,''
  arXiv:1906.12122 [nucl-th].
  \bibitem{Dri19} 
  C.~Drischler, W.~Haxton, K.~McElvain, E.~Mereghetti, A.~Nicholson, P.~Vranas and A.~Walker-Loud,
  ``Towards grounding nuclear physics in QCD,''
  arXiv:1910.07961 [nucl-th].
\bibitem{KGE12} H. Krebs, A. Gasparyan, and E. Epelbaum, 
{Phys. Rev. C} {\bf 85}, 054006 (2012).
\bibitem{Ent15a}
D. R. Entem, N. Kaiser, R. Machleidt, and Y. Nosyk,
{Phys. Rev. C} {\bf 91}, 014002 (2015).
\bibitem{Ent15b}
D. R. Entem, N. Kaiser, R. Machleidt, and Y. Nosyk,
{Phys. Rev. C} {\bf 92}, 064001 (2015).
\bibitem{Hof15}
M. Hoferichter, J. Ruiz de Elvira, B. Kubis, and U.-G. Mei\ss ner,
{Phys. Rev. Lett.} {\bf 115}, 192301 (2015).
\bibitem{Hof16}
M. Hoferichter, J. Ruiz de Elvira, B. Kubis, and U.-G. Mei\ss ner,
{Phys. Rep.} {\bf 625}, 1 (2016).
\bibitem{BS66} R. Blankenbecler and R. Sugar, 
{Phys.\ Rev.} {\bf 142}, 1051 (1966).
\bibitem{KSW96}
D. B. Kaplan, M. Savage, and M. B. Wise,   
{Nucl. Phys.} {\bf B478}, 629 (1996).
\bibitem{KSW98a}
D. B. Kaplan, M. Savage, and M. B. Wise,   
{Phys. Lett.} {\bf B424}, 390 (1998).
\bibitem{KSW98b}
D. B. Kaplan, M. Savage, and M. B. Wise,   
{Nucl. Phys.} {\bf B534}, 329 (1998).
\bibitem{FMS00} 
  S.~Fleming, T.~Mehen and I.~W.~Stewart,
  Nucl.\ Phys.\ A {\bf 677}, 313 (2000).
\bibitem{NTK05} 
  A.~Nogga, R.~G.~E.~Timmermans, and U.~van Kolck,
  Phys.\ Rev.\ C {\bf 72}, 054006 (2005).
\bibitem{Bir06}
M. C. Birse, 
{Phys. Rev. C} {\bf 74}, 014003 (2006).
\bibitem{Bir07}
M. C. Birse, 
Phys. Rev. C {\bf 76}, 034002 (2007).
\bibitem{Bir08}
M. C. Birse, 
Phys. Rev. C {\bf 77}, 047001 (2008).
\bibitem{Bir11}
M. C. Birse, 
  Phil.\ Trans.\ Roy.\ Soc.\ Lond.\ A {\bf 369}, 2662 (2011).
\bibitem{LY12}
B. Long and C. Yang, 
Phys. Rev. C {\bf 86}, 024001 (2012).
\bibitem{Lon16}
B. Long, Int. J. Mod. Phys. E {\bf 25}, 1641006 (2016).
\bibitem{Val11} 
  M.~P.~Valderrama,
  Phys.\ Rev.\ C {\bf 83}, 024003 (2011).
\bibitem{Val11a} 
  M.~Pavon Valderrama,
  Phys.\ Rev.\ C {\bf 84}, 064002 (2011).
\bibitem{Val16}
M. P. Valderrama, Int. J. Mod. Phys. E {\bf 25}, 1641007 (2016).
\bibitem{Val16a} 
  M.~Pavon Valderrama, M.~Sanchez Sanchez, C.~J.~Yang, B.~Long, J.~Carbonell and U.~van Kolck,
  Phys.\ Rev.\ C {\bf 95}, 054001 (2017).
\bibitem{EGM17} 
  E.~Epelbaum, J.~Gegelia and U.~G.~Mei\ss ner,
  Nucl.\ Phys.\ B {\bf 925}, 161 (2017).
\bibitem{Kon17} 
  S.~K\H{o}nig, H.~W.~Grie\ss hammer, H.-W.~Hammer, and U.~van Kolck,
  Phys.\ Rev.\ Lett.\  {\bf 118}, 202501 (2017).
  \bibitem{Epe18}
   E.~Epelbaum, A.~M.~Gasparyan, J.~Gegelia and U.~G.~Mei\ss ner,
  Eur.\ Phys.\ J.\ A {\bf 54}, 186 (2018).
  \bibitem{Kol19}
U. van Kolck, ``Nuclear Effective Field Theories: The Whole Enchilada'',
talk given at the INT program {\it Nuclear Structure at the Crossroads}, July 1 - August 2, 2019.
\bibitem{Val19}
Manuel Pavon Valderrama,
``Scattering amplitudes versus potentials in nuclear effective field theory:
is there a potential compromise?,''
arXiv:1902.08172 [nucl-th].
\bibitem{EGM00} E. Epelbaum, W. Gl\"ockle, and U.-G. Mei\ss ner,
{Nucl.\ Phys.} {A671}, 295 (2000).
\bibitem{EM03} D. R. Entem and R. Machleidt,
{Phys.\ Rev.\ C} {\bf 68}, 041001 (2003).
\bibitem{EGM05} E. Epelbaum, W. Gl\"ockle, and U.-G. Mei\ss ner,
{Nucl.\ Phys.} {A747}, 362 (2005).
\bibitem{Eks13}
A. Ekstr\H{o}m {\it et al.}, {Phys. Rev. Lett.} {\bf 110}, 192502 (2013).
\bibitem{Gez14}
A.~Gezerlis, I.~Tews, E.~Epelbaum, M.~Freunek, S.~Gandolfi, K.~Hebeler, A.~Nogga, and A.~Schwenk,
{Phys.\ Rev.\ C} {\bf 90}, 054323 (2014).
  \bibitem{Pia15}
 M.~Piarulli, L.~Girlanda, R.~Schiavilla, R.~Navarro P\'{e}rez, J.~E.~Amaro, and E.~Ruiz Arriola,
{Phys.\ Rev.\ C} {\bf 91},  024003 (2015).
  \bibitem{Pia16}
 M.~Piarulli, L.~Girlanda, R.~Schiavilla, A. Kievsky, A. Lovato, L. E. Marcucci, Steven C. Pieper, M. Viviani, and R. B. Wiringa,
{Phys.\ Rev.\ C} {\bf 94},  054007 (2016).
\bibitem{Eks15} 
  A.~Ekstr\H{o}m {\it et al.},
  Phys.\ Rev.\ C {\bf 91}, 051301 (2015)
\bibitem{EKM15}
E. Epelbaum, H. Krebs, U.-G. Mei\ss ner,
{Eur. Phys. J. A} {\bf 51}, 53 (2015);
{Phys. Rev. Lett.} {\bf 115}, 122301 (2015).
  \bibitem{PAA15}
 R.~Navarro P\'{e}rez, J.~E.~Amaro, and E.~Ruiz Arriola,
{Phys.\ Rev.\ C} {\bf 91}, 054002 (2015), and references to the {\it comprehensive} work by the Granada group therein.
\bibitem{Car16} 
  B.~D.~Carlsson {\it et al.},
  Phys.\ Rev.\ X {\bf 6}, 011019 (2016)
\bibitem{RKE18}
P.~Reinert, H.~Krebs and E.~Epelbaum,
  Eur.\ Phys.\ J.\ A {\bf 54}, 86 (2018).
  \bibitem{Eks18} 
  A.~Ekstr\H{o}m, G.~Hagen, T.~D.~Morris, T.~Papenbrock and P.~D.~Schwartz,
  Phys.\ Rev.\ C {\bf 97}, 024332 (2018)
\bibitem{Lep97}
G. P. Lepage,
How to Renormalize the Schr\"odinger Equation,
arXiv:nucl-th/9706029.
\bibitem{Mar13b}
E. Marji {\it et al.}, Phys. Rev. C {\bf 88}, 054002 (2013).
\bibitem{Sto93} V.\ G.\ J.\ Stoks, R.\ A.\ M.\ Klomp, 
M.\ C.\ M.\ Rentmeester, and J.\ J.\ de Swart, 
{ Phys.\ Rev.\ C} {\bf 48}, 792 (1993).
\bibitem{SP07}  W. J. Briscoe, I. I. Strakovsky, and R. L. Workman,
SAID Partial-Wave Analysis Facility, Data Analysis Center,
The George Washington University,  solution SP07 (Spring 2007).
\bibitem{BKM97}
V. Bernard, N. Kaiser, and Ulf-G. Mei\ss ner,
Nucl. Phys. A {\bf 615}, 483 (1997).
\bibitem{Kol94} 
U. van Kolck, {Phys. Rev. C} {\bf 49}, 2932 (1994).
\bibitem{Epe02b} 
E. Epelbaum, A. Nogga, W. Gl\"ockle, H. Kamada, U.-G. Mei\ss ner, and H. Witala,
{Phys. Rev. C} {\bf 66}, 064001 (2002).
\bibitem{Kai00a} N. Kaiser, {Phys.\ Rev.\ C} {\bf 61}, 014003 (2000).
\bibitem{Kai00b} N. Kaiser, {Phys.\ Rev.\ C} {\bf 62}, 024001 (2000).
\bibitem{Ber08}
V. Bernard, E. Epelbaum, H. Krebs, and Ulf-G. Mei\ss ner,
{Phys. Rev. C} {\bf 77}, 064004 (2008).
\bibitem{Ber11}
V. Bernard, E. Epelbaum, H. Krebs, and Ulf-G. Mei\ss ner,
{Phys. Rev. C} {\bf 84}, 054001 (2011).
\bibitem{Epe07}
E. Epelbaum, {Eur. Phys. J.} {\bf A34}, 197 (2007).
\bibitem{KGE13}
H. Krebs, A. Gasparyan, and E. Epelbaum, 
{Phys. Rev. C} {\bf 87}, 054007 (2013).  
\bibitem{GKV11}
L. Girlanda, A. Kievsky, M. Viviani, 
{Phys. Rev. C} {\bf 84}, 014001 (2011).
\bibitem{EM03a}
D. R. Entem and R. Machleidt, unpublished.
\bibitem{Jen11}
U. D. Jentschura, A. Matveev, C. G. Parthey, J. Alnis, R. Pohl, Th. Udem, N. Kolachevsky, 
and T. W. H\H{a}nsch,
Phys. Rev. A {\bf 83}, 042505 (2011).
\bibitem{Cau02}
E. Caurier, P. Navratil, W. E. Ormand, and J. P. Vary,
{Phys. Rev. C} {\bf 66}, 024314 (2002).
 \bibitem{Heb11} 
  K.~Hebeler, S. K. Bogner, R. J. Furnstahl, A. Nogga, and A.~Schwenk,
{Phys.\ Rev.\ C} {\bf 83}, 031301(R) (2011).
\bibitem{Sam12} 
  F.~Sammarruca, B.~Chen, L.~Coraggio, N.~Itaco, and R.~Machleidt,
  Phys.\ Rev.\ C {\bf 86}, 054317 (2012).
  \bibitem{Cor14}
  L. Coraggio, J. W. Holt, N. Itaco, R. Machleidt, L. E. Marcucci, and F. Sammarruca,  
{Phys. Rev. C} {\bf 89}, 044321 (2014).
  \bibitem{Sam15}
  F. Sammarruca, L. Coraggio, J. W. Holt, N. Itaco, R. Machleidt, and L. E. Marcucci,   
{Phys. Rev. C} {\bf 91}, 054311 (2015).
\bibitem{MS16}
R. Machleidt and F. Sammarruca, {Phys. Scr.} {\bf 91}, 083007 (2016).
 \bibitem{FHK99} 
  J.~L.~Friar, D.~Huber and U.~van Kolck,
  Phys.\ Rev.\ C {\bf 59}, 53 (1999).
  \bibitem{CH01} 
  S.~A.~Coon and H.~K.~Han,
  Few Body Syst.\  {\bf 30}, 131 (2001).
\bibitem{Nav07} 
  P.~Navratil,
{ Few Body Syst.}  {\bf 41}, 117 (2007).
\bibitem{Nog06}
A. Nogga, P. Navratil, B. R. Barrett, and J. P. Vary,
{Phys.\ Rev.\ C} {\bf 73}, 064002 (2006).
\bibitem{Mar12} 
  L.~E.~Marcucci, A.~Kievsky, S.~Rosati, R.~Schiavilla, and M.~Viviani,
{Phys.\ Rev.\ Lett.}  {\bf 108}, 052502 (2012).
   \bibitem{NRQ10} 
  P.~Navratil, R.~Roth, and S.~Quaglioni,
{Phys.\ Rev.\ C} {\bf 82}, 034609 (2010).
\bibitem{Viv13}
M. Viviani, L. Girlanda, A. Kievsky, and L. E. Marcucci, 
{Phys. Rev. Lett.} {\bf 111}, 172302 (2013).
\bibitem{Nav07a} 
  P.~Navratil, V.~G.~Gueorguiev, J.~P.~Vary, W.~E.~Ormand and A.~Nogga,
  Phys.\ Rev.\ Lett.\  {\bf 99}, 042501 (2007).
\bibitem{Rot11} 
  R.~Roth, J.~Langhammer, A.~Calci, S.~Binder and P.~Navratil,
  Phys.\ Rev.\ Lett.\  {\bf 107}, 072501 (2011).
\bibitem{Rot12} 
  R.~Roth, S.~Binder, K.~Vobig, A.~Calci, J.~Langhammer and P.~Navratil,
  Phys.\ Rev.\ Lett.\  {\bf 109}, 052501 (2012).
\bibitem{Hag12a}
H. Hagen, M. Hjorth-Jensen, G. R. Jansen, R.  Machleidt, and T. Papenbrock,
{Phys. Rev. Lett.} {\bf 108}, 242501 (2012).
\bibitem{Hag12b}
H. Hagen, M. Hjorth-Jensen, G. R. Jansen, R.  Machleidt, and T. Papenbrock,
{Phys. Rev. Lett.} {\bf 109}, 032502 (2012).
 \bibitem{BNV13} 
  B.~R.~Barrett, P.~Navratil, and J.~P.~Vary,
{Prog.\ Part.\ Nucl.\ Phys.}  {\bf 69}, 131 (2013).
  \bibitem{Her13} 
  H.~Hergert, S.~K.~Bogner, S.~Binder, A.~Calci, J.~Langhammer, R.~Roth, and A.~Schwenk,
{Phys.\ Rev.\ C} {\bf 87}, 034307 (2013).
  \bibitem{Hag14a} 
  G.~Hagen, T.~Papenbrock, M.~Hjorth-Jensen, and D.~J.~Dean,
{Rept.\ Prog.\ Phys.}  {\bf 77}, 096302 (2014).
\bibitem{Bin14} 
  S.~Binder, J.~Langhammer, A.~Calci and R.~Roth,
  Phys.\ Lett.\ B {\bf 736}, 119 (2014).
\bibitem{HJP16} 
  G.~Hagen, G.~R.~Jansen and T.~Papenbrock,
  Phys.\ Rev.\ Lett.\  {\bf 117}, 172501 (2016).
\bibitem{Sim16} 
  J.~Simonis, K.~Hebeler, J.~D.~Holt, J.~Menendez and A.~Schwenk,
  Phys.\ Rev.\ C {\bf 93}, 011302 (2016)
  \bibitem{Sim17} 
  J.~Simonis, S.~R.~Stroberg, K.~Hebeler, J.~D.~Holt, and A.~Schwenk,
{Phys. Rev. C} {\bf 96}, 014303 (2017)
 \bibitem{Mor18} 
T. D. Morris,  J.~Simonis, S.~R.~Stroberg, C. Stumpf, G. Hagen, J. D. Holt, 
G. R. Jansen, T. Papenbrock, R. Roth, and A.~Schwenk,
  Phys. Rev. Lett. {\bf 120}, 152503 (2018).
  \bibitem{Som19} 
  V.~Soma, P.~Navratil, F.~Raimondi, C.~Barbieri and T.~Duguet,
  arXiv:1907.09790 [nucl-th].
   \bibitem{HS10} 
  K.~Hebeler and A.~Schwenk,
{Phys.\ Rev.\ C} {\bf 82}, 014314 (2010).
  \bibitem{Hag14b} 
  G.~Hagen, T.~Papenbrock, A.~Ekstr\H{o}m, K.~A.~Wendt, G.~Baardsen, S.~Gandolfi, M.~Hjorth-Jensen, and C.~J.~Horowitz,
{Phys.\ Rev.\ C} {\bf 89}, 014319 (2014).
  \bibitem{Cor13} 
L. Coraggio, J. W. Holt, N. Itaco, R. Machleidt, and F. Sammarruca,  
{Phys. Rev. C} {\bf 87}, 014322 (2013).
  \bibitem{Gol14}
J.~Golak {\it et al.},
{Eur.\ Phys.\ J.\ A} {\bf 50}, 177 (2014).
  \bibitem{Kru13} 
  T. Kr\H{u}ger, I.~Tews, K Hebeler, and A.~Schwenk,
{Phys.\ Rev.\ C} {\bf 88}, 025802 (2013).
  \bibitem{Dri16} 
 C. Drischler, A. Carbone, K. Hebeler, and A.~Schwenk,
{Phys.\ Rev.\ C}  {\bf 94}, 054307 (2016).
\bibitem{Heb15} 
  K.~Hebeler, H.~Krebs, E.~Epelbaum, J.~Golak and R.~Skibinski,
  Phys.\ Rev.\ C {\bf 91}, 044001 (2015).
\bibitem{DHS19} 
  C.~Drischler, K.~Hebeler and A.~Schwenk,
  Phys.\ Rev.\ Lett.\  {\bf 122}, 042501 (2019)
\bibitem{Hop19} 
  J.~Hoppe, C.~Drischler, K.~Hebeler, A.~Schwenk and J.~Simonis,
  Phys.\ Rev.\ C {\bf 100}, 024318 (2019).
\bibitem{Rot19}
  T.~H\H{u}ther, K.~Vobig, K.~Hebeler, R.~Machleidt, and R.~Roth,
  Family of Chiral Two- plus Three-Nucleon Interactions for Accurate Nuclear Structure Studies,
  arXiv:1911.04955 [nucl-th].
 \bibitem{Gir19}
L. Girlanda, A. Kievsky, M. Viviani, and L. E. Marcucci,
Phys. Rev. C {\bf 99}, 054003 (2019).
\bibitem{Shi95}
S. Shimizu {\it et al.}, Phys. Rev. C 52, 1193 (1995).
\bibitem{Roz06}
D. Rozpedzik, J. Golak, R. Skibinski, H. Witala, W. Gl\"ockle, E. Epelbaum,  A. Nogga, and H. Kamada,
{ Acta Phys. Polon.} {\bf B37}, 2889 (2006);
arXiv:nucl-th/0606017.
\bibitem{Kai12}
N. Kaiser, { Eur. Phys. J. A} {\bf 48}, 135 (2012).
\bibitem{KM16}
N. Kaiser and R. Milkus, {Eur. Phys. J. A} {\bf 52}, 4 (2016).
hys. Rev. C {\bf 92}, 064001 (2015).
\bibitem{FPW15} 
R.~J. Furnstahl, D.~R. Phillips, and S. Wesolowski,
J. Phys. G {\bf 42}, 034028 (2015).
\end{thebibliography}
%

\end{document}